\pdfoutput=1
\documentclass[11pt]{article}
\usepackage{arxiv}
\usepackage[hidelinks]{hyperref}
\usepackage[utf8]{inputenc}
\usepackage[T1]{fontenc}
\usepackage{orcidlink}
\usepackage{subcaption}
\usepackage[group-separator={,}]{siunitx}
\usepackage{graphics}
\usepackage{graphicx}
\DeclareGraphicsExtensions{.pdf,.png}
\graphicspath{ {./figures} }

\usepackage{authblk}

\newcommand{\NumberOfDevicesPreCOVID}{135271}
\newcommand{\NumberOfDevicesCOVID}{35894}

\newcommand{\UniqueDevicesInTheTwoDataSet}{161294}

\newcommand{\TrunkLength}{27.2}
\newcommand{\MotorwayLength}{12.6}

\usepackage{glossaries}

\makeglossaries
\newacronym{BCR}{BCR}{Barrier Crossing Ratio}
\newacronym{GIS}{GIS}{Geographic Information System}
\newacronym{GPS}{GPS}{Global Positioning System}
\newacronym{HCSO}{HCSO}{Hungarian Central Statistical Office}
\newacronym{OLS}{OLS}{Ordinary Least Squares}
\newacronym{OSM}{OSM}{OpenStreetMap}
\newacronym{POI}{POI}{Point of Interest}
\newacronym{SAD}{SAD}{Symmetric Area Difference}
\newacronym{WGS84}{WGS 84}{World Geodetic System, also known as EPSG:4326}

\usepackage{caption}
\usepackage{booktabs}
\usepackage{tabularx}

\usepackage[style=numeric,sorting=none,defernumbers=true]{biblatex}
\usepackage{xurl}
\title{Quantifying Barriers of Urban Mobility}

\author[1,2,*]{Gergő Pintér \orcidlink{0000-0003-4731-3816}}
\author[1,2,3,4]{Balázs Lengyel \orcidlink{0000-0001-5196-5599}}
\affil[1]{ANETI Lab, Corvinus University of Budapest, Budapest, 1093, Hungary and HUN-REN Centre for Economic and Regional Studies, Budapest, 1097, Hungary}
\affil[2]{Corvinus Institute for Advanced Studies, Corvinus University of Budapest, Budapest, 1093, Hungary}
\affil[3]{Institute of Economics, HUN-REN Centre for Economic and Regional Studies, Budapest, 1097, Hungary}
\affil[4]{Institute for Data Analytics and Information Systems, Corvinus University of Budapest, Budapest, 1093, Hungary}

\affil[*]{Corresponding author: \href{mailto:gergo.pinter@uni-corvinus.hu}{gergo.pinter@uni-corvinus.hu}}
\date{}

\begin{document}

\flushbottom
\maketitle

\begin{abstract}
    Barriers in cities, such as administrative boundaries, natural obstacles, railways or major roads are thought to induce segregation. However, the empirical knowledge about this phenomenon is limited. Here, we present a network science framework to assess barriers to urban mobility along their hierarchy, across residential areas and visited amenities. Using GPS mobility data, we construct a network of blocks from the sequence of individual stays in a major European city. A community detection algorithm allows us to partition this network into non-overlapping areas of dense mobility clusters, in which the effect of transportation hubs can be tuned with a parameter. We apply the Symmetric Area Difference index to quantify the overlap between these mobility clusters and the polygons of urban area separated by barriers. Reducing the effect of transportation hubs results in smaller scale mobility clusters that fit better to lower rank administrative or road barriers compared to their higher rank pairs. We find that characteristic urban barriers can replace each other in dividing mobility clusters of different scales. Next, we define the Barrier Crossing Ratio, the fraction of barrier crossings that bridge mobility clusters. The decomposition of this indicator by origins and destinations suggests a significantly higher impact of barriers on those who live closer to the city center and smaller impact on visits to complex amenities. These results contribute to the ongoing discourse on urban segregation, emphasizing the importance of barriers to urban mobility in shaping interactions and mixing.
\end{abstract}

\begin{refsection}
\section*{Introduction}
\label{sec:introduction}

%

Mixing people is essential for cities' economic progress; yet, most interaction happen within neighborhoods \cite{jacobs2016death}. Residential segregation is known to worsen economic opportunities and health outcomes of the poor \cite{chetty2016effects, diez2001investigating}, is associated with high crime levels \cite{sampson1999systematic}, and has been found to hinder economic growth of cities \cite{li2013residential}.
Building on recent developments in urban mobility research that leverages high resolution location data retrieved from mobile devices \cite{gonzalez2008nature, alessandretti2020nature, schlapfer2021universal}, segregation inquiry is increasingly moving to understand mixing beyond residential areas. This literature has documented in detail that social mixing varies by locations and amenities in the city \cite{athey2021estimating, fan2023diversity, yabe2023behavioral, juhasz2023amenity, dong2020epj, wang2018pnas} and is influenced by mobility policies \cite{abbiasov202215, nilforoshan2023human}. However, the role of urban barriers, despite their direct impact on mobility \cite{jin2021identifying}, has remained an unsolved puzzle.

Major roads are great examples of this puzzle, as these were built to increase accessibility, but roads connecting distant places can be obstacles for local mobility \cite{jesus2022barrier}. Research in ``Road Ecology'' investigates the impact of roads on wildlife \cite{coffin2007roadkill}, and has led to different types of mitigation measures in the last decades to reduce the number of roadkill, including building tunnels and bridges for mammals \cite{villalobos2022wildlife} or amphibians \cite{helldin2019effectiveness}. The barrier effect (also known as ``community severance'') is also examined in human environments, but are mainly conducted via surveys \cite{anciaes2020comprehensive,emmanouilidis2022settlements, jesus2022barrier} and focus on a specific part of a city \cite{van2020missing, jesus2022barrier}.
Further characteristic physical barriers like railways, parks or rivers 
are natural borders between neighborhoods \cite{park2015neighborhood} and can amplify segregation by hindering mixing and interaction across separated areas \cite{ananat2011wrong, toth2021naturecomm, aiello2024urban}.
However, the questions of which barriers limits mixing, where in the city and for whom, have not been answered.

In this paper, we contribute a network science framework to evaluate the role of barriers in urban mobility. The basis of the approach is a widely used community detection procedure that, in our case, groups together locations by maximizing the density of mobility flows within groups \cite{blondel2008fast, fortunato202220}. Compared to other approaches that evaluate barriers by quantifying the probability of mobility for pairs of locations, community detection reveals clustering relationships among locations acknowledging that individuals often are present in more than two locations during the day and their mobility flows contains more than one network edge \cite{jin2021identifying}. Community detection on spatial networks can reveal the role of barriers by analyzing how clusters of such complex mobility flows fit to these boundaries. The method can be fine-tuned to decrease the impact of large mobility hubs and increase the importance of local mobility flows that can help find the scales of clustered mobility links that have the best fit to urban barriers \cite{alessandretti2020nature, expert2011uncovering}.


The method is demonstrated with an empirical exercise based on a \acrshort{GPS} mobility data set that contains \num{\UniqueDevicesInTheTwoDataSet} mobile phone devices in Budapest, Hungary over 6 months before and 6 months during the COVID-19 pandemic. From the geolocation and time-stamp of pings, we construct a mobility network, in which nodes are blocks where individuals spent at least 15 minutes and edges are weighted by the number of individual mobility events between the blocks during the day. We first assess the role of various administrative and physical barriers as obstacles to mobility with a gravity model. Next, we run the Louvain community detection algorithm \cite{blondel2008fast} on the mobility network that helps us identify dense mobility clusters. By increasing the resolution parameter of the algorithm, we mitigate the dominance of mobility hubs and group blocks into clusters of relatively smaller sizes. Then, we identify polygons of residential areas bounded by characteristic barriers and calculate the \acrfull{SAD} indicator that is a global indicator to quantify how mobility clusters align with boundaries. Finally, we compute the \acrfull{BCR} to capture the relationship between mobility events across urban barriers and detected mobility clusters that can be decomposed by origins and destinations. This procedure and the novel indicators enable us to quantify the impact of urban barriers on mobility across the hierarchies of barriers, across residential locations and visited amenities.

We find that administrative barriers, rivers, and roads significantly decrease mobility but their impact varies across the city. Increasing the resolution parameter, we observe an improved alignment -- measured by the SAD indicator -- with lower rank administrative barriers and secondary roads, while the fit with higher rank administrative boundaries and major roads improves only up to a certain point. We demonstrate that this is due to a shift among barrier as we mitigate the role of transportation hubs, characteristics urban barriers replace each other in dividing mobility clusters. Next, we decompose the BCR indicator by home locations and find that the impact of barriers depends on the residential areas and visited amenity locations. Mobility clusters of urban dwellers fit to all barriers significantly better than the mobility of residents who commute to the city from its agglomeration suggesting that barriers have a higher effect on local mobility than on those that link distant places. The measure also implies that barriers are not likely to impact mobility to amenities that are rarely found elsewhere in the city, except the river that is the most significant boundary of mobility. Repeating the exercise for Nagoya, Japan using a distinct data set, we find that fitting the mobility clusters to higher versus lower level urban barriers can reveal general patterns across cities. However, the method is also able to show differences that we show by contrasting barrier impact by home location.


\section*{Results}
\label{sec:results}

\subsection*{Empirical framework}
\begin{figure}[t]
    \centering
    \begin{subfigure}[t]{0.325\linewidth}
        \includegraphics[width=\linewidth]{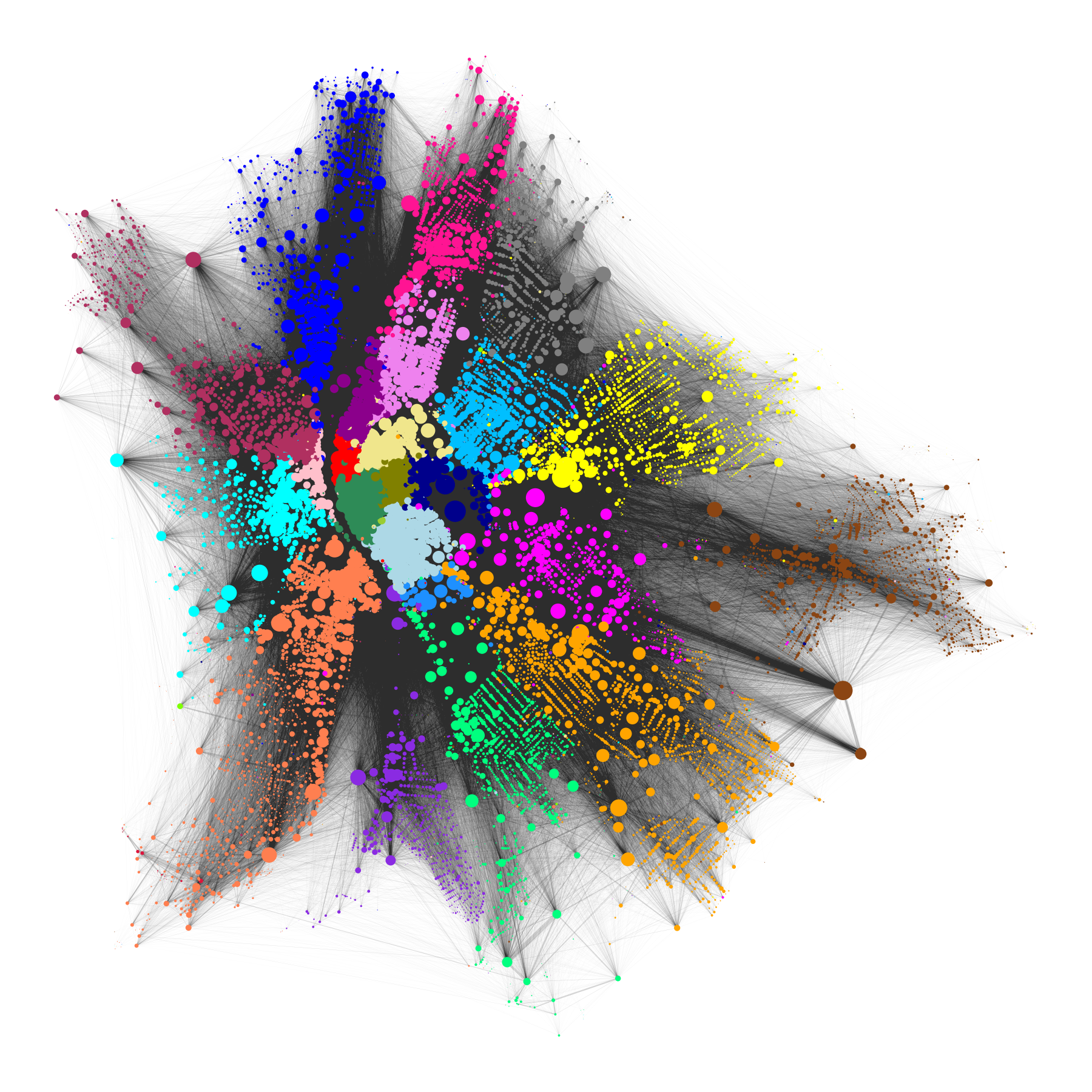}
        \captionsetup{position=bottom,justification=centering}
        \caption{}
        \label{fig:network}
    \end{subfigure}
    \hfill
    \begin{subfigure}[t]{0.325\linewidth}
        \includegraphics[width=\linewidth]{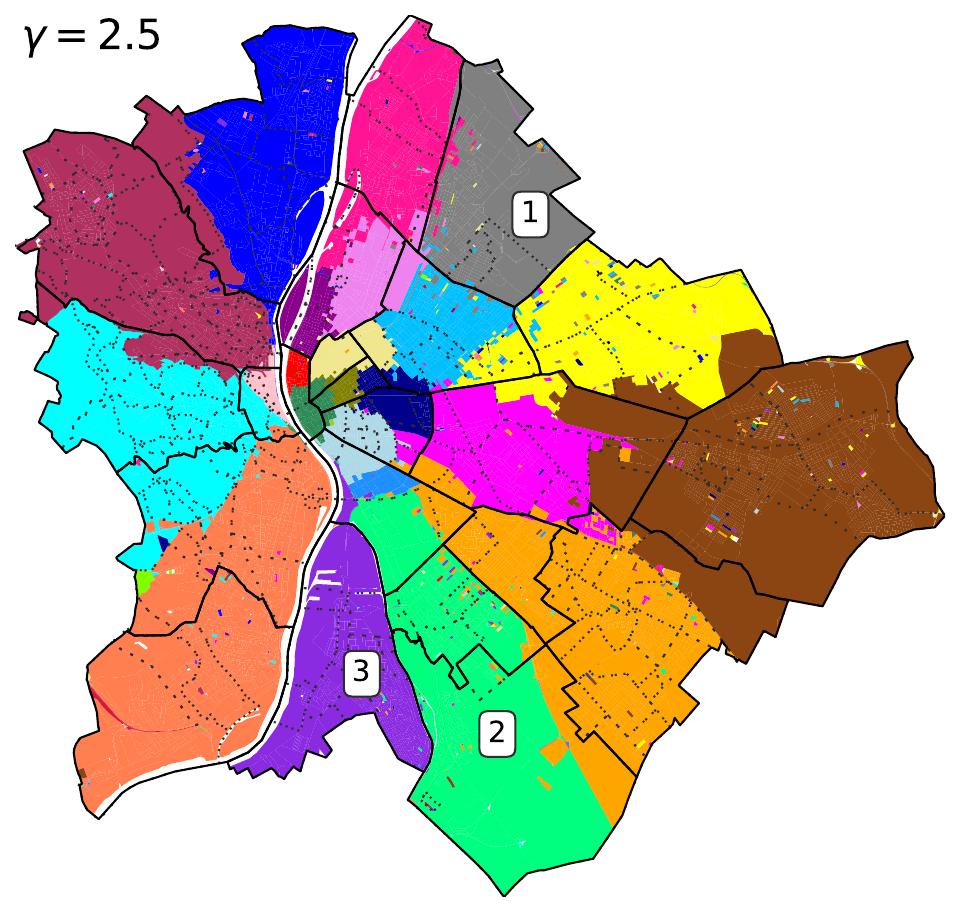}
        \captionsetup{position=bottom,justification=centering}
        \caption{}
        \label{fig:louvain_resolution2.5_two_level_districts}
    \end{subfigure}
    \hfill
    \begin{subfigure}[t]{0.325\linewidth}
        \includegraphics[width=\linewidth]{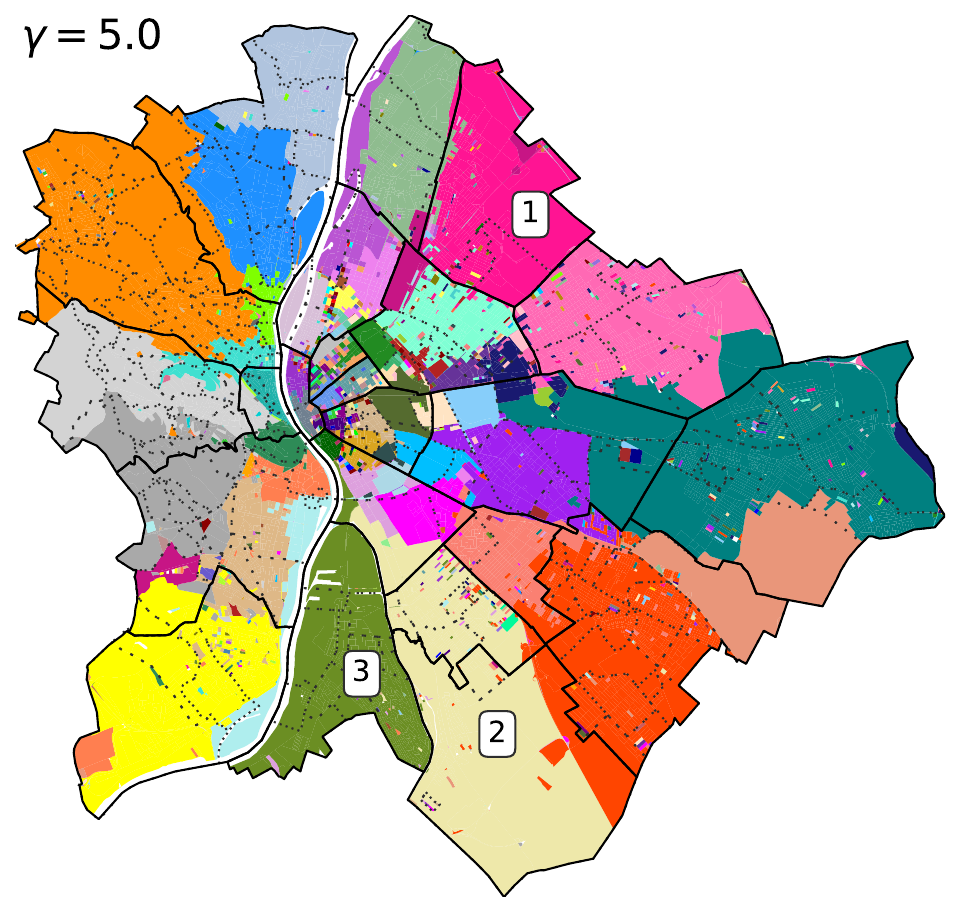}
        \captionsetup{position=bottom,justification=centering}
        \caption{}
        \label{fig:louvain_resolution5.0_two_level_districts}
    \end{subfigure}
    \vfill
    \vspace{.25em}
    \begin{subfigure}[t]{0.325\linewidth}
        \includegraphics[trim={3.0cm 0cm 0 0cm},clip,width=\linewidth]{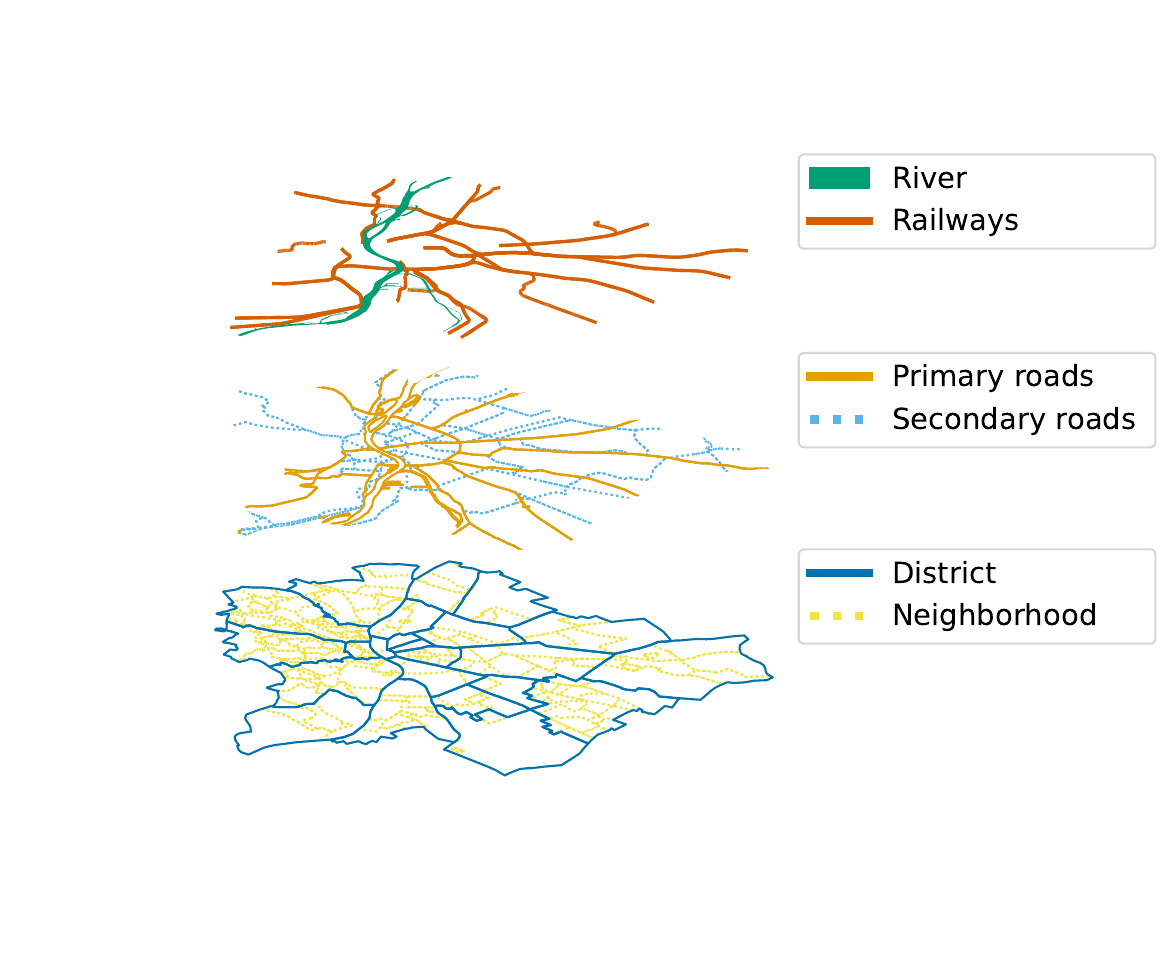}
        \captionsetup{position=bottom,justification=centering}
        \caption{}
        \label{fig:barriers_3d}
    \end{subfigure}
    \hfill
    \begin{subfigure}[t]{0.325\linewidth}
        \includegraphics[width=\linewidth]{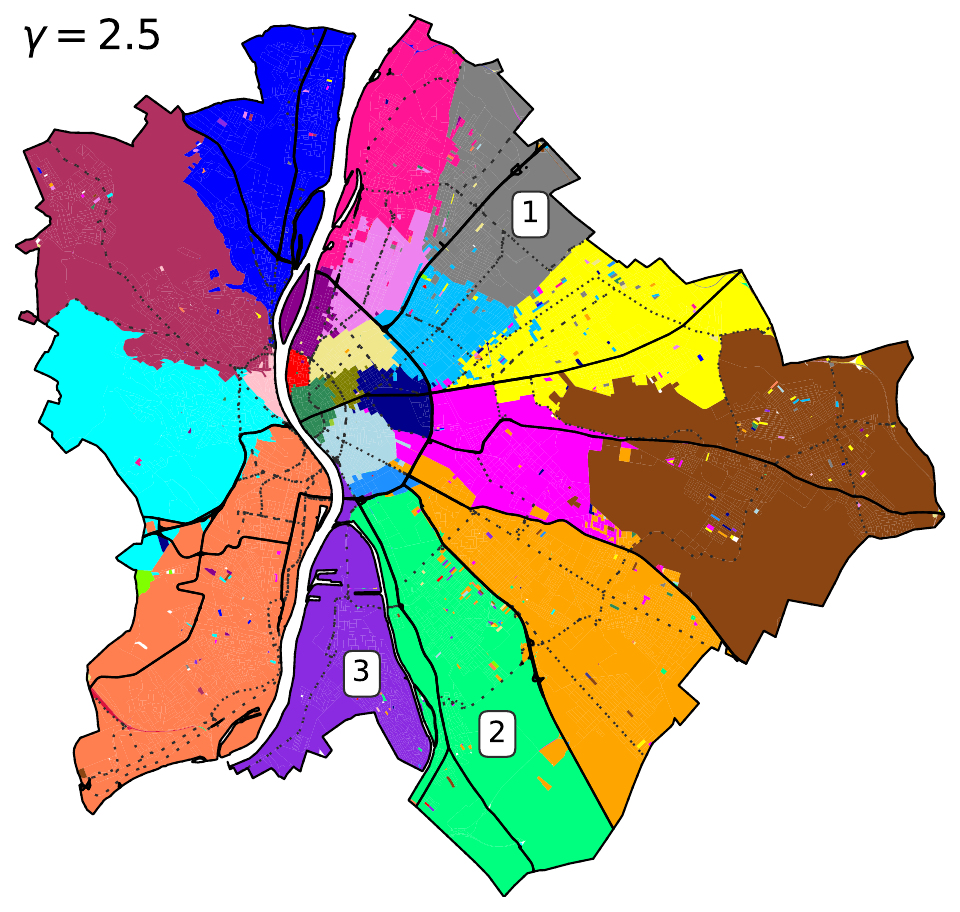}
        \captionsetup{position=bottom,justification=centering}
        \caption{}
        \label{fig:louvain_resolution2.5_two_level_barriers}
    \end{subfigure}
    \hfill
    \begin{subfigure}[t]{0.325\linewidth}
        \includegraphics[width=\linewidth]{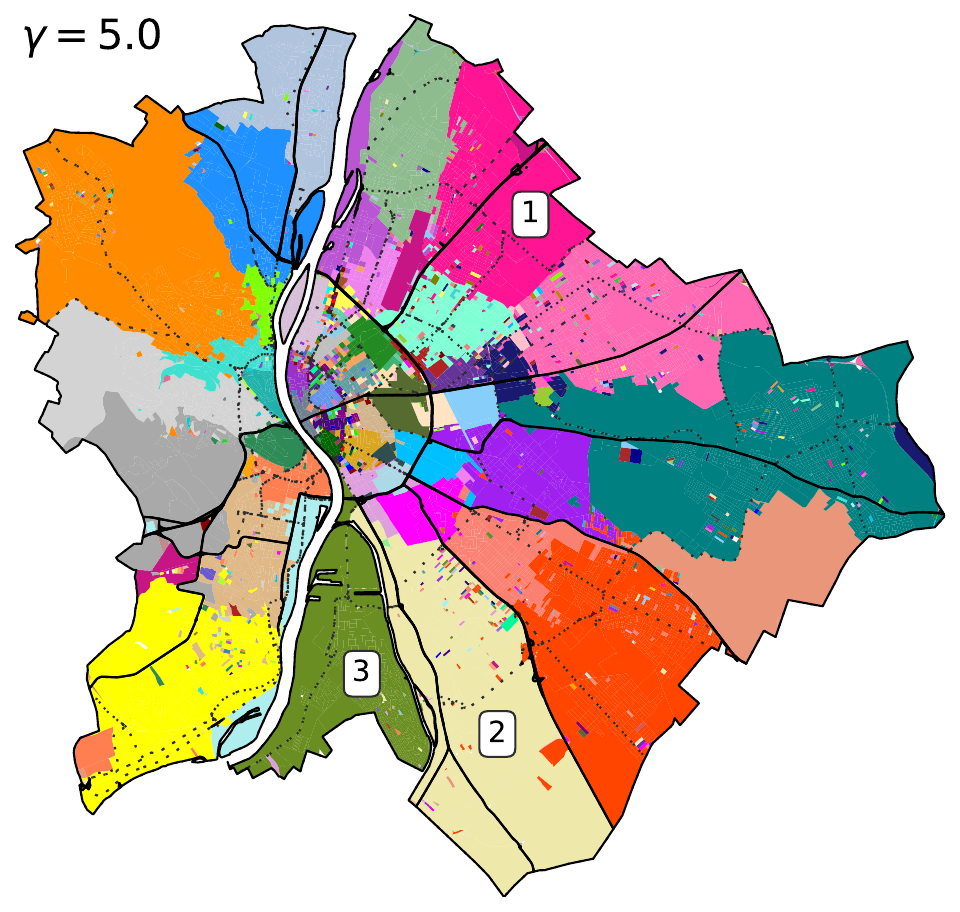}
        \captionsetup{position=bottom,justification=centering}
        \caption{}
        \label{fig:louvain_resolution5.0_two_level_barriers}
    \end{subfigure}
    \caption{
        Illustration of the mobility network and the research problem.
        The observed mobility network (\textbf{\subref{fig:network}}), where the node sizes are proportional to the degree of a node and the colors represent the detected communities at resolution 2.5.
        The ``layers'' of the barriers (\textbf{\subref{fig:barriers_3d}}) analyzed in the study include relatively rare natural (river) and built (railways) barriers and relatively frequent ones that also have hierarchical orders like roads (primary and secondary roads) or boundaries administrative urban units (districts and neighborhoods).
        Mobility clusters detected with resolution parameter $\gamma=2.5$ (\textbf{\subref{fig:louvain_resolution2.5_two_level_districts}}, \textbf{\subref{fig:louvain_resolution2.5_two_level_barriers}})
        and $\gamma=5$ (\textbf{\subref{fig:louvain_resolution5.0_two_level_districts}}, \textbf{\subref{fig:louvain_resolution5.0_two_level_barriers}})
        compared to administrative boundaries (\textbf{\subref{fig:louvain_resolution2.5_two_level_districts}}, \textbf{\subref{fig:louvain_resolution5.0_two_level_districts}}) and to roads (\textbf{\subref{fig:louvain_resolution2.5_two_level_barriers}}, \textbf{\subref{fig:louvain_resolution5.0_two_level_barriers}}).
    }
    \label{fig:stream}
\end{figure}

Our GPS mobility data contains \num{\NumberOfDevicesPreCOVID} devices in Budapest that we aggregate to a 6 months period starting from September 2019 and \num{\NumberOfDevicesCOVID} devices that are aggregated to 6 months time window starting from November 2020 that contains the second and third waves of COVID-19 contagions in Hungary (see Figure~\ref{si:fig:deaths} of the Supplementary Section~\ref{si:sec:covid}). The data consist of the location coordinates and time stamp of pings generated by mobile apps. The analysis is restricted to those mobile phones that have at least 20 pings, similarly to \cite{juhasz2023amenity}.
Then, we construct a mobility network that connects blocks across the city following two steps. First, we identify blocks where individuals spend at least 15 minutes by a stop-detection algorithm \cite{aslak2020infostop} that clusters the ping sequences of single mobile phones in space and time \cite{juhasz2023amenity}. Second, we link those blocks that individuals visit subsequently during the day and generate a network in which the edge weight is the number of individual mobility flows over a 6-month period, either before the COVID-19 pandemic (see a representation of this network in Figure~\ref{fig:stream}a) or during the COVID-19 restrictions.
Next, we identify home location of every phone as the most frequently visited stop in the evening and during nighttime. A home is identified for the entire period only if it appears to be home for at least 10 days.
Finally, to capture barriers, we apply the openly available OpenStreetMap data for Budapest that contains polygons of rivers, railroads, and hierarchies of administrative boundaries and roads (Figure~\ref{fig:stream}d). Details of mobility and urban barrier data, the cleaning process, and the network generation procedure are explained in \nameref{sec:materials_and_methods}.

Previous research investigated the impact of barriers on days of mobility links \cite{jin2021identifying}. This approach has limitations in capturing the full complexity of mobility networks that can be important since mobility during the day typically include more stops that can form triangles, and more complex patterns. Here, we apply a widely used community detection method on the mobility network that can provide new insights by detecting dense mobility clusters that contain most of the mobility events and can represent silos of highly likely mobility links. We use the Louvain algorithm \cite{blondel2008fast} that partitions nodes by maximizing the modularity index that is calculated using the formula (Eq.~\ref{eq:modularity_index}).

\begin{equation}
 Q=\dfrac{1}{2m}\sum_{ij}\Big[ A_{ij}-\gamma\dfrac{k_{i}k_{j}}{2m} \Big]\delta(c_i,c_j),
 \label{eq:modularity_index}
\end{equation}

where $A_{ij}$ is the weight of the link between block $i$ and $j$, $k_i$ is the sum of the weights of links of block $i$,
$c_i$ and $c_j$ denote the community of blocks $i$ and $j$, $\delta$ is equal to 1 if $c_i=c_j$ and zero otherwise and $m$ is the sum of the weights in the network.
Increasing the resolution parameter $\gamma>0$, one can get smaller communities, while a small $\gamma$ provides large communities. In our mobility network, this procedure mitigates the role of flows between large mobility hubs and enables the emergence of local mobility clusters. Previous research has shown that community detection on spatial networks forms spatially segregated partitions \cite{expert2011uncovering, sobolevsky2013delineating, lengyel2015geographies, grauwin2017identifying} and increasing the resolution parameter has been applied to partition labor mobility networks into occupations with similar skill requirements \cite{adam2018detecting} and also to find the scale of administrative boundaries that fragment commuting networks across cities \cite{o2022modular}.

However, detecting the role of barriers in urban mobility networks requires further analyzes, since a variety of boundaries may hinder mobility and thus partition the network, while their impact may differ across the city in non-trivial ways. In our Budapest case, for example, administrative barriers define communities in the Northeastern part of the city where major roads have seemingly no impact on the community structure at $\gamma=2.5$ and $\gamma=5$ (Figure~\ref{fig:stream} e and f/marker 1). Marker 1 denotes a community that fits to the District 15 in both resolutions but is also crossed by freeway M3, a multilane road with physical separation from its surrounding that connect downtown of Budapest to cities in East Hungary.
On the contrary, a major road separates mobility communities in the Southeast of the city (Figure~\ref{fig:stream}/marker 2).
The bright green community, denoted with marker 2, does not fit to administrative boundaries but is perfectly limited by the freeway M5 and by the river Danube. Finally, physical and administrative boundaries might align, as we see in the case of District 21 in Budapest (Figure~\ref{fig:stream} b, c, e, f/ marker 3). This district that is depicted by marker 3 is bounded by the river Danube and its fork and thus, it constitutes a community in the mobility network not only at the two displayed values of the resolution parameter but in further settings too (see Supplementary Section \ref{si:sec:symmetric_area_difference_tendency} for details).
Note that the features of a physical barrier can affect its community-forming power as Figure~\ref{fig:stream} show lower order roads (displayed by dotted lines) seem to have no impact on communities at lower resolution, but higher order roads do (e.g., Figure~\ref{fig:stream}/marker 2 or 3). As expected, the communities detected with higher resolution parameter are smaller. However, this trend is more true in the downtown, while communities with the markers 1-3 remained almost the same.

Regarding the heterogeneity of the barrier impact, it is unclear how long-distance versus local mobility is affected. To provide a better understanding, we first compare the role of higher- versus lower rankings of physical or administrative barriers. Next, we test whether those who commute or travel from further away are less likely to be affected by barriers. We finally complement the analysis by checking how visited locations differentiate the barrier impact.

\begin{figure}[!b]
    \centering
    \begin{subfigure}[t]{0.4\linewidth}
        \includegraphics[width=\linewidth]{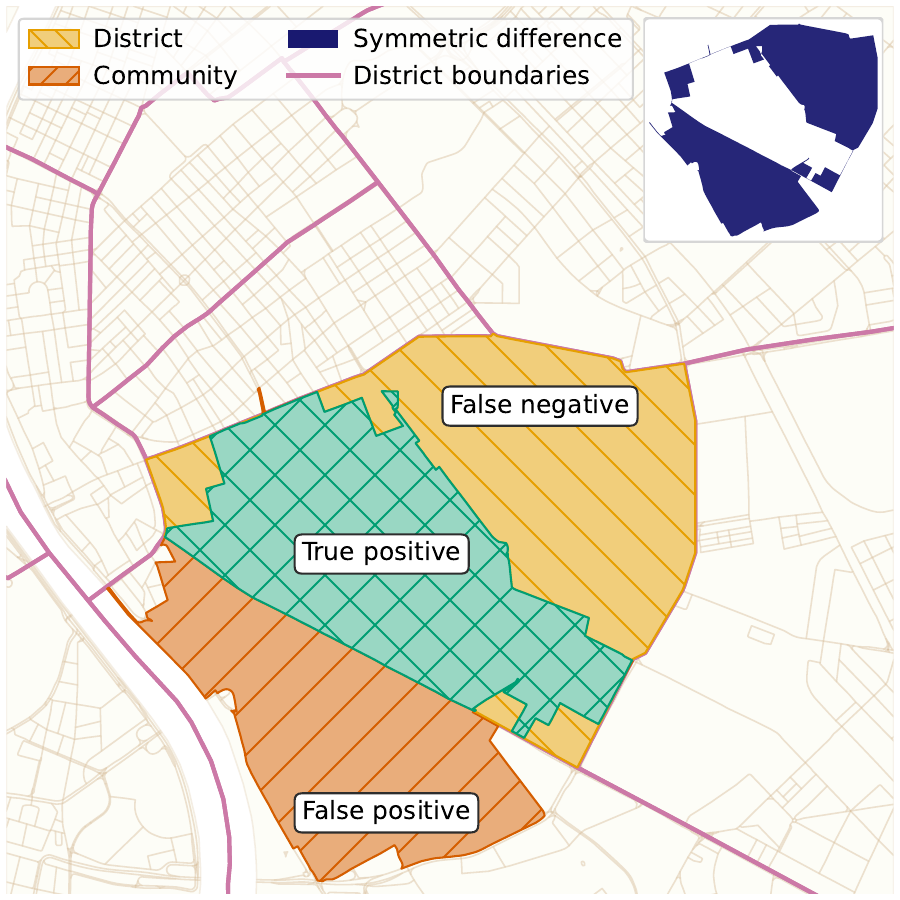}
        \captionsetup{position=bottom,justification=centering}
        \caption{}
        \label{fig:community_overlaps_district}
    \end{subfigure}
~
    \begin{subfigure}[t]{0.4\linewidth}
        \includegraphics[width=\linewidth]{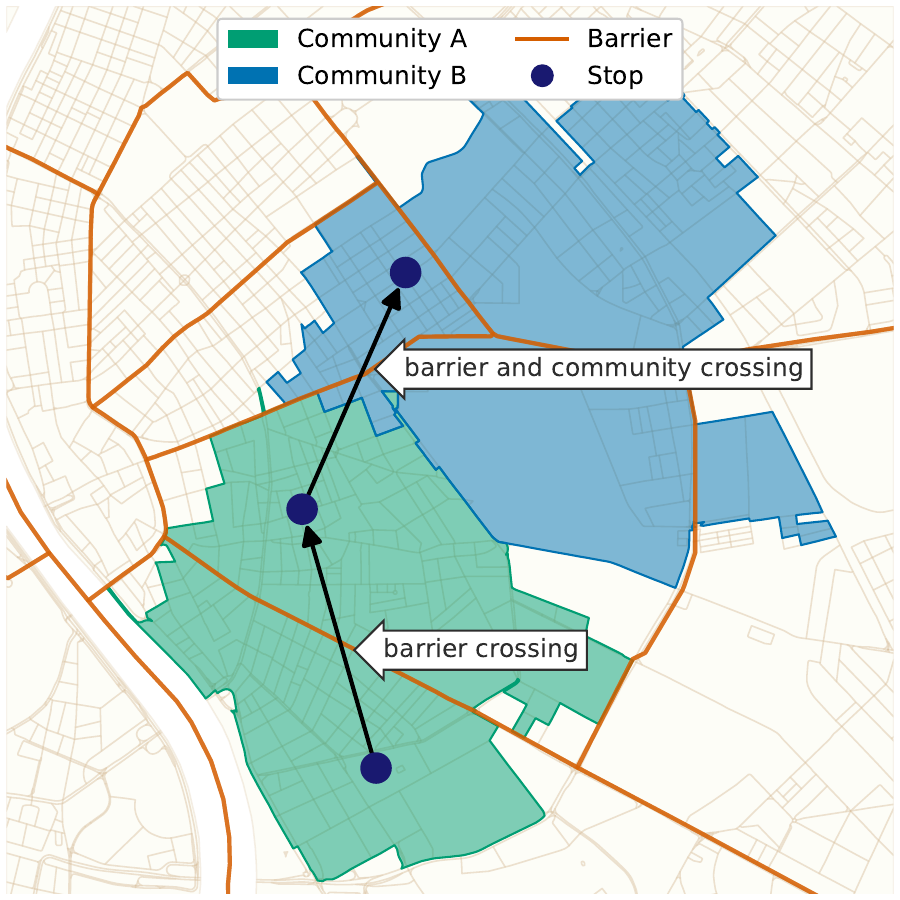}
        \captionsetup{position=bottom,justification=centering}
        \caption{}
        \label{fig:barrier_and_community_crossing}
    \end{subfigure}
    \caption{
    Illustration of indicators used to assess the fit of detected mobility communities to urban barriers. The \acrfull{SAD} (\textbf{\subref{fig:community_overlaps_district}}) measures the sum of the areas that are not overlapping between detected mobility clusters and the areas delineated by the given barrier.
        The \acrfull{BCR} quantifies the fraction of those trips that cross barriers and mobility clusters as well among all barrier crossing movements (\textbf{\subref{fig:barrier_and_community_crossing}}).
    }
    \label{fig:communities_and_barriers}
\end{figure}

We propose two indicators to evaluate how administrative and physical boundaries fragment mobility networks by quantifying the fit of mobility clusters detected by the Louvain method at various levels of the resolution parameter. The \acrfull{SAD} index, $SAD_\gamma$ captures the lack of overlap between the polygons defined by mobility clusters and the polygons delineated by characteristic urban barriers for every resolution $\gamma$ (Figure~\ref{fig:communities_and_barriers}a). Formally,

\begin{equation}
    \label{eq:SAD}
    SAD_\gamma=\dfrac{1}{n}\sum_{p_{\gamma}q} (A_{p_\gamma} \cup A_q) \setminus  (A_{p_\gamma} \cap A_q),
\end{equation}

where $A_p$ and $A_q$ are the area sizes of the overlapping mobility cluster polygons $p$ and urban area polygons $q$ and $n$ is the number of iterations of the Louvain community detection at resolution $\gamma$. Lower values of this index mean a better fit of the detected mobility clusters to urban boundaries.

Next, we define the \acrfull{BCR} index, $BCR_\gamma$ measures the relationship between those mobility events that cross barriers and those that cross barriers and detected mobility clusters as well (Figure~\ref{fig:communities_and_barriers}b). We calculate the fraction of barrier and cluster crosses among all barrier crossings, and take the inverse of this ratio. This indicator is formalized by

\begin{equation}
    \label{eq:BCR}
BCR_\gamma = \dfrac{1}{n}\frac{\sum_{m} \text{{CB}}(M_k^i, U)}{\sum_{m} \left(\text{{CB}}(M_k^i, U) \times \text{{CC}}(M_k^i, C_{\gamma})\right)},
\end{equation}

where $m$ is the total number of mobility edges, $CB$ is a binary function that evaluates to 1 if $M_k^i$, the $k^{th}$ mobility edge from block $i$, crosses an urban barrier $U$ and 0 otherwise, while the function $CC$ takes the value of 1 if $M_k^i$ crosses mobility clusters and $n$ is the number of Louvain iterations at resolution $\gamma$. Since barrier crossings are constant, a decreasing value of the indicator along increasing $\gamma$ signals that border crossings align with mobility cluster crossings.

Both measures are calculated for $n=10$ iterations of the community partitioning for all decimal values of $\gamma \in [1,10]$. Further details about the indices can be found in the \nameref{sec:materials_and_methods}.

\subsection*{Assessing the community fit to urban barriers with Symmetric Area Difference}
\label{sec:fitting_with_sad}

\begin{figure}[!b]
    \centering
    \begin{subfigure}[t]{0.325\linewidth}
        \includegraphics[width=\linewidth]{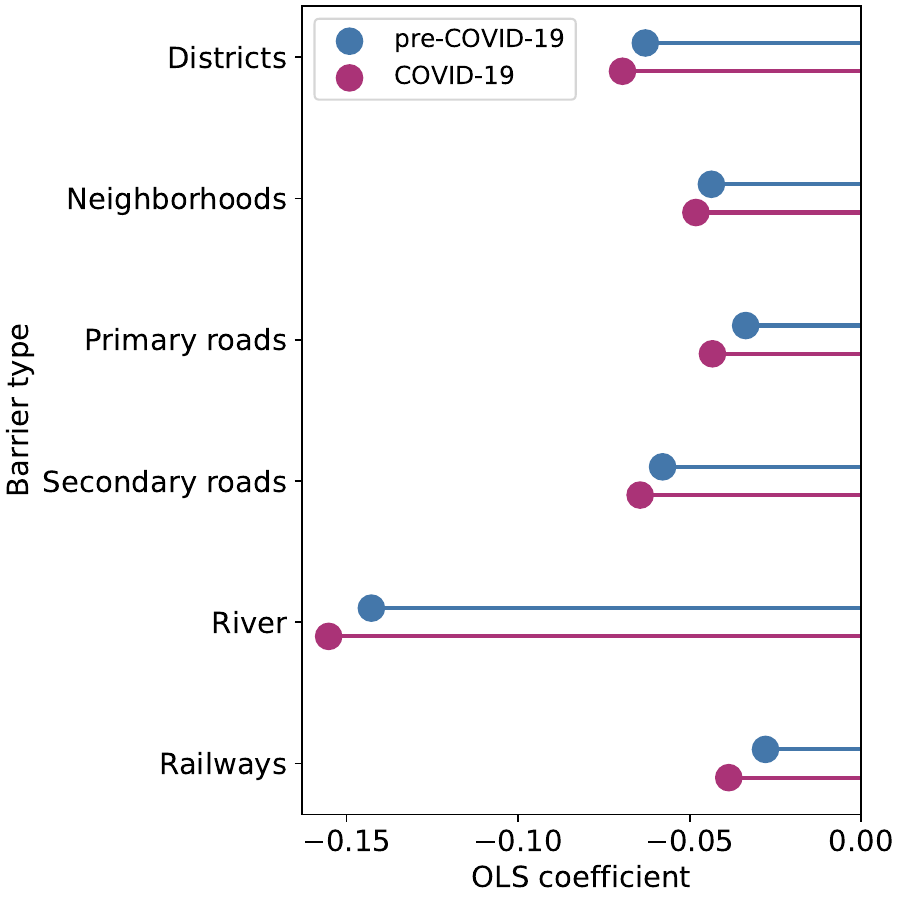}
        \captionsetup{position=bottom,justification=centering}
        \caption{}
        \label{fig:precovid_vs_covid}
    \end{subfigure}
    \hfill
    \vspace{0.25em}
    \begin{subfigure}[t]{0.325\linewidth}
        \includegraphics[width=\linewidth]{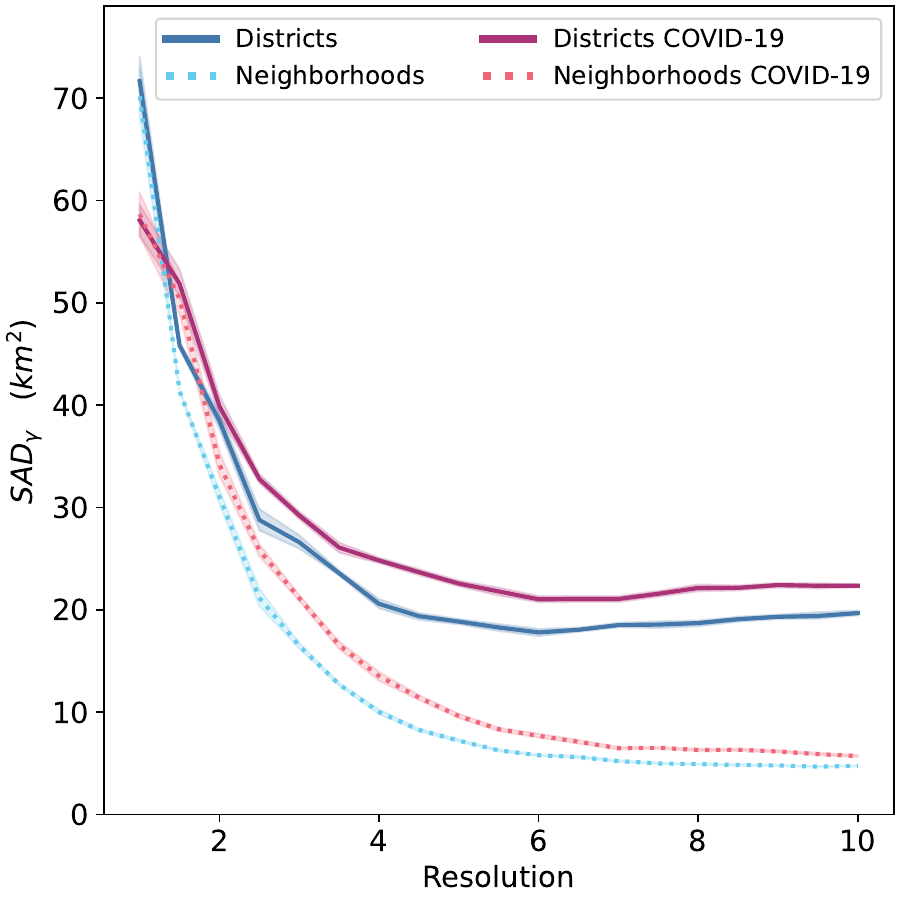}
        \captionsetup{position=bottom,justification=centering}
        \caption{}
        \label{fig:area_symmdiff_administrative}
    \end{subfigure}
    \hfill
    \begin{subfigure}[t]{0.325\linewidth}
        \includegraphics[width=\linewidth]{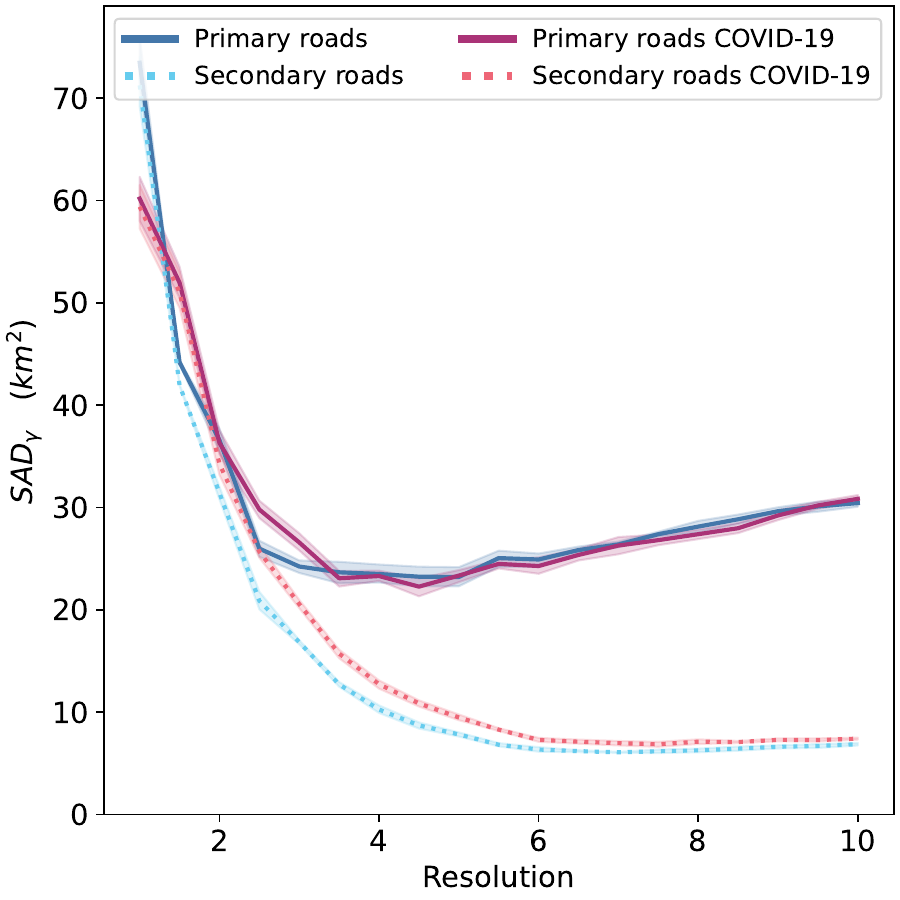}
        \captionsetup{position=bottom,justification=centering}
        \caption{}
        \label{fig:area_symmdiff_barriers}
    \end{subfigure}
    \caption{
        Barrier effect on urban mobility in the gravity and the mobility network approach, preceding and during COVID-19 restrictions. The coefficients of ordinary least squares (\acrshort{OLS}) regression measure barrier impact on flows between two locations (\textbf{\subref{fig:precovid_vs_covid}}). In (\textbf{\subref{fig:area_symmdiff_administrative}}, \textbf{\subref{fig:area_symmdiff_barriers}}), the lines and shaded area represent the mean and the 95 $\%$ confidence interval of the \acrfull{SAD} indicator as a measure of barrier fit of mobility clusters detected in ten iterations of the Louvain method for every $\gamma$. The improving fit to administrative boundaries along the increasing resolution parameter (\textbf{\subref{fig:area_symmdiff_administrative}}) signals that lower-rank boundaries tend to align with local mobility clusters. This is the case for secondary roads as well but the optimal fit to primary roads at $\gamma \approx 4$ suggests that emerging local mobility clusters are less bounded by primary roads (\textbf{\subref{fig:area_symmdiff_barriers}}).
    }
    \label{fig:symdiff}
\end{figure}

To present a benchmark for urban barrier assessment, we first investigate their impact on urban mobility by testing a widely-used gravity model, in which the volume of mobility flow between block $i$ and $j$ is estimated by their Euclidean distance, the sum of their incoming and outgoing flows and the number of barriers between them (for more details see \nameref{sec:materials_and_methods}). We find that river Danube has the strongest impact on mobility and railways have the weakest (Figure~\ref{fig:precovid_vs_covid}). Further, there are barriers that have hierarchical relations within the category, such as administrative boundaries can be defined at the higher rank of districts or at the lower rank of neighborhoods; while roads can be primary or secondary. In the gravity model, we find no clear pattern whether higher or lower rank barriers in these hierarchies have a stronger impact on mobility. While boundaries of districts have a bigger impact than neighborhood borders, the coefficient of secondary roads is stronger than that of primary roads. Yet, the gravity model measures an average impact of barriers disregarding the locations of mobility links. On the contrary, a network of mobility links is denser in the center of cities. By finding smaller scales of mobility clusters in high-density areas, our approach provides a better way to understand hierarchies of barriers because we can fit mobility clusters to lower-rank administrative barriers too.

Indeed, assessing the fit of mobility clusters to urban barriers with the $SAD_\gamma$ indicator, we find that barrier hierarchies have a more intuitive imprint on mobility than in the gravity model. As $\gamma$ increases in its low ranges, $SAD_\gamma$ decreases for both district and neighborhood boundaries (Figure ~\ref{fig:area_symmdiff_administrative}). However, the fit to district boundaries reaches a maximum at higher values of $\gamma$, while to the fit to neighborhoods further improves as the algorithm finds smaller and smaller groups of blocks that tend to represent mobility clusters at the scale of neighborhoods. A similar trend is found in terms of road hierarchie as $\gamma$ grows, community detection fits to primary and secondary roads, but at high values of $\gamma$ the technique provides an improving fit to the secondary roads only (Figure ~\ref{fig:area_symmdiff_barriers}).

Comparing the period that precedes the COVID-19 pandemic with the period that contains the second and third waves of the virus diffusion, the increasing coefficients of the gravity model inform us about an increasing impact of all types of urban barriers on mobility (Figure~\ref{fig:precovid_vs_covid}). However, according to the $SAD$ indicator, the fit of detected communities gets worse for both levels of administrative boundaries and for secondary roads as well. The worsening fit during the pandemic might be due to the drastic drop of mobility links (ca. 33\% of the links are present). However, the growth of $SAD$ values at given $\gamma$ values are not drastic. In the case of primary roads, one can observe that the fit does not differ between the two periods at $\gamma>3.5$ that also includes the best fit we find. Thus, the community detection method provides stable results for primary roads across very different periods in terms of mobility habits.

In the case of primary roads, we find a characteristic level of the resolution parameter ($\gamma \approx 4$) that provides the best fit of communities. Up to this resolution parameter, the fit measured by $SAD_\gamma$ improves but the fit gets worse at higher values of $\gamma$. This optimal fit arises due to local mobility clusters that emerge after we decrease the dominance of mobility links between large transportation hubs by increasing $\gamma$ (see Equation ~\ref{eq:modularity_index}).

\begin{figure}[!t]
    \centering
    \begin{subfigure}[t]{0.16\linewidth}
        \includegraphics[width=\linewidth]{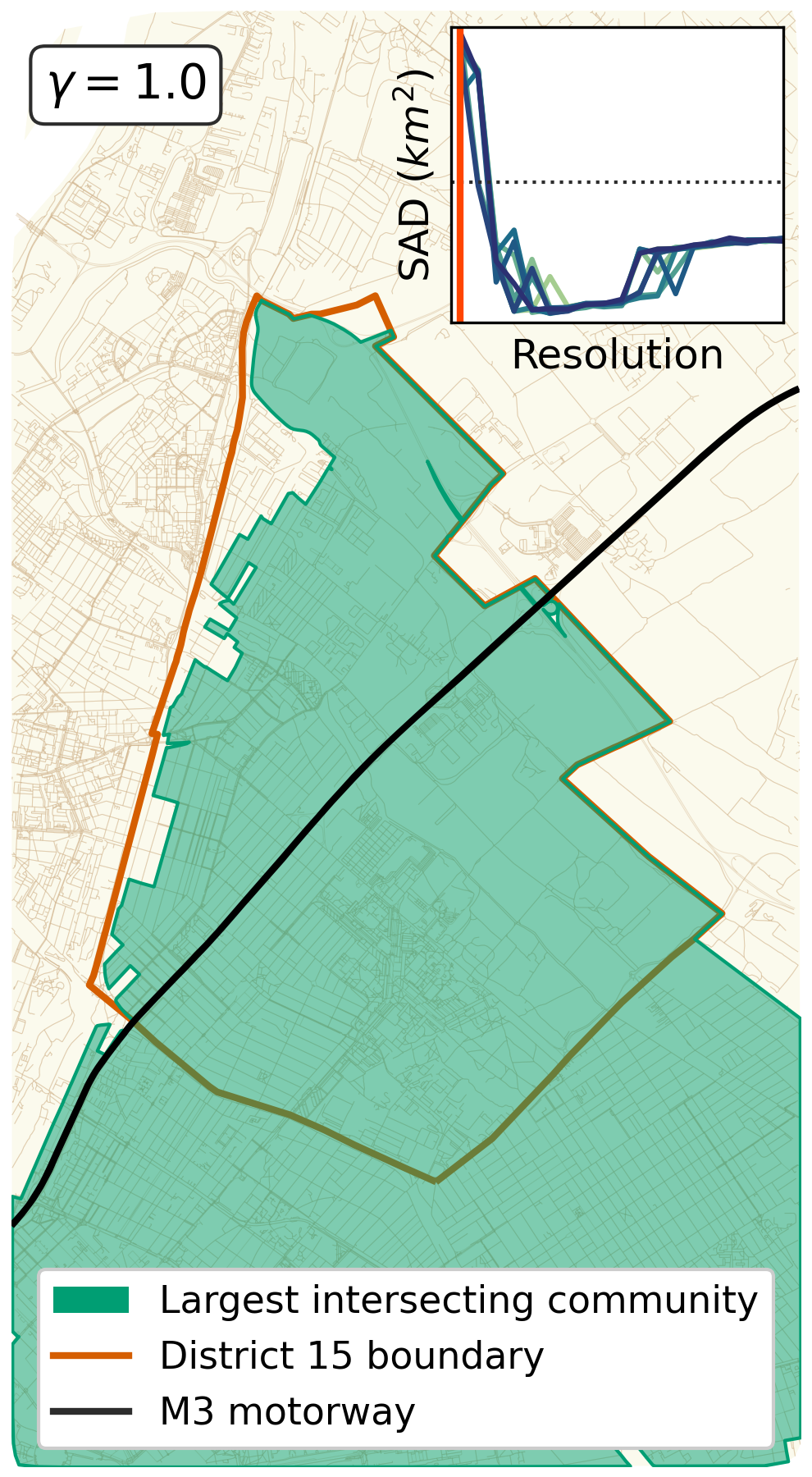}
        \captionsetup{position=bottom,justification=centering}
        \caption{}
        \label{fig:d15_r1.0}
    \end{subfigure}
    \hfill
    \vspace{0.25em}
    \begin{subfigure}[t]{0.16\linewidth}
        \includegraphics[width=\linewidth]{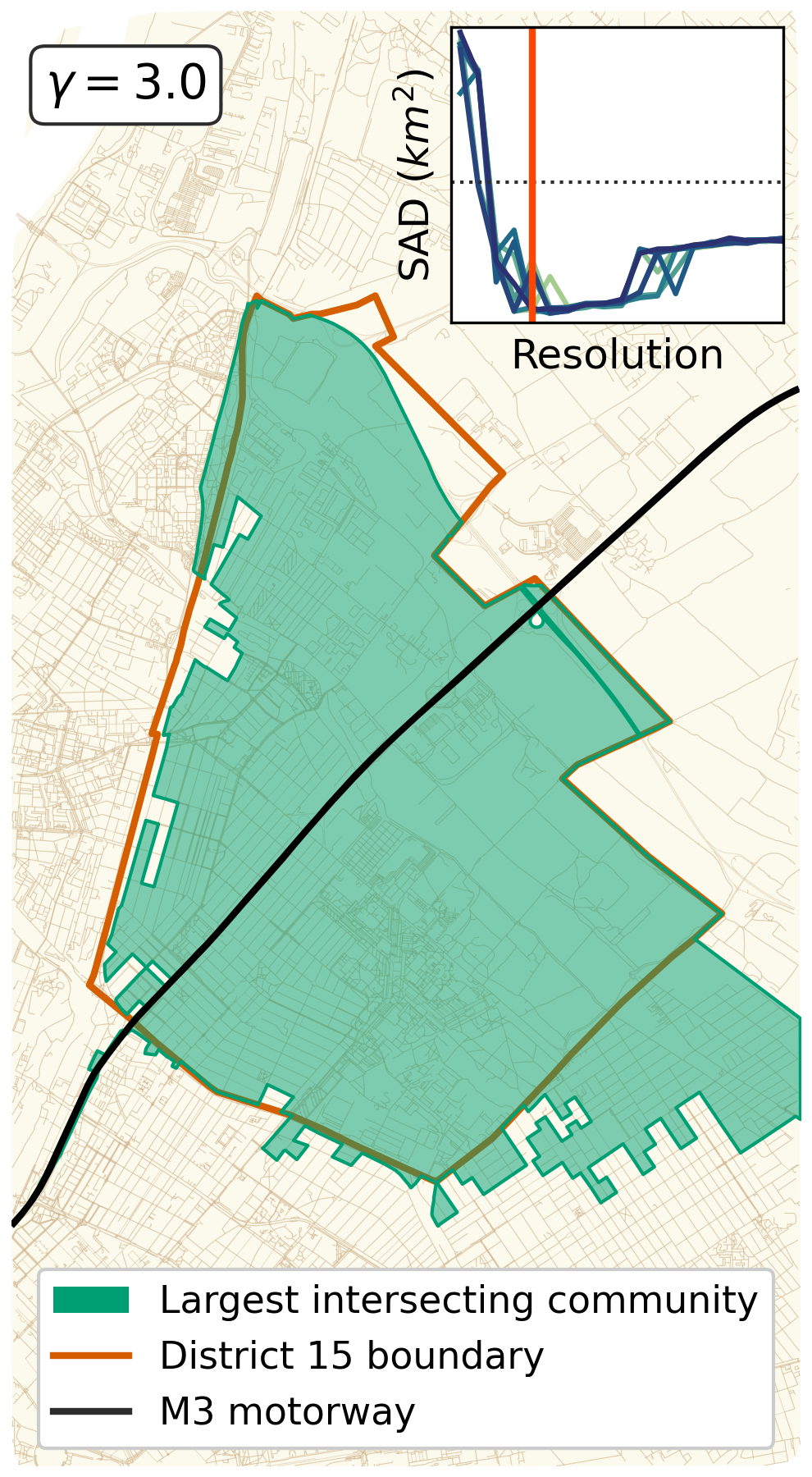}
        \captionsetup{position=bottom,justification=centering}
        \caption{}
        \label{fig:d15_r3.0}
    \end{subfigure}
    \hfill
    \begin{subfigure}[t]{0.16\linewidth}
        \includegraphics[width=\linewidth]{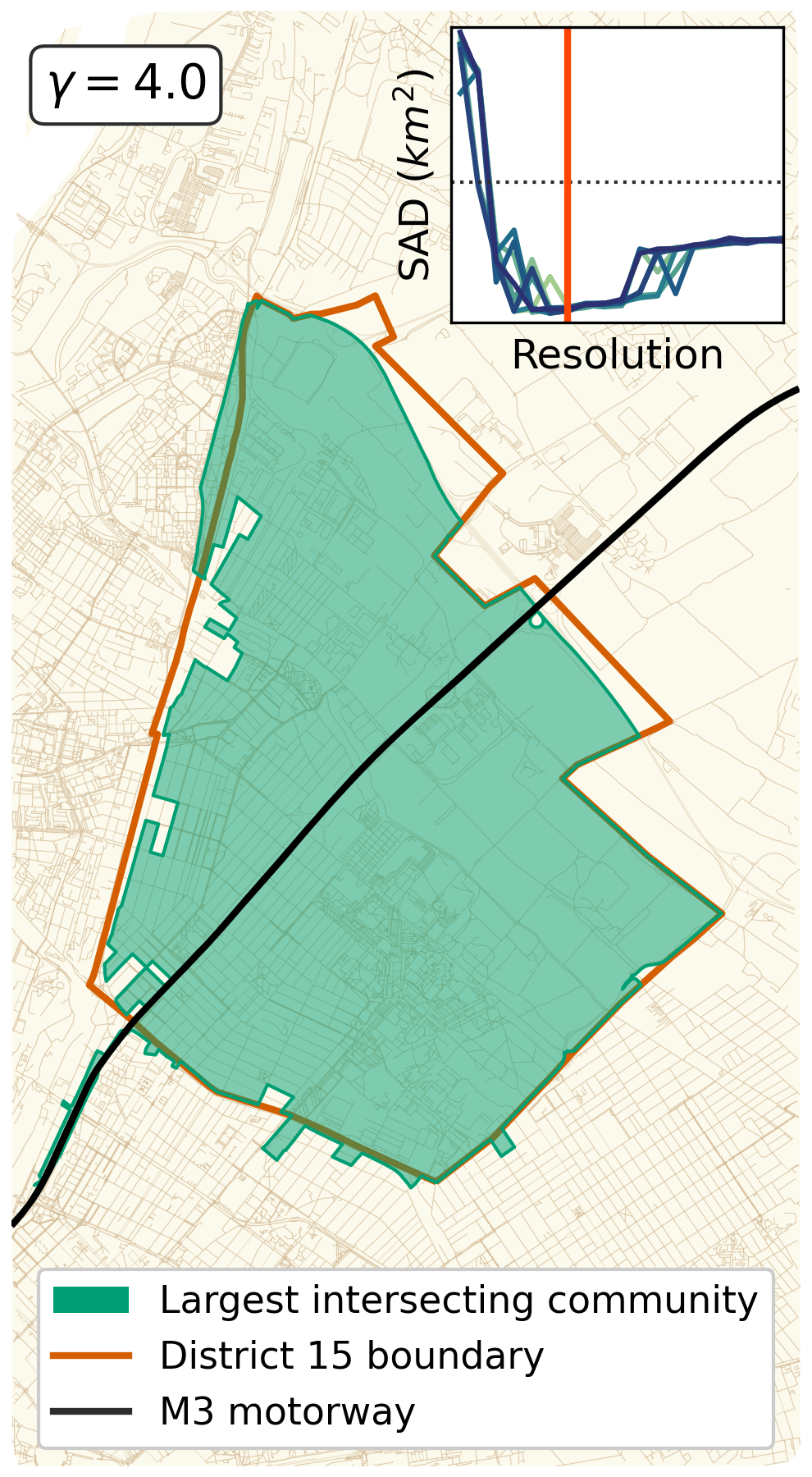}
        \captionsetup{position=bottom,justification=centering}
        \caption{}
        \label{fig:d15_r4.0}
    \end{subfigure}
    \hfill
    \begin{subfigure}[t]{0.16\linewidth}
        \includegraphics[width=\linewidth]{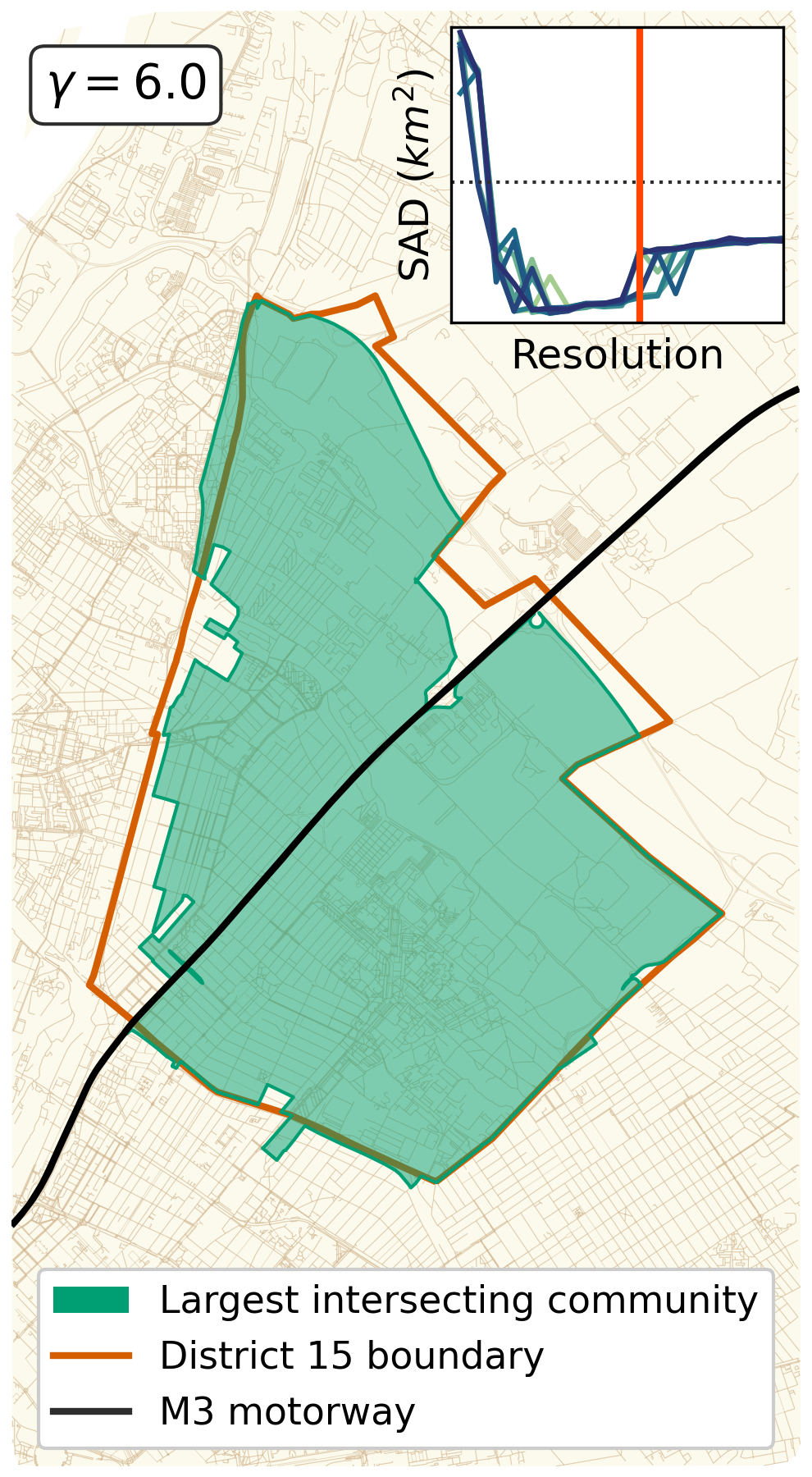}
        \captionsetup{position=bottom,justification=centering}
        \caption{}
        \label{fig:d15_r6.0}
    \end{subfigure}
    \hfill
    \begin{subfigure}[t]{0.16\linewidth}
        \includegraphics[width=\linewidth]{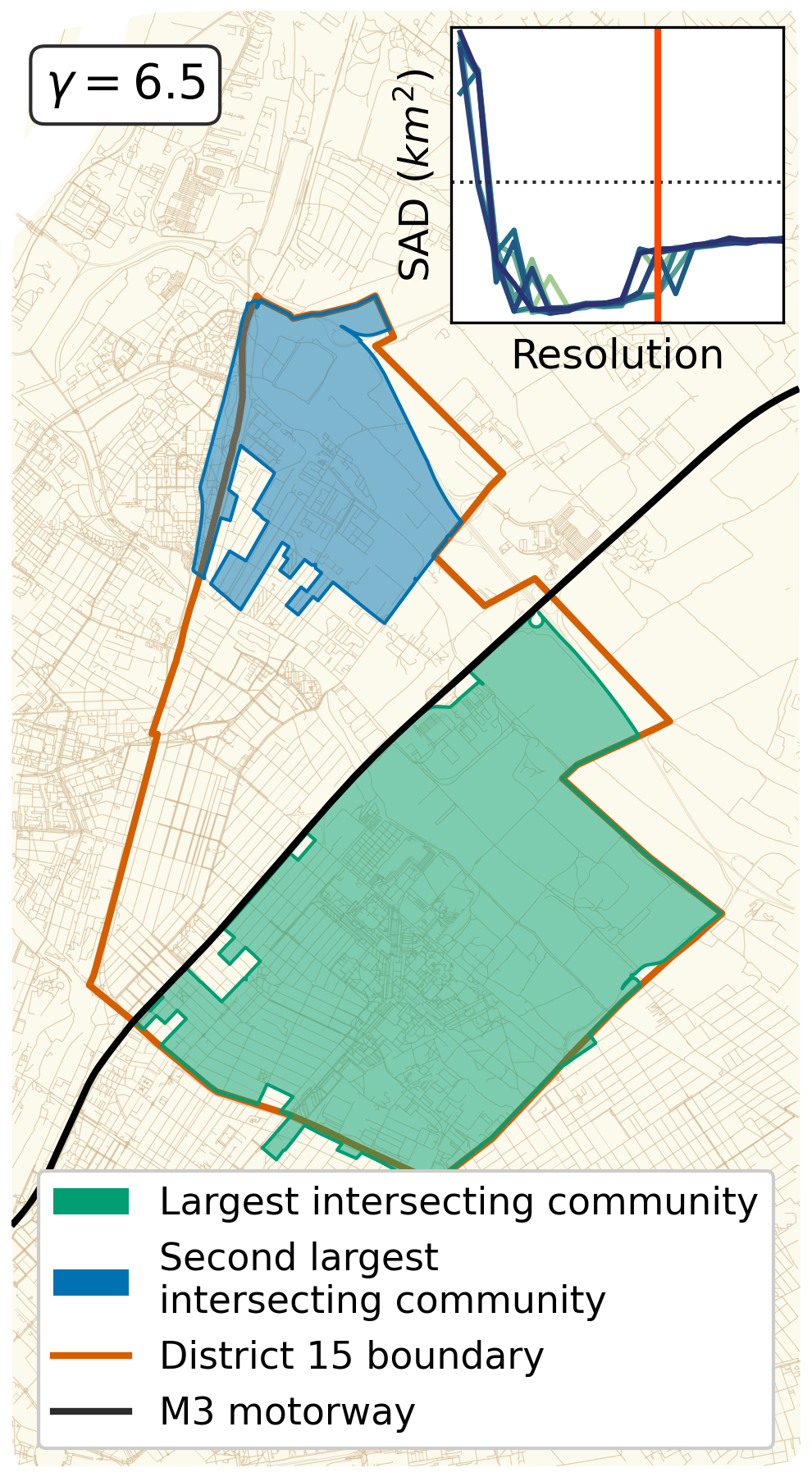}
        \captionsetup{position=bottom,justification=centering}
        \caption{}
        \label{fig:d15_r6.5}
    \end{subfigure}
    \hfill
    \begin{subfigure}[t]{0.16\linewidth}
        \includegraphics[width=\linewidth]{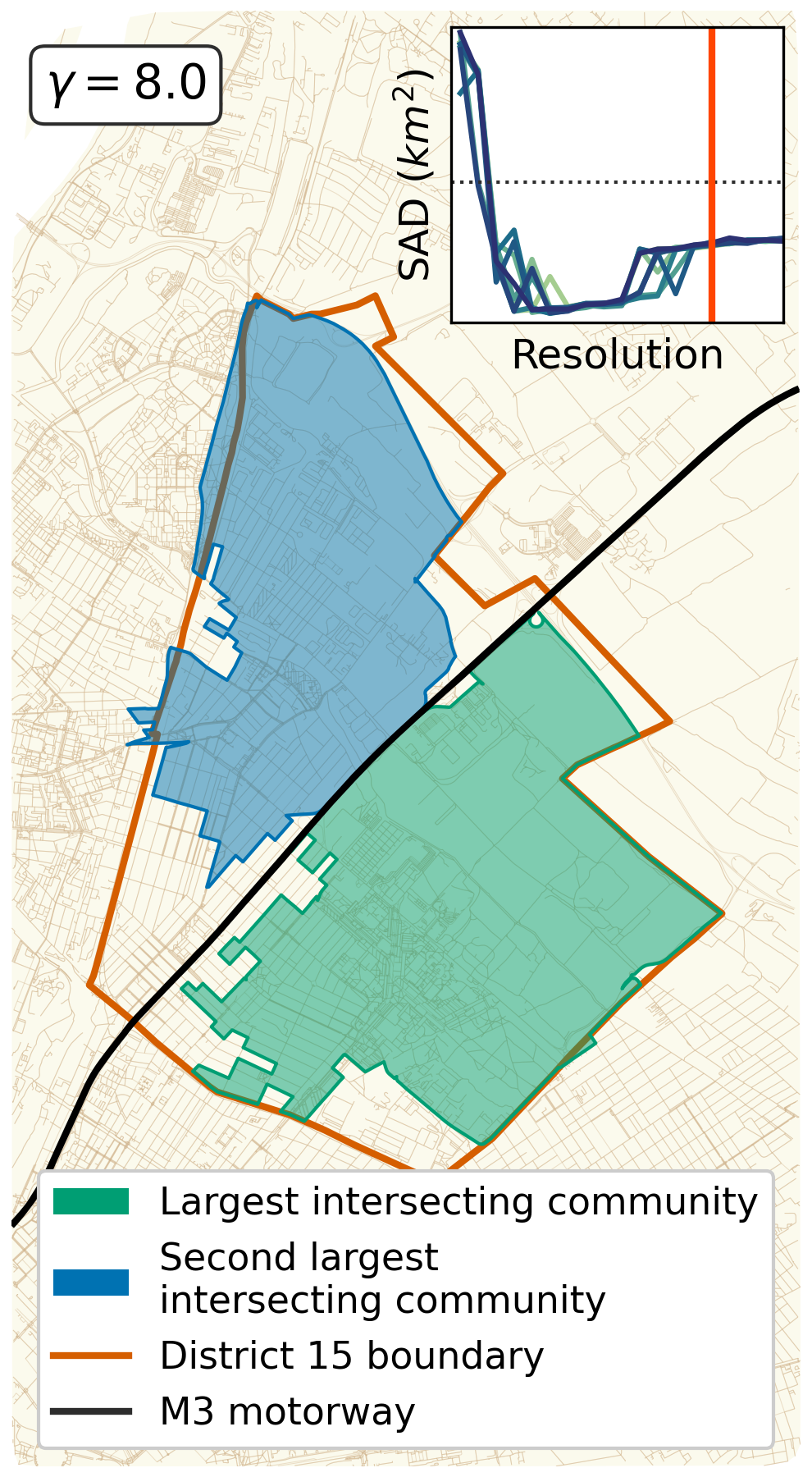}
        \captionsetup{position=bottom,justification=centering}
        \caption{}
        \label{fig:d15_r8.0}
    \end{subfigure}
    \caption{
        The process of community fitting to District 15 and inset figures show the \acrfull{SAD} in respect of the resolution.
        Note that the largest intersecting community at $\gamma = 1.0$ (\textbf{\subref{fig:d15_r1.0}}) is about twice as big as the district.
        As the resolution increases the community matches the district boundary (\textbf{\subref{fig:d15_r3.0}}, \textbf{\subref{fig:d15_r4.0}}), which looses the community forming power at about $\gamma = 6.0$, as the M3 motorway gradually cuts in half the community (\textbf{\subref{fig:d15_r6.0}}--\textbf{\subref{fig:d15_r8.0}}).
    }
    \label{fig:district15}
\end{figure}

A potential explanation of the worsening fit in Figure \ref{fig:area_symmdiff_barriers} can be that local mobility events cross primary roads. However, it may also be that the role of a certain barrier weakens because another one replaces it. Indeed, Figure \ref{fig:district15} illustrates an example where such a replacement happens using the case of marker 1 in Figure~\ref{fig:stream}e. The fit to the administrative boundary of District 15 is weak at $\gamma=1$ when the mobility cluster covers a relatively large area (Figure \ref{fig:d15_r1.0}) but becomes reasonably good in the range of $\gamma \in (3;6)$. Further increasing the parameter, we find that the smaller scale mobility cluster fits to the administrative border in the south but is limited in the north by the major road that cuts the district into two. The local mobility cluster is bounded by a combination of barriers that worsens the fit of a single category and suggests that it is not just single barriers that impact mobility but they can influence mixing in more complex configurations.

The fit of mobility clusters to the railways and the river can be found in Supplementary Section \ref{si:sec:sad_railways_and_river}. The fit of railways during COVID restrictions quantified by $SAD_\gamma$ is of the same level with districts or primary roads but fits worse at large $\gamma$ in the pre-COVID period. The $SAD_\gamma$ values of the fit to the river decrease as $\gamma$ increases.

Repeating the analysis on a distinct data set from Nagoya, Japan \cite{yabe2024yjmob100k}, we find that the $SAD_\gamma$ indicator describes the role of administrative barriers in urban mobility similarly (see Supplementary Section \ref{si:sec:yjmob100k}). By increasing the resolution parameter $\gamma$, the fit of mobility clusters to administrative barriers improves, while it is significantly better for lower-rank boundaries for all $\gamma$.

\subsection*{Barrier Crossing Ratio by home locations and amenity mix at destinations}
\label{sec:bcr_by_residence}

\begin{figure}[!t]
    \centering
    \begin{subfigure}[t]{0.285\linewidth}
        \includegraphics[width=\linewidth]{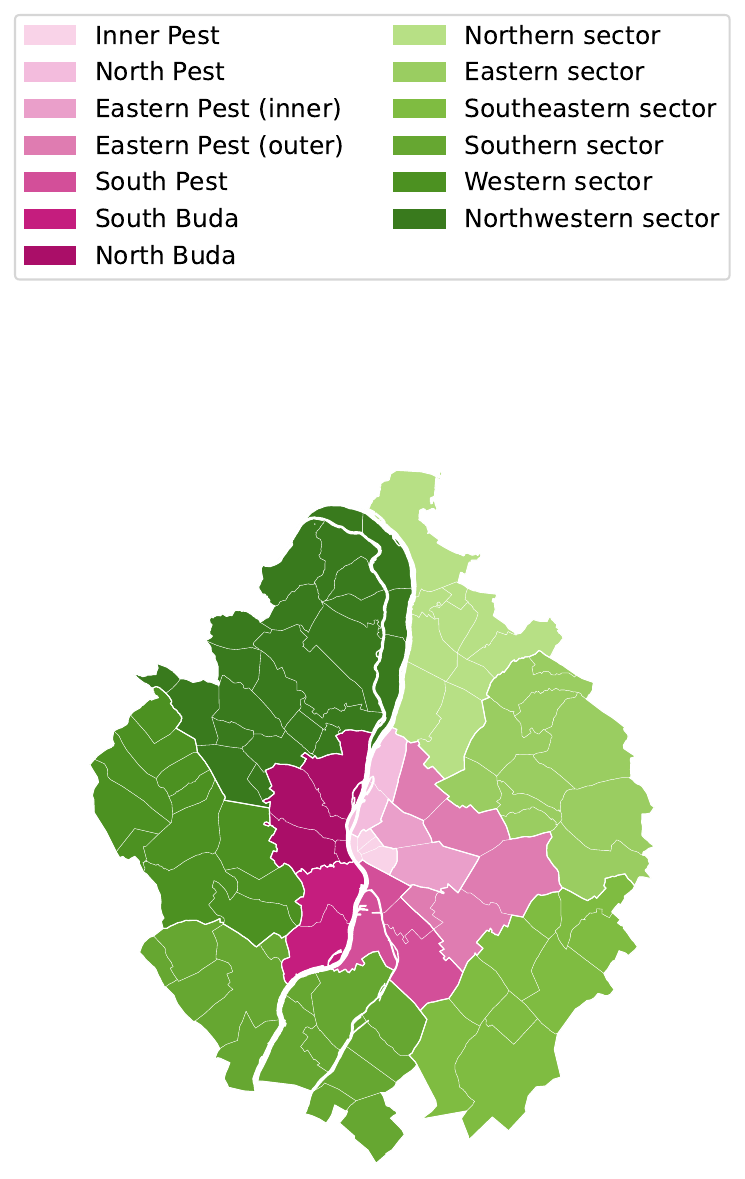}
        \captionsetup{position=bottom,justification=centering}
        \caption{}
        \label{fig:map}
    \end{subfigure}
    \hfill
    \begin{subfigure}[t]{0.7\linewidth}
        \includegraphics[width=\linewidth]{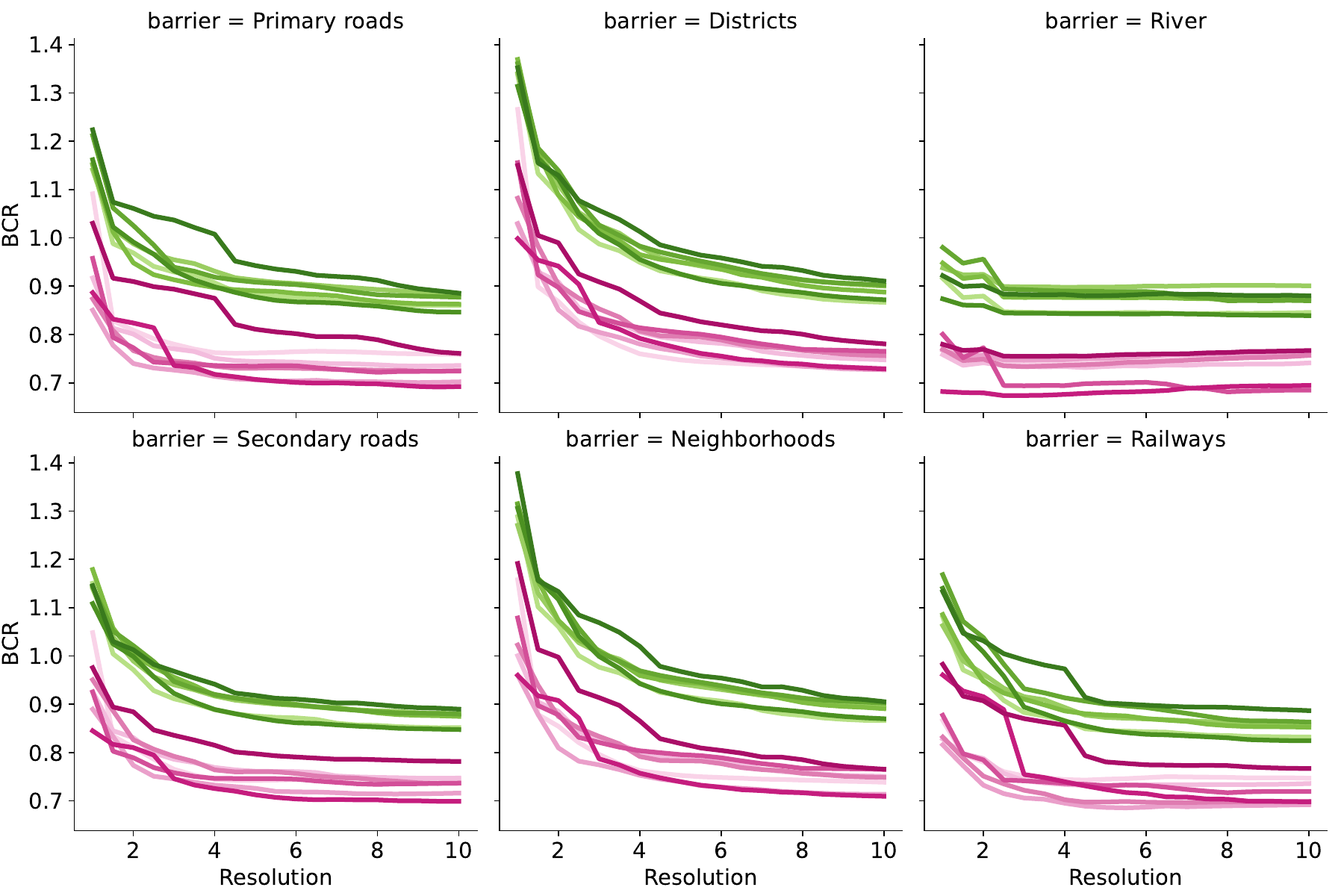}
        \captionsetup{position=bottom,justification=centering}
        \caption{}
        \label{fig:by_barrier}
    \end{subfigure}

    \caption{
        Barrier effect based on the users' home location groups.
        The \acrshort{HCSO} defines seven district groups in Budapest and six sectors of the agglomeration (\textbf{\subref{fig:map}}).
        From the Budapest mobility of the inhabitants of these areas, thirteen networks were built and evaluated for all barrier type with the $BCR_{\gamma}$ index and increasing the $\gamma$ parameter (\textbf{\subref{fig:by_barrier}}).
    }
    \label{fig:coeff_by_home}
\end{figure}

Urban barriers might have a divergent impact on neighborhoods as individuals daily mobility depends largely on the locations of work, schools, and services that are unevenly distributed in and around cities. Thus, in the next step of the analysis, we assess the role of barriers in mobility by decomposing the set of individuals by areas of residence and by visited amenities. To do that, we leverage the home locations of individuals identified in \cite{juhasz2023amenity} by the device location during the 8pm-8am daily periods and the amenity portfolio of stop locations measured in the same paper.
As many people commute from the agglomeration to the capital \cite{pinter2022commuting}, not only the inhabitants of Budapest are taken into consideration, but people living in the larger agglomeration as well.
We use the urban area definition of \acrfull{HCSO} to distinguish seven district groups within the city of Budapest and sort municipalities into six sectors of the Budapest agglomeration to group home locations \cite{ksh2018budapest} (visualized in Figure~\ref{fig:map}).
Then, we use the mobility network across Budapest blocks described in Figure ~\ref{fig:stream} to quantify the tendency that individual mobility events cross urban barriers and detected mobility clusters, using the \acrlong{BCR} (Eq.~\ref{eq:BCR}) that can be decomposed by origins and destinations of mobility events.

We find a striking difference between the \acrshort{BCR} indicator of those dwellers who live in the city and those who live in its agglomeration (Figure ~\ref{fig:coeff_by_home}). In every barrier category, agglomeration sectors have higher \acrshort{BCR} values than sectors of the city, indicating that commuters from the agglomeration cross relatively more urban barriers without crossing boundaries of mobility clusters than residents of city sectors do. Except the river, the value of \acrshort{BCR} is above 1 for agglomeration sectors for the lowest $\gamma$ values in the case of all other barriers. This indicates that the number of total barrier crossings is higher than those crossings that are across barriers and mobility clusters at the same time. Increasing $\gamma$ provides a better fit of mobility clusters to barriers, in which the differences between urban and agglomeration sectors are stable as $\gamma$ grows.

Thus, the detected mobility clusters fits urban barriers better for those who live close to these actual barriers than for those who live farther away. Naturally, people living in the agglomeration have different mobility patterns than urban dweller their vast majority use cars when commuting to the city but also optimize home and work location so that they usually commute to the closest part of the capital. For example, people living in the Northwestern sector visit mostly North Buda and only a small portion of their visits are located in the inner Eastern Pest (see the sectors in Figure \ref{fig:map}and Supplementary Section~\ref{si:sec:stop_distribution} for details).
However, North Buda and South Buda behave like agglomeration sectors at low levels of $\gamma$ in the case of Primary roads and Railways but perform a relatively good fit to barriers at high $\gamma$.

The Danube as barrier has the strongest impact on mobility. Interestingly, the ratio of barrier crossings and mobility cluster crossings is nearly constant across values of $\gamma$ for most sectors, meaning that local mobility flows hardly cross the river. In other words, crossing the river almost automatically means crossing mobility clusters too. This is especially true for residents of the agglomeration: 90\% of their crossings over the river leads to other mobility clusters.
The Supplementary Section~\ref{si:sec:bcr_by_residence_plus} contains further analyses in respect of the residence-based mobility networks, including another evaluation of the \acrshort{BCR} where separate networks were generated for the urban areas. This approach takes into consideration that commuting distance impacts mobility patterns that can be captured by different mobility networks. Yet, these results also confirm that urban barriers in Budapest have a stronger impact on urban dwellers than on commuters.

Interestingly, we see a different pattern for Nagoya (Supplementary Section \ref{si:sec:yjmob100k}) where we could evaluate the fit of mobility clusters to municipality boundaries and primary roads using the $BCR_{\gamma}$ index. Residents of the city have higher $BCR_{\gamma}$ values suggesting that they cross urban barriers that are also barriers of mobility clusters more often than residents of agglomeration towns. Thus, local contexts have an important role in distinguishing the impact of urban barriers on dweller groups that further research should focus on.

\begin{figure}[!t]
    \centering
    \begin{subfigure}[t]{0.495\linewidth}
        \includegraphics[width=\linewidth]{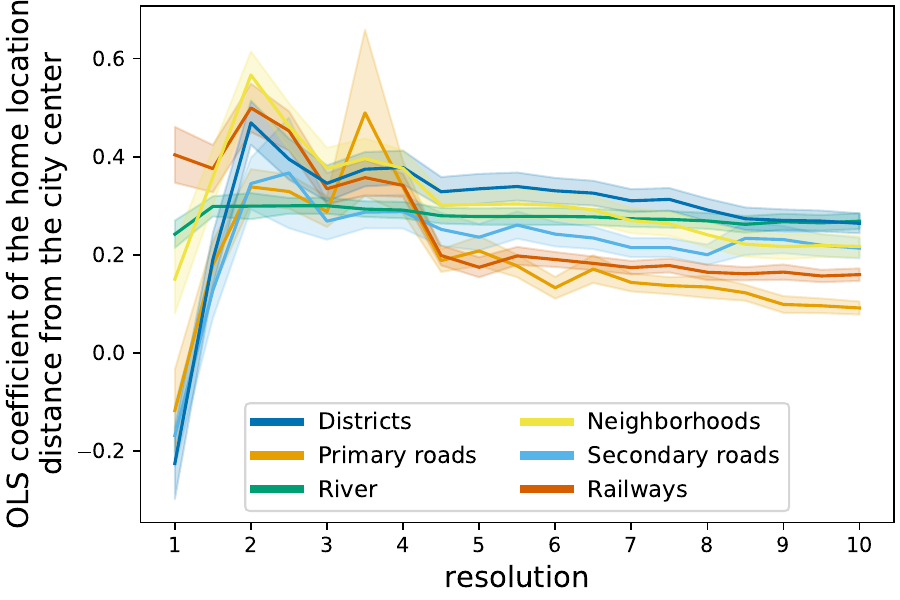}
        \captionsetup{position=bottom,justification=centering}
        \caption{}
        \label{fig:ibcr_distance_log10}
    \end{subfigure}
    \hfill
    \begin{subfigure}[t]{0.495\linewidth}
        \includegraphics[width=\linewidth]{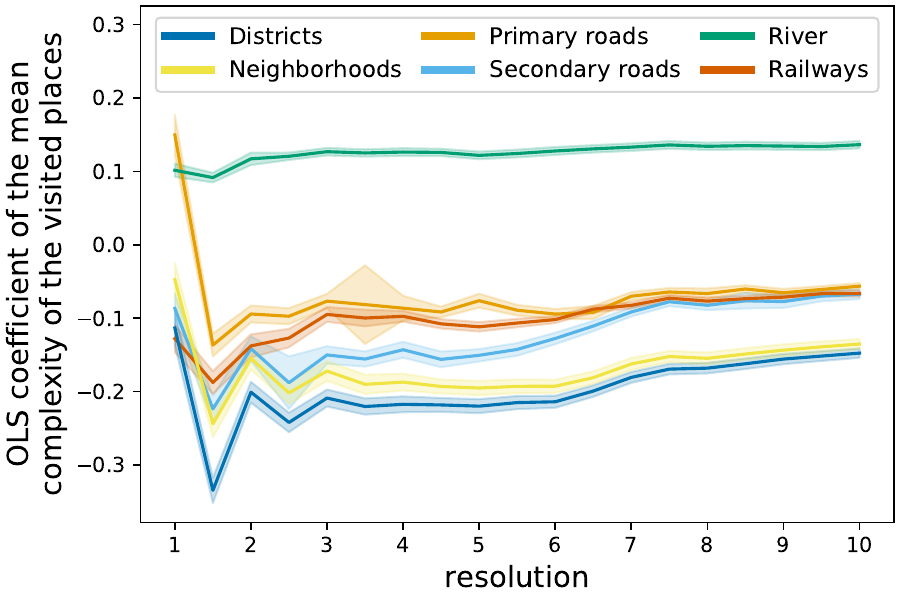}
        \captionsetup{position=bottom,justification=centering}
        \caption{}
        \label{fig:ibcr_avg_visitation_complexity}
    \end{subfigure}

    \caption{
        Urban barrier impact on individuals by distance from the center and by visited amenities. We depict the relationship between Barrier Crossing Ration and the distance between the inhabitants home location and the city center (\textbf{\subref{fig:ibcr_distance_log10}}) and the mean location complexity of the visited places (\textbf{\subref{fig:ibcr_avg_visitation_complexity}}). Solid lines represent estimation coefficients and the shaded are denotes standard errors.
    }
    \label{fig:complexity_ols}
\end{figure}


Next, we calculate the \acrfull{BCR} indicator for individuals, instead of areas. Since the mobility network is aggregated from individual trips, one can compare the barrier and mobility cluster crossings at the individual level in the same ways as it is introduced in Section \nameref{sec:crossing_ratio}. The individual \acrfull{BCR} captures the average \acrshort{BCR} of individual citizen $c$ in the $n^{th}$ run of the community detection with resolution $\gamma$. The indicator takes a lower value for those individual citizens who cross barriers by visiting mobility clusters and takes a larger value for those who cross barriers without leaving the mobility cluster.

This measure enables us to investigate how the role of barriers depends on home locations and also on visited amenities in a multivariate analysis. We regress \acrshort{BCR} with an \acrfull{OLS} linear regression, using the formula:

\begin{equation}
        log BCR_{\gamma, n}^{c} = \alpha + \beta_1\log{D_c} + \beta_2\log{AC_j} + \epsilon,
    \label{eq:bcr}
\end{equation}

where $D_c$ is the geographical distance of the home location of citizen $c$ from the city center and $AC_j$ is the complexity of amenity portfolio of the visited $j$ location as proposed by \cite{juhasz2023amenity} and $\epsilon$ is the error term. Amenity complexity is high in those locations that host diverse services that are difficult to find elsewhere in the city. Locations with complex amenity portfolios have been found to promote social mixing because they serve a diverse demand and can attract visitors from the whole city \cite{juhasz2023amenity}.

The findings reveal significant relationship between the distance, amenity complexity and \acrlong{BCR} that also seems to be stable $\gamma > 1.5$ for all types of barriers (Figure \ref{fig:complexity_ols}). As expected, $D_c$ correlates positively with \acrshort{BCR} (Figure \ref{fig:ibcr_distance_log10}) confirming our previous finding that those who live farther away from the center are more likely to cross barriers without crossing mobility clusters because the fit of the detected cluster is worse for them. However, we find a negative correlation between $AC_j$ and \acrshort{BCR} for most types of barriers. This result suggests that visiting complex amenities that are in other mobility clusters, individuals usually cross urban barriers too. The river seems to be an exception here as well, but as discussed above, there are rarely any mobility clusters that extend across the river; thus, \acrshort{BCR} calculation in this case could only consider visits to two islands that are sharing mobility clusters with districts on the Pest side.
The stable coefficients of $D_c$ and $AC_j$ over $\gamma$ signals that relationship with \acrshort{BCR} is not an artifact of community sizes; instead, these correlations capture a general impact of barriers on mobility, considering complex structures of mobility networks.
The \acrshort{OLS} tables are in the supplementary section \ref{si:sec:complexity_ols_tables}.

\section*{Discussion}
\label{sec:discussion}


Urban barriers are obstacles of social interactions and even the built urban infrastructure, which is aimed to facilitate access, can hinder mobility. Here we apply a network science framework that provides a new understanding of this phenomenon and enables us to investigate how urban barriers delineate clusters of mobility considering complex network structures. Our results demonstrate that besides natural and administrative boundaries, major roads within cities can separate individuals. We explore that mobility clusters can be impacted by combinations of barrier types. We also discover that barriers has a smaller impact on mobility to complex amenity locations because these places attract visits from many neighborhoods. In the case of Budapest, urban barriers better align with mobility clusters of urban dwellers and less so for residents who live in the larger metropolitan area and commute to the city. However, we find a different pattern in the case of Nagoya, where barriers have a bigger impact on commuters. Thus, the question why the barrier impact come to be and how it differs across groups of citizens is open. Further research should therefor investigate the role of local economic environment, the transportation system, commuting trends, social segregation, and urban geography in general to better understand what makes urban barriers impact mobility.


The strong impact of barriers has wide-ranging implications for the ongoing discussion of social mixing in cities \cite{moro2021naturecomm, juhasz2023amenity, bokanyi2021universal, athey2021estimating, fan2023diversity}. Similarly to the role of mobility restrictions that have been documented to increase urban segregation \cite{yabe2023behavioral, napoli2023socioeconomic}, urban barriers can hinder the mixing of individuals from various socio-economic strata, even if they live only on the opposite sides of roads or railroads \cite{ ananat2011wrong}. Such separation can foster segregation in social networks and can lead to growing inequalities \cite{toth2021naturecomm}. Therefore, the impact of barriers on mobility should be considered in recent proximity-based urban developments - that are often referred to as the 15-minute city concept - that aim to foster short distance active mobility and inclusion in cities. We need a better understanding how improving access to amenities \cite{xu2020deconstructing, abbiasov202215}, which are reachable with active mobility, can overcome urban barriers like major roads that still channel the flows of larger distance car traffic.




Our study is not without limitations. As the mobile positioning data used in this study does not have information about the means of transport, it is not possible to differentiate between pedestrians, bikers, public transport passengers or car drivers.
A possible future direction is to use additional data sources, e.g. from bike sharing services, to analyze the barrier effect on bikers.
Such advances could help improve the bicycle lane infrastructure and detect possibly dangerous spots that behave string barriers for bikers.
Using further data from public transport usage could help us understand how public transport contributes to the emergence of mobility clusters.
Future work should investigate the circumstances of community formation in mobility networks across a variety of cities. To highlight the separation power of built infrastructure, we have to compare different trajectories of urban evolution that has led to the dominance of administrative boundaries or, on the contrary, to the dominance of roads in delineating mobility clusters. Finally, we need further research that focuses on mixing of strata across urban barriers and the types of mobility that can decrease experienced segregation given the boundaries in cities.

\section*{Materials and Methods}
\label{sec:materials_and_methods}

\subsection*{Mobility Data}
\label{sec:mobility_data}

The mobility data was collected by a data aggregator company from various unspecified smartphone applications.
Some applications -- e.g., navigation -- leaves frequent traces leaving a continuous trajectory, but other applications generated sporadic traces.
A record contains a timestamp, a user ID, and a \acrshort{GPS} location.
The data covers whole Hungary. We limit the analysis to mobility events in Budapest. The observation period is six months from September 2019 and another six months starting from November 2020.
The raw pings were processed by applying the Infostop algorithm \cite{aslak2020infostop}.
It can detect the stationary points of individual movements and cluster pings around locations, where the individual stopped.
The algorithm gives each stop a label indicating a place that can reoccur along the trajectory of the user.
The process is detailed in a previous work of the authors \cite{juhasz2023amenity}. This procedure has enabled us to identify home locations by detecting stops between the 8pm-8am hours. In the \acrshort{BCR} exercise, we limit the observations to those users who have home locations in Budapest or in municipalities in the Budapest agglomeration.

\subsection*{Barrier Data}
\label{sec:varrier_data}

In this paper, administrative, natural and infrastructural barriers are also considered, which were obtained from \acrfull*{OSM}.
As for the administrative boundaries, administrative level 9 (districts) and 10 (neighborhoods) is utilized.
Budapest has 23 districts and 207 urban neighborhoods.
Note that neighborhoods are in between districts and census tracts in the spatial hierarchy.
The \acrlong*{OSM} classifies the roads based on their hierarchy.
For this study, the motorways, the primary and secondary roads are considered.
In the \acrshort{OSM} terminology, trunks are between motorways and primary roads, however they are not significant in Budapest.
Motorways are merged with the primary roads, since they cannot enclose areas on their own.
Illustrated in Figure~\ref{fig:barriers_3d} by solid orange lines.
Secondary roads are less important than primary roads in the \acrshort{OSM} highways hierarchy and they limit smaller areas than primary roads.
The primary roads divide the city into some larger parts, and with the secondary roads into smaller parts.
A script was developed to create polygons from the city parts enclosed by the roads, for details see Section~\ref{si:sec:enclosures} in the Supplementary Materials.

\subsection*{Mobility Network and Community Detection}
\label{sec:mobility_network_and_community_detection}

When all the lower order roads are also considered during the polygon creation, the enclosures denote blocks.
Then, the places from the mobility data are mapped to the blocks, which will be the nodes of a network.
Two blocks are connected by an edge if a user had consecutive places between the given blocks within a day.
The places are considered only within a day because of the sporadic nature of the data.
It is undesired to connect a user's the last position of a day to the first position of the next day.
Naturally, a threshold could be introduced as if a user is active in the night, a valid mobility trajectory is broken at midnight.
This, however, is a possible direction of future enhancements.

The Louvain community detection \cite{blondel2008fast} is applied to the stop-network with different resolution values, that clusters the blocks into communities.
Due to the nondeterministic nature of the community detection algorithm, it was executed ten times for every resolution value.
The stability of the community detection across executions was examined using Cramér's V, which can be found in the supplementary section \ref{si:sec:cramers_v}.
Then, the values of the difference executions were aggregated.

\subsection*{Gravity Model}
\label{sec:gravity_model}

The gravity model \cite{krings2009urban} is a widely used method for ac­tivity distance and border effect measuring \cite{jin2021identifying}.
Originally, the gravity model for mobility flow is formalized a $ Mob_{i,j} = k \frac{P_i^{\alpha} P_j^{\beta}}{r_{i,j}^{\gamma}} $, where $Mob_{i,j}$ is the mobility between spatial unit $i$ and $j$, $P_i$ is the population in unit $i$, and $r_{i,j}$ is the distance between spatial units $i$ and $j$, and $k$, $\alpha$, $\beta$, and $\gamma$ are constant parameters \cite{jin2021identifying}.

The original formula is extended with the number of crossed barriers, one by one to measure the effect of the given barrier type.
When counting the crossed barriers between two blocks, the beeline is considered between the centroid of the block as the mobility data does not contain any information about the way of transport.
So, every barrier is counted -- by type -- which the straight crosses.
As for the distance ($r_{i,j}$), the haversine (great-circle) distance is used between the two block centroids.
The model (\ref{eq:gravity}) was evaluated using the \acrfull{OLS} regression. The barrier variables have been inserted separately to the regression model due to their high correlation.

\begin{equation}
    \begin{split}
        log Mob_{i,j} = \alpha\log{P_i} + \beta\log{P_j} + \gamma\log{D_{i,j}} + & \delta\log{road^p} \\
        & \delta\log{road^s}\\
        & \delta\log{river}\\
        & \delta\log{railways}\\
        & \delta\log{district}\\
        & \delta\log{neighborhood}
    \end{split}
    \label{eq:gravity}
\end{equation}

\subsection*{Symmetric Area Difference}
\label{sec:measuring_similarity}

It is a known feature of the Louvain community detection algorithm that by increasing the resolution parameter, the size of the communities decreases.
The hypothesis was that the community sizes did not only decrease arbitrarily but converge to the barriers.
To confirm this assumption, the similarity of the communities and the enclosures had to be measured.
The area of the blocks clustered as a community, but outside the target polygon (e.g., district or enclosure) and the area of the target polygon that is not covered by the community is summarized as a measure.
In other words, the area of the false positive and the false negative blocks, which is the symmetric difference of the two polygons (Figure~\ref{fig:community_overlaps_district}).

The changes in the community areas are evaluated in respect of the administrative and infrastructural barriers during the change of the resolution parameter with the symmetric area difference between the clustered blocks (i.e., community) and the administrative districts.

\subsection*{Barrier Crossing Ratio}
\label{sec:crossing_ratio}

The barrier crossings are counted for every barrier type, also used in the gravity model.
It is also counted if a barrier crossing
The count of the barrier crossings that also represent a community crossing (Figure~\ref{fig:barrier_and_community_crossing}) also determined.
The quotient of these two values is called crossing ratio, calculated as $\frac{BC^{obs}}{CC^{obs}}$, where BC is the barrier crossing and CC is the barrier crossing that is also a community crossing.
As the resolution parameter of the Louvain community detection algorithm is increased, the communities become smaller, which means that the trips crosses communities with higher probability.
Barrier crossing does not depend on the resolution, it is a constant, while the CC is increasing as the communities become smaller.
Thus, the quotient is decreasing, which can be seen in Figure~\ref{si:fig:crossing_ratio} and \ref{si:fig:crossing_ratio_covid} of the Supplementary Materials.


\section*{Acknowledgement}

The authors acknowledge the help of Endre Borza with the initial data processing and the stop detection as a part of \cite{juhasz2023amenity}. Useful comments have been received at the NetMob23 Conference in Madrid, the NetSci23 Conference in Vienna.

\section*{Data and Code Availability}
\label{sec:data_availability}

The barrier data is obtained from OpenStreetMap; these data are copyrighted by the OpenStreetMap contributors and licensed under the Open Data Commons Open Database License (ODbL).

The observed mobility networks will be available via Zenodo under ODbL, and the code, that will be published in GitHub, the results are reproducible.
However, the raw mobile positioning data is not publicly available.


\printbibliography[title=References]
\end{refsection}

\clearpage
\section*{Supplementary Materials}

\renewcommand{\thefigure}{S\arabic{figure}}
\renewcommand{\thetable}{S\arabic{table}}
\renewcommand{\thesection}{S\arabic{section}}
\setcounter{figure}{0}
\setcounter{section}{0}

\begin{refsection}
    \section{COVID-19 data set}
\label{si:sec:covid}

Six months of mobility data is used before the COVID-19 pandemic (between September 2019 and February 2020) to analyze the barrier effect in urban mobility.
As a part of the epidemic prevention strategy, restrictions have been issued, which limited to the mobility.

To compare the barrier effect during the pandemic, another 6-month interval was selected from the available mobile positioning data.
In Hungary, the second and third waves took place between the autumn of 2020 and the spring of 2021.

\begin{figure}[ht!]
    \centering
        \includegraphics[width=0.95\linewidth]{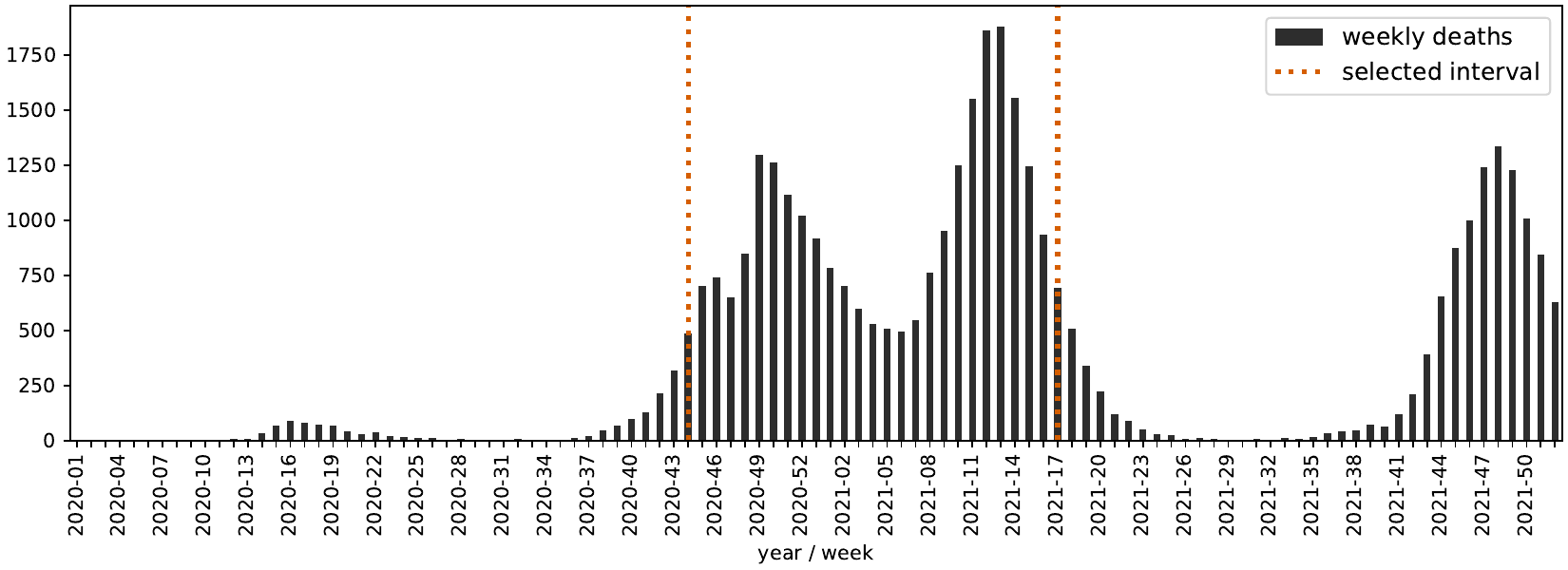}
    \caption{
        COVID-19 mortality in Hungary between January 2020 and December 2021.
        The selected interval is considered the second and third wave in Hungary.
        Data is from European Centre for Disease Prevention and Control \cite{ecdc2023data}.
    }
    \label{si:fig:deaths}
\end{figure}

The strictest protocols were issued between 11 November 2020 and the late May 2021, including curfew in nighttime.
To match the pandemic data set as close as possible to the pre-pandemic data set, the six months between November 2020 and April 2021 have been selected.
Figure~\ref{si:fig:deaths} shows the selected interval in comparison to the weekly number of deaths during the pandemic.

\clearpage
\section{Stability of the community detection}
\label{si:sec:cramers_v}

\begin{figure}[ht!]
    \centering
    \begin{subfigure}[t]{0.495\linewidth}
        \includegraphics[width=\linewidth]{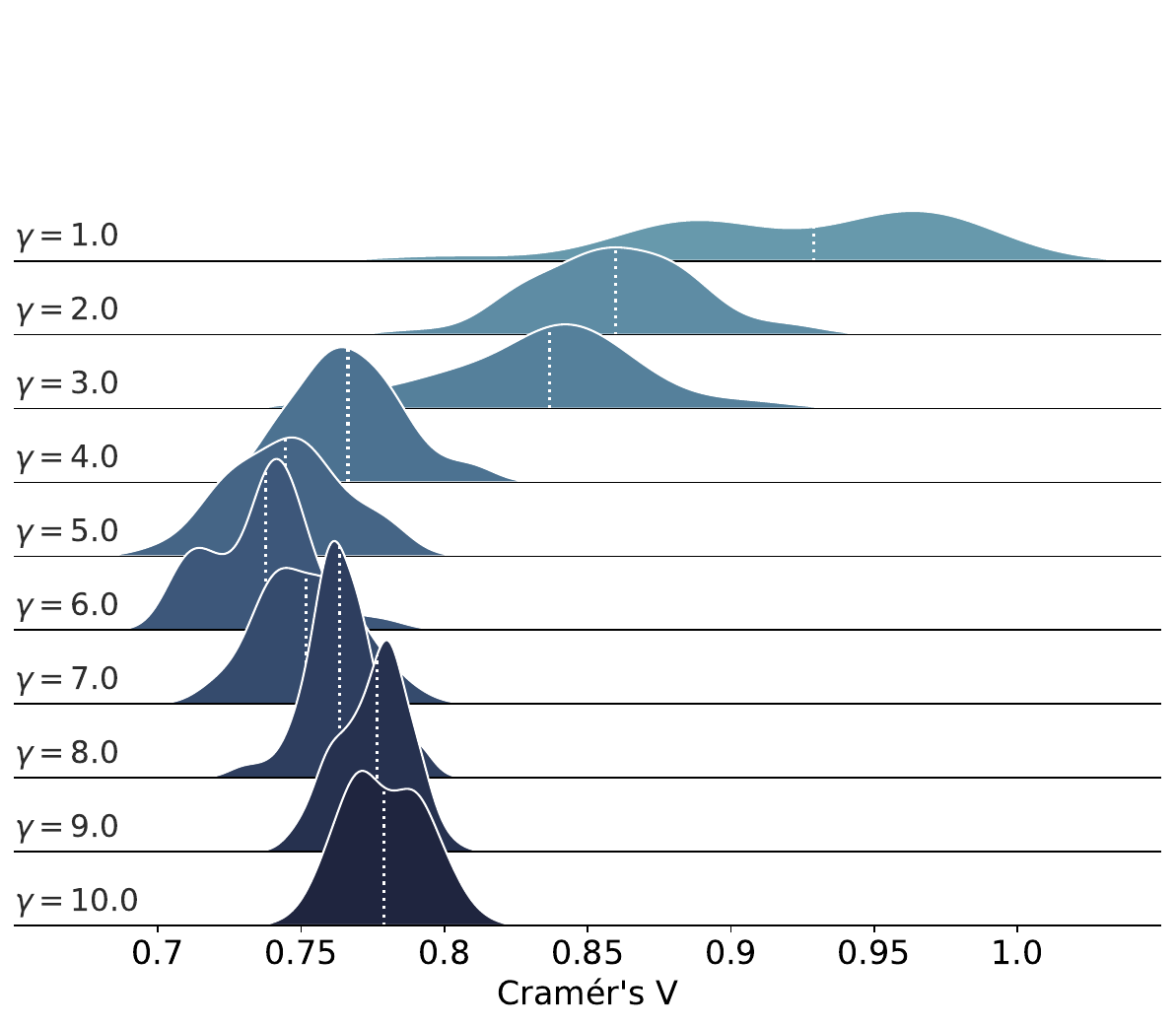}
        \captionsetup{position=bottom,justification=centering}
        \caption{}
        \label{si:fig:cramers_v}
    \end{subfigure}
    \begin{subfigure}[t]{0.495\linewidth}
        \includegraphics[width=\linewidth]{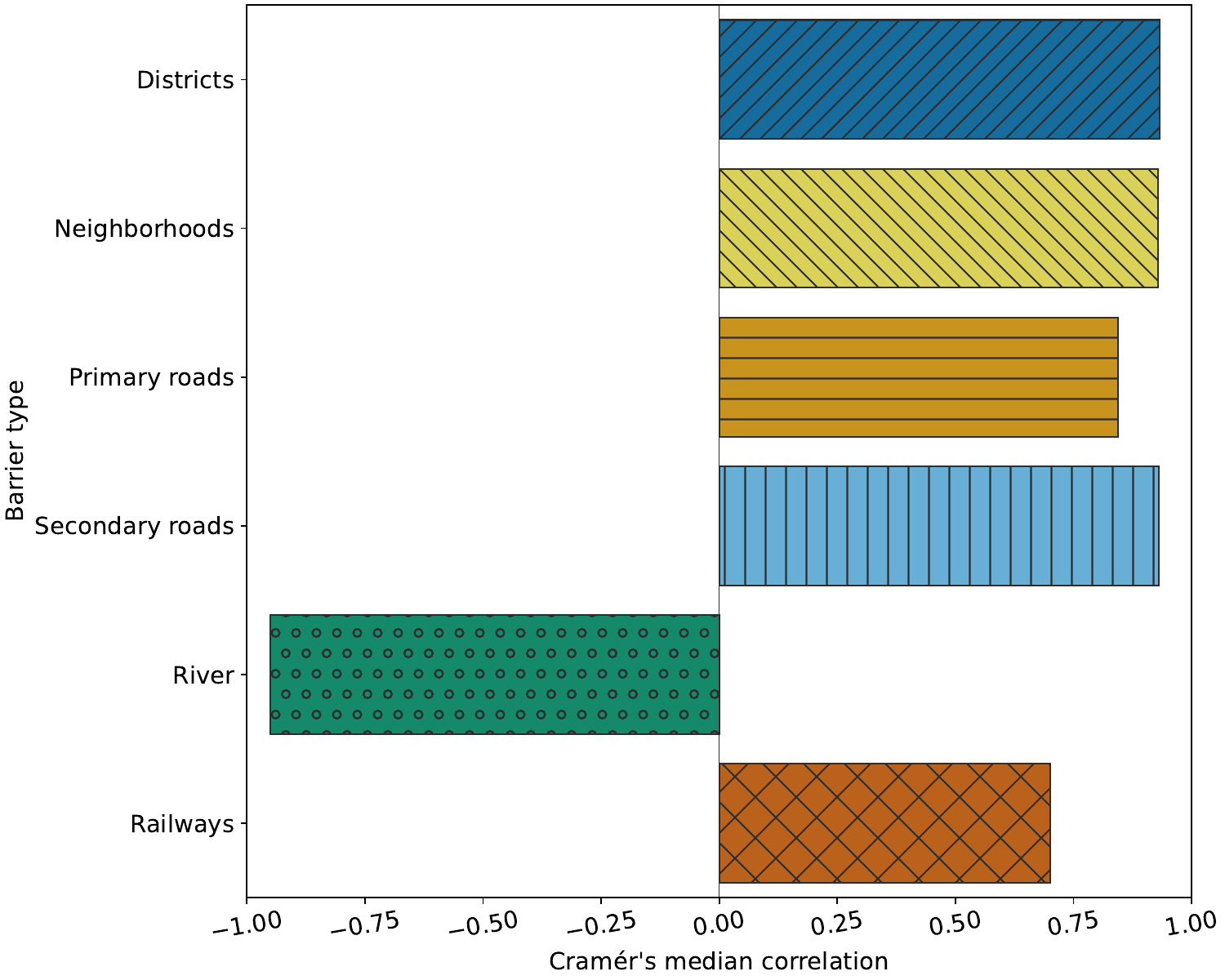}
        \captionsetup{position=bottom,justification=centering}
        \caption{}
        \label{si:fig:cramers_v_corr_with_sad}
    \end{subfigure}
    \caption{
        Cramér's V in contrast to the resolution, showing median values with white dotted lines (\textbf{\subref{si:fig:cramers_v}}), and the correlation (Pearson's R) between the V median and the \acrlong{SAD} (\textbf{\subref{si:fig:cramers_v_corr_with_sad}}).
    }
    \label{si:fig:cramers_v_frame}
\end{figure}

To test the stability of the community detection, the Cramér's V was applied to the communities by resolution.
Figure~\ref{si:fig:cramers_v} shows the kernel density plots of the Cramér's V values by resolution, with the medians using white dotted lines.
At low resolution the median V value is high, but strongly distributed.
The median decreases until $\gamma = 6$, then it starts converging to \num{0.78}, while the distribution gets narrower.
Furthermore, the trend of the medians is analogous to the \acrfull{SAD}.
Figure~\ref{si:fig:cramers_v_corr_with_sad} show the correlations (Pearson's R) between the medians and the \acrshort{SAD} by barrier type.
There are strong positive correlations between the barrier types, except the river where the correlation is strong, but negative as the river behaves differently (Figure~\ref{si:fig:area_symmdiff_river_covid} in contrast to Figure~\ref{fig:symdiff}).

\clearpage
\section{Enclosures}
\label{si:sec:enclosures}

\begin{figure}[ht!]
    \centering
    \begin{subfigure}[t]{0.325\linewidth}
        \includegraphics[width=\linewidth]{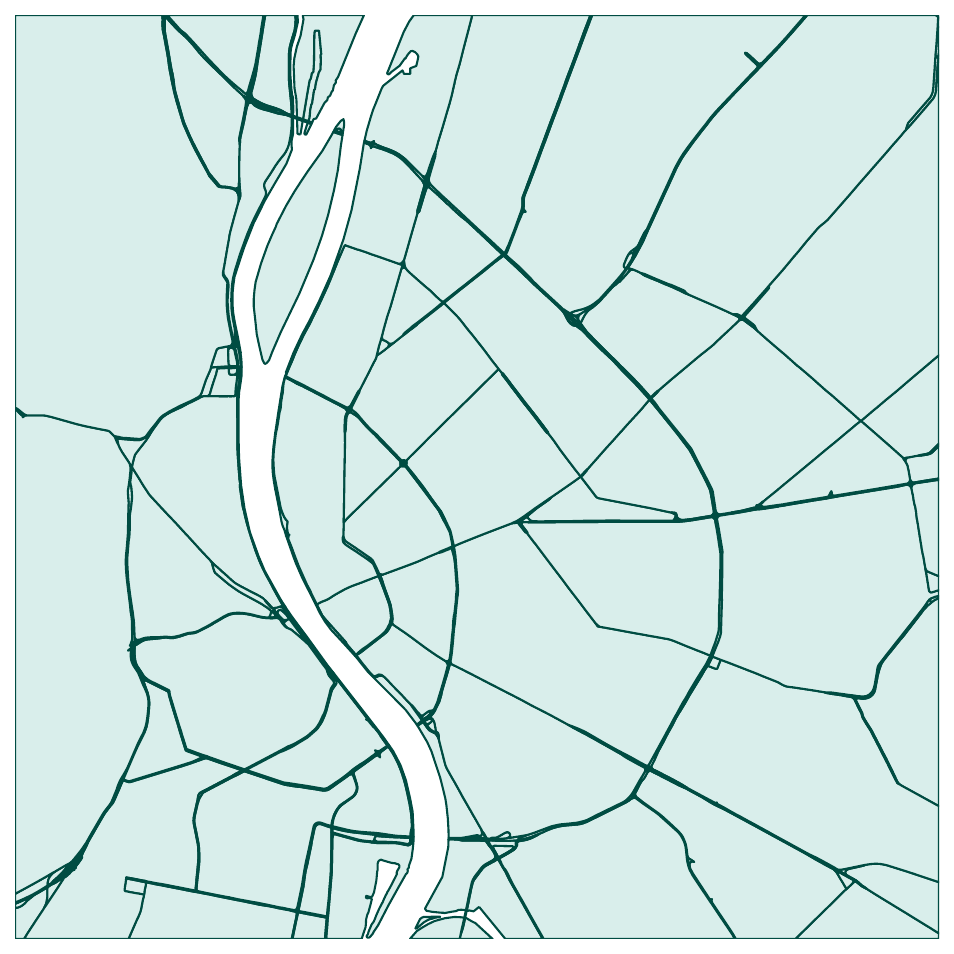}
        \captionsetup{position=bottom,justification=centering}
        \caption{}
        \label{si:fig:enclosures_mps}
    \end{subfigure}
    \begin{subfigure}[t]{0.325\linewidth}
        \includegraphics[width=\linewidth]{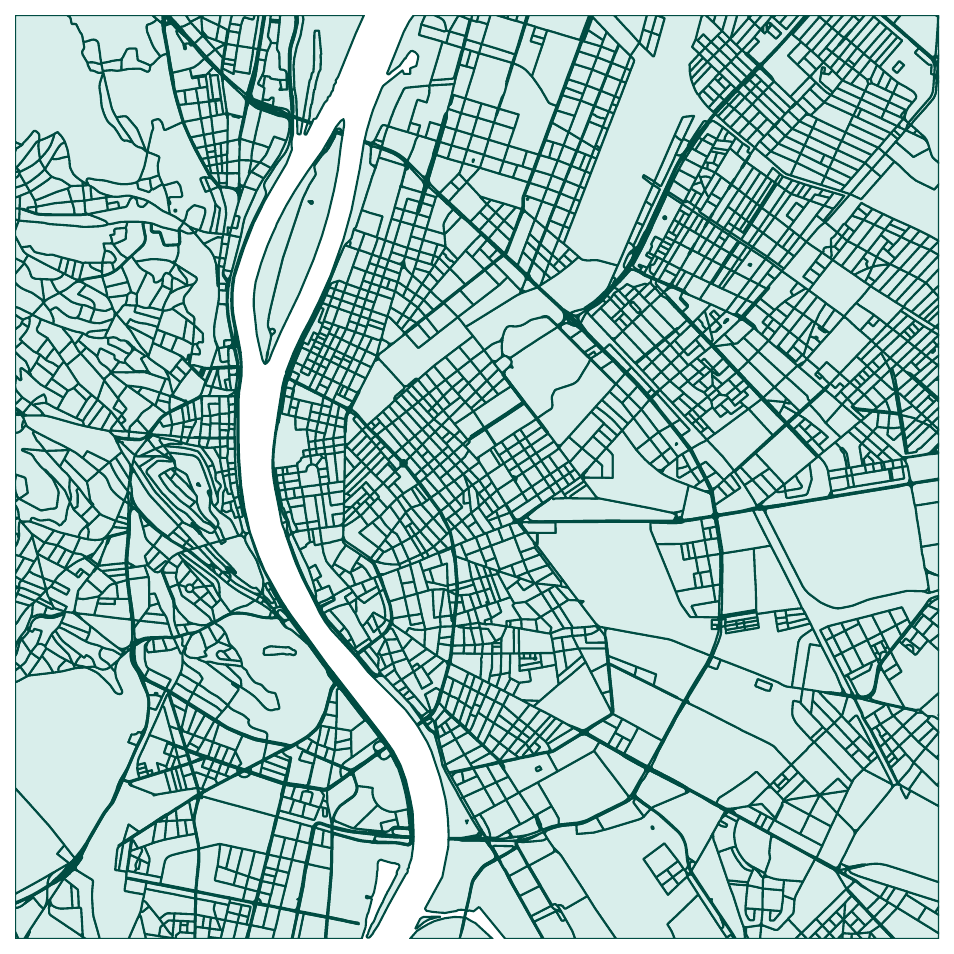}
        \captionsetup{position=bottom,justification=centering}
        \caption{}
        \label{si:fig:house_blocks}
    \end{subfigure}
    \begin{subfigure}[t]{0.325\linewidth}
        \includegraphics[width=\linewidth]{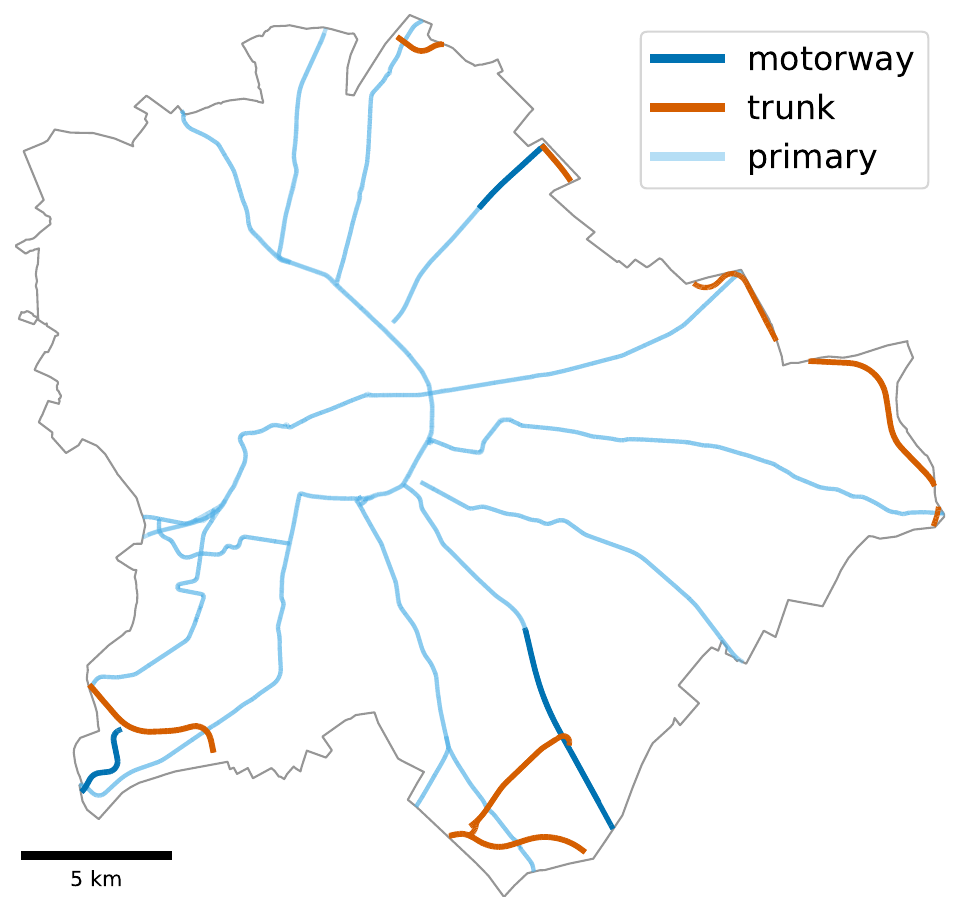}
        \captionsetup{position=bottom,justification=centering}
        \caption{}
        \label{si:fig:trunk_vs_motorway}
    \end{subfigure}
    \caption{
        Enclosure polygons generated from primary and secondary roads (\textbf{\subref{si:fig:enclosures_mps}}) and all roads (\textbf{\subref{si:fig:house_blocks}}), focusing to Budapest downtown.
        The `motorway', `trunk' and `primary' road types within Budapest based on \acrshort*{OSM} data (\textbf{\subref{si:fig:trunk_vs_motorway}}).
    }
    \label{si:fig:enclosures}
\end{figure}

A Python script was developed to obtain enclosed areas from \acrfull{OSM} with different type of roads and other potential barriers, based on \cite{urbangrammarai}.
To select the roads that were considered as barriers, the classification system of \acrshort{OSM} is used.
The script can be parameterized which road types should be included.
The road types are taken into consideration in three levels: (i) ``primary'' including \acrshort*{OSM} motorway and \acrshort*{OSM} primary types, (ii) ``secondary'' which extends the primary with the \acrshort*{OSM} primary type (Figure~\ref{si:fig:enclosures_mps}), and (iii) every road (Figure~\ref{si:fig:house_blocks}).
At the third level, the concept is extended to every lower-order streets until the enclosures became blocks.

Note that trunks are omitted from the analysis in this study.
Trunks and motorways are not significant within the administrative boundaries of Budapest.
There are actually more trunks (\num{\TrunkLength}~km) than freeway (\num{\MotorwayLength}~km) but freeways continue as primary roads whereas trunk are on the periphery (see Figure~\ref{si:fig:trunk_vs_motorway}) as mostly part of the M0 ring freeway.

\clearpage
\section{Symmetric Area Difference Changes}
\label{si:sec:symmetric_area_difference_tendency}

\begin{figure}[ht!]
    \centering
    \begin{subfigure}[t]{0.475\linewidth}
        \includegraphics[width=\linewidth]{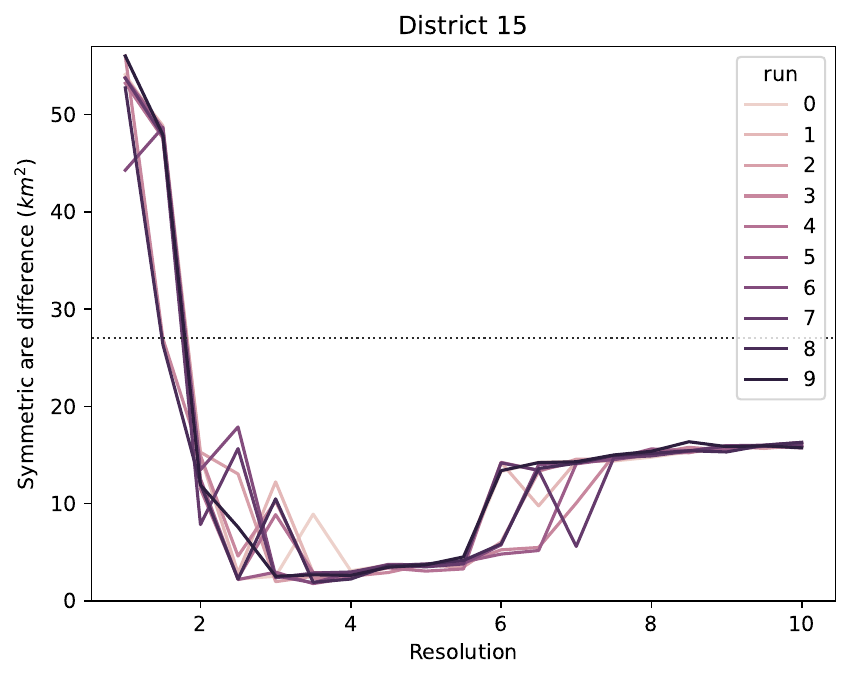}
        \captionsetup{position=bottom,justification=centering}
        \caption{}
        \label{si:fig:sad_15}
    \end{subfigure}
    \begin{subfigure}[t]{0.475\linewidth}
        \includegraphics[width=\linewidth]{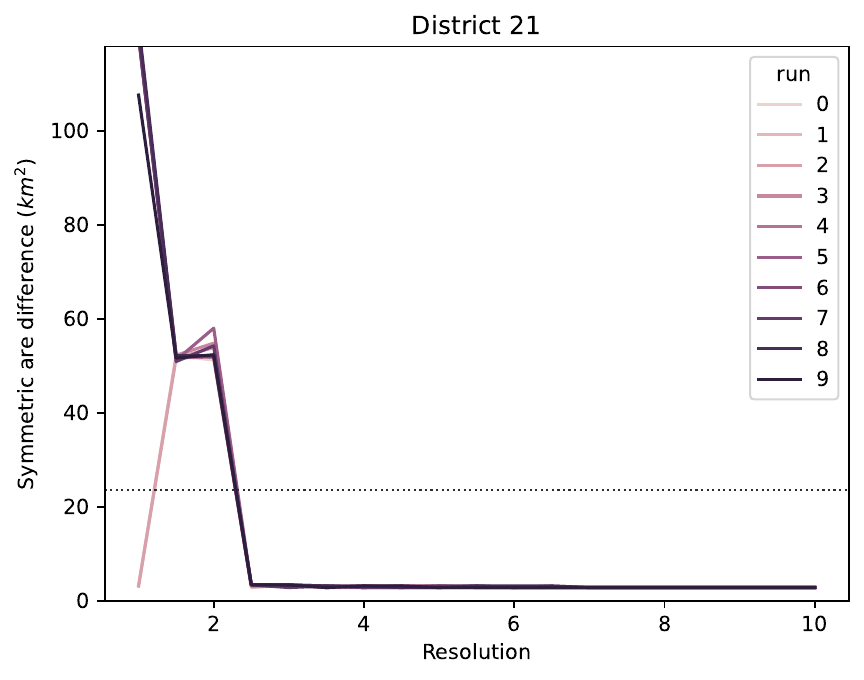}
        \captionsetup{position=bottom,justification=centering}
        \caption{}
        \label{si:fig:sad_21}
    \end{subfigure}
    \caption{
        The changes in symmetric area differences by the resolution parameter of the Louvain community detection, in the cases of District 15 (\textbf{\subref{si:fig:sad_15}}) and District 21 (\textbf{\subref{si:fig:sad_21}}).
        The dotted line represents the area of the district as a reference.
    }
    \label{si:fig:symmetric_area_difference_tendency}
\end{figure}

Figure~\ref{si:fig:symmetric_area_difference_tendency} shows the changes in symmetric area differences \acrshort{SAD} by the resolution parameter ($\gamma$) of the Louvain community detection, for two selected districts.
As stated in the main text, District 21 is on an island, which makes its boundaries very stable (Figure~\ref{si:fig:sad_21}).
The symmetric area difference is minimal from resolution 2.5, and does not change later.
In the case of District 15, it reaches the minimum at about $\gamma=2.5$ (later during some runs), but it remains more or less stable only until $\gamma=5.5$ to $\gamma=6.5$, with some alternation between the community detection runs (Figure~\ref{si:fig:sad_15}).
After that point, the community splits up, and the district loses its community-forming power.

Note that the midstream of river Danube is a natural district border between Buda- and Pest-side districts, but that also means that some water surface is part of the administrative area of the districts.
As water surfaces are not classified to communities, they were removed from the area of the affected districts.

\clearpage
\section{SAD of Railways and River Barriers}
\label{si:sec:sad_railways_and_river}

The main text introduces analysis on the community fit to urban barriers with \acrfull{SAD} focusing on administrative barriers and the road infrastructure, before and during the COVID-19 pandemic.

As for the railways, the fit is not so strong, and even worse before the pandemic.
It has a minimum around $\gamma = 3$.
Interestingly, during the pandemic the barriers effect regarding the railways are strengthened, which is the opposite effect as it was in respect of the roads and the administrative boundaries (Figure~\ref{fig:area_symmdiff_barriers} and \ref{fig:area_symmdiff_administrative} of the main text).

\begin{figure}[ht!]
    \centering
    \begin{subfigure}[t]{0.25\linewidth}
        \includegraphics[width=\linewidth]{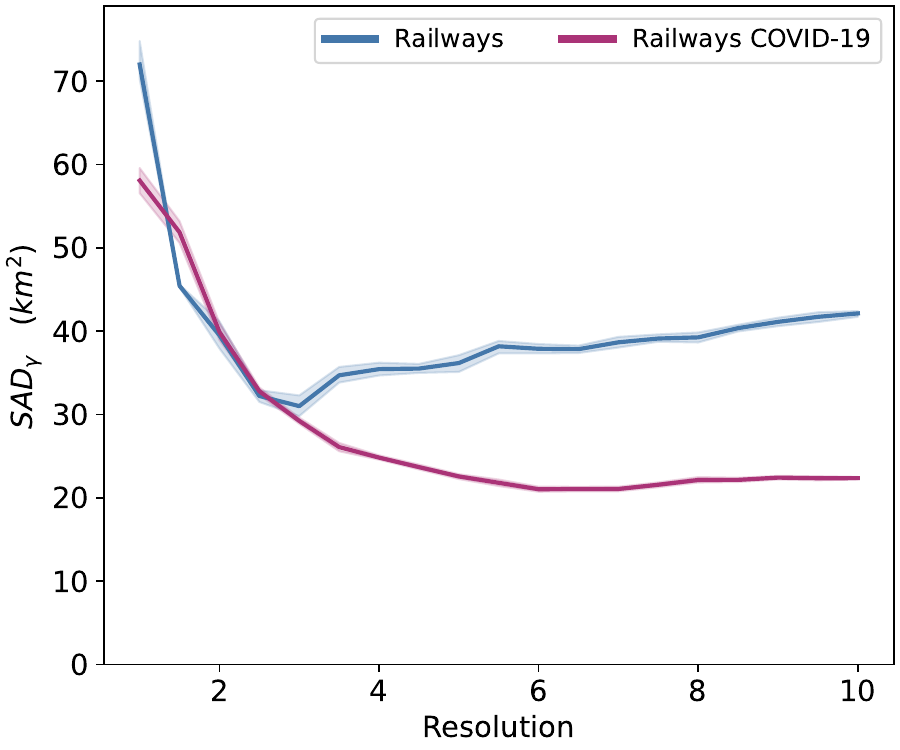}
        \captionsetup{position=bottom,justification=centering}
        \caption{}
        \label{si:fig:area_symmdiff_railways_covid}
    \end{subfigure}
    \begin{subfigure}[t]{0.25\linewidth}
        \includegraphics[width=\linewidth]{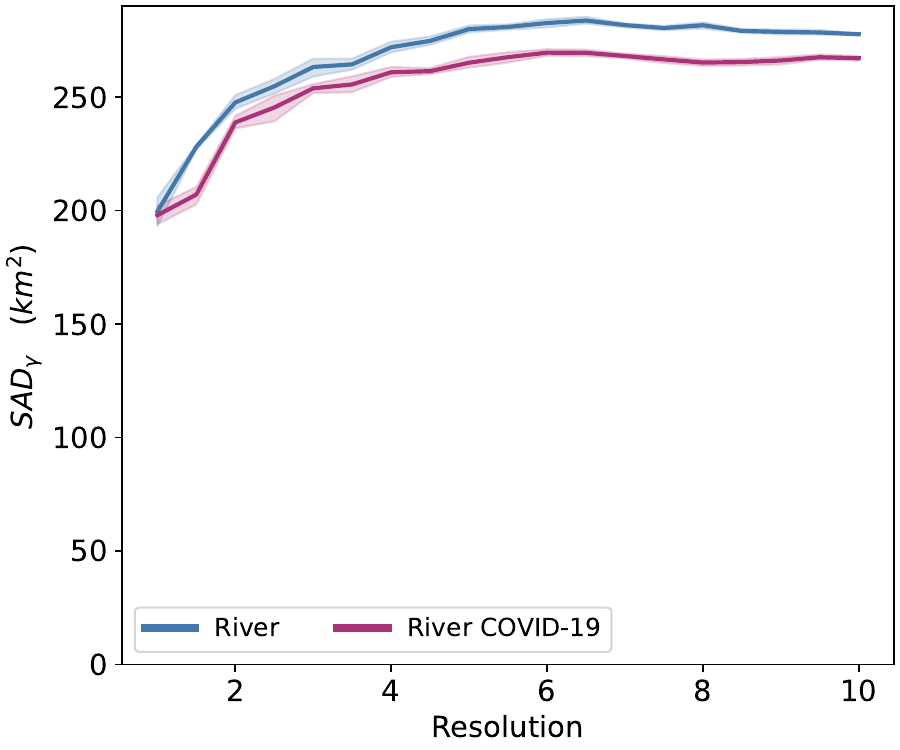}
        \captionsetup{position=bottom,justification=centering}
        \caption{}
        \label{si:fig:area_symmdiff_river_covid}
    \end{subfigure}
    \hfill
    \begin{subfigure}[t]{0.23\linewidth}
        \includegraphics[width=\linewidth]{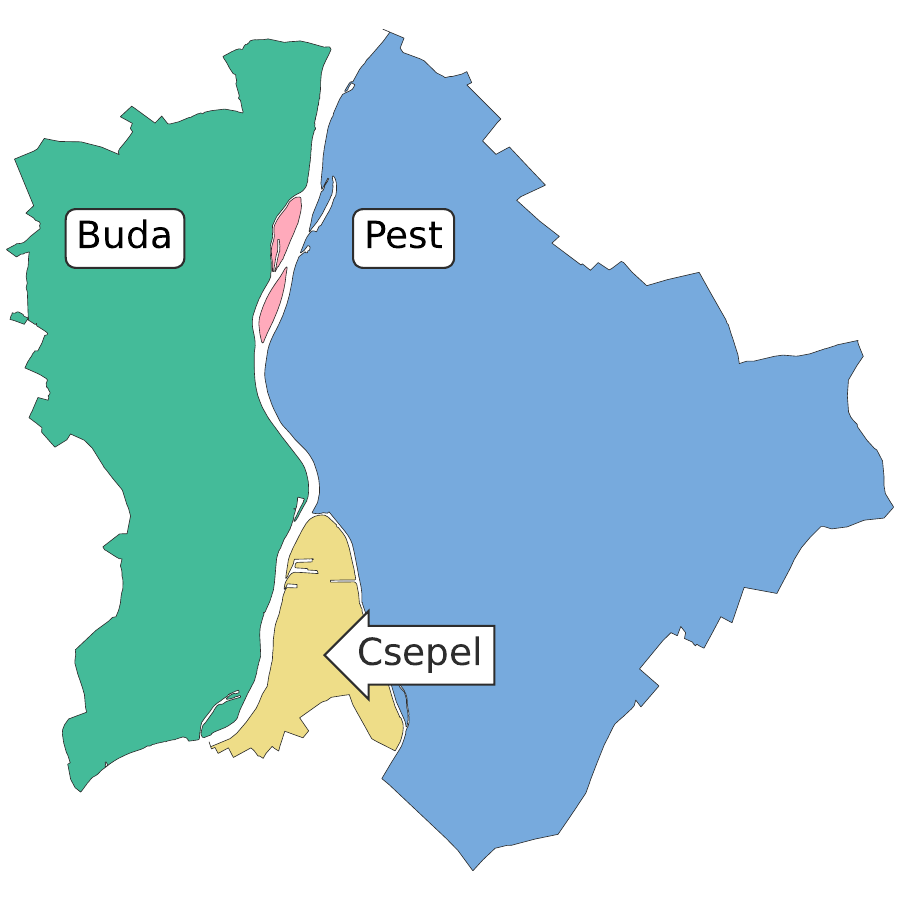}
        \captionsetup{position=bottom,justification=centering}
        \caption{}
        \label{si:fig:river_map}
    \end{subfigure}
    \begin{subfigure}[t]{0.23\linewidth}
        \includegraphics[width=\linewidth]{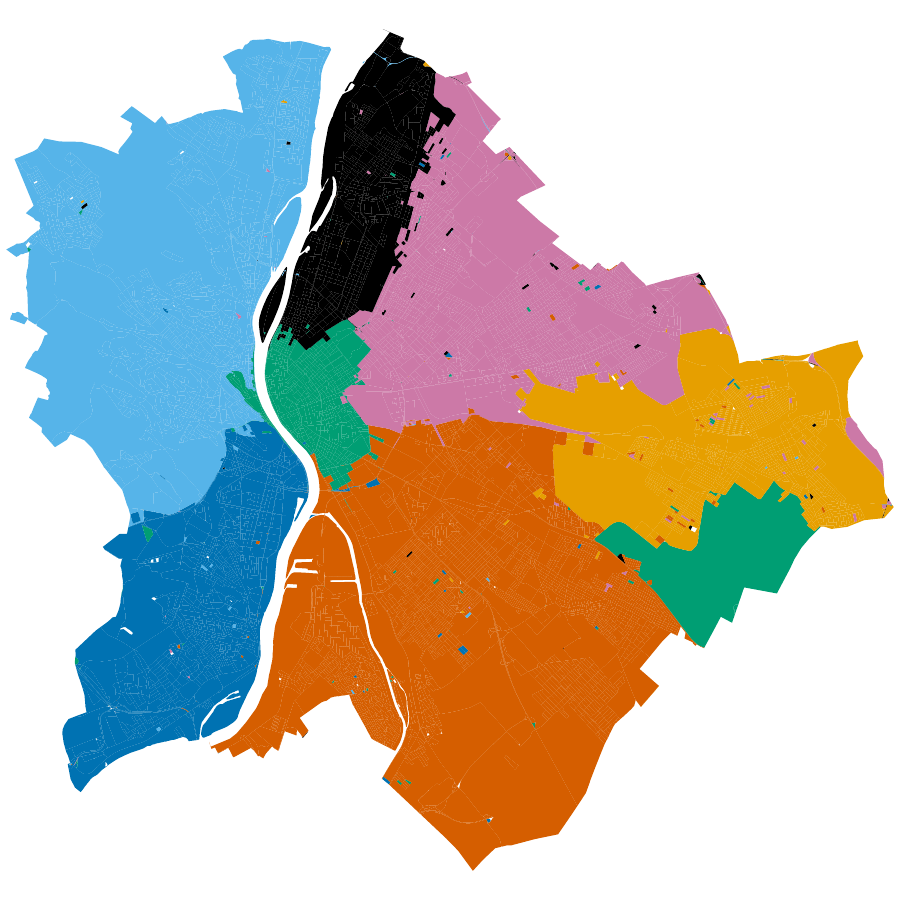}
        \captionsetup{position=bottom,justification=centering}
        \caption{}
        \label{si:fig:r1}
    \end{subfigure}
    \caption{
        The \acrfull{SAD} regarding the railway lines (\textbf{\subref{si:fig:area_symmdiff_railways_covid}}) and the river Danube (\textbf{\subref{si:fig:area_symmdiff_river_covid}}).
        In (\textbf{\subref{si:fig:area_symmdiff_railways_covid}}, \textbf{\subref{si:fig:area_symmdiff_river_covid}}), the lines and shaded area represent the mean and the 95 $\%$ confidence interval of the \acrshort{SAD} indicator as a measure of barrier fit of mobility clusters detected in ten iterations of the Louvain community detection algorithm.
        The river Danube divides Budapest into three main parts, illustrated in (\textbf{\subref{si:fig:river_map}}).
        Also, (\textbf{\subref{si:fig:r1}}) shows a possible community partition at $\gamma = 1$ with the Louvain community detection.
    }
    \label{si:fig:sad_railways_and_river}
\end{figure}

Figure~\ref{si:fig:area_symmdiff_river_covid} show the \acrshort{SAD} of the river enclosures.
The river Danube divides Budapest to three parts (Figure~\ref{si:fig:river_map}), excluding Margaret Island and Óbuda Island: Buda, Pest and Csepel (district 21).
Although river midstream is the boundary between the Buda and Pest-side districts and definitely a strong natural barrier, the river Danube on its own does not create complete communities, except for District~21 (see Figure~\ref{si:fig:sad_21}).
It is because Buda and Pest does not form a community, even at $\gamma = 1$ there are multiple communities both in Buda and Pest (Figure~\ref{si:fig:r1}).

Also note that the $\gamma = 1$ communities are roughly approximate the \acrshort{HCSO} district group partitioning (Figure~\ref{si:fig:map_orig_comm}), except that Csepel (District 21) is belongs to the South-Pest area and there is a larger green area at the eastern part, where the airport is located, that is in the same community as the Inner-Pest.
There is not much residential are near the airport that can explain the size of that community part, but the airport has strong transportation ties to the city center.

\clearpage
\section{Barrier Crossing Ratio by home locations}
\label{si:sec:bcr_by_residence_plus}

\begin{figure}[!ht]
    \centering
    \begin{subfigure}[t]{0.285\linewidth}
        \includegraphics[width=\linewidth]{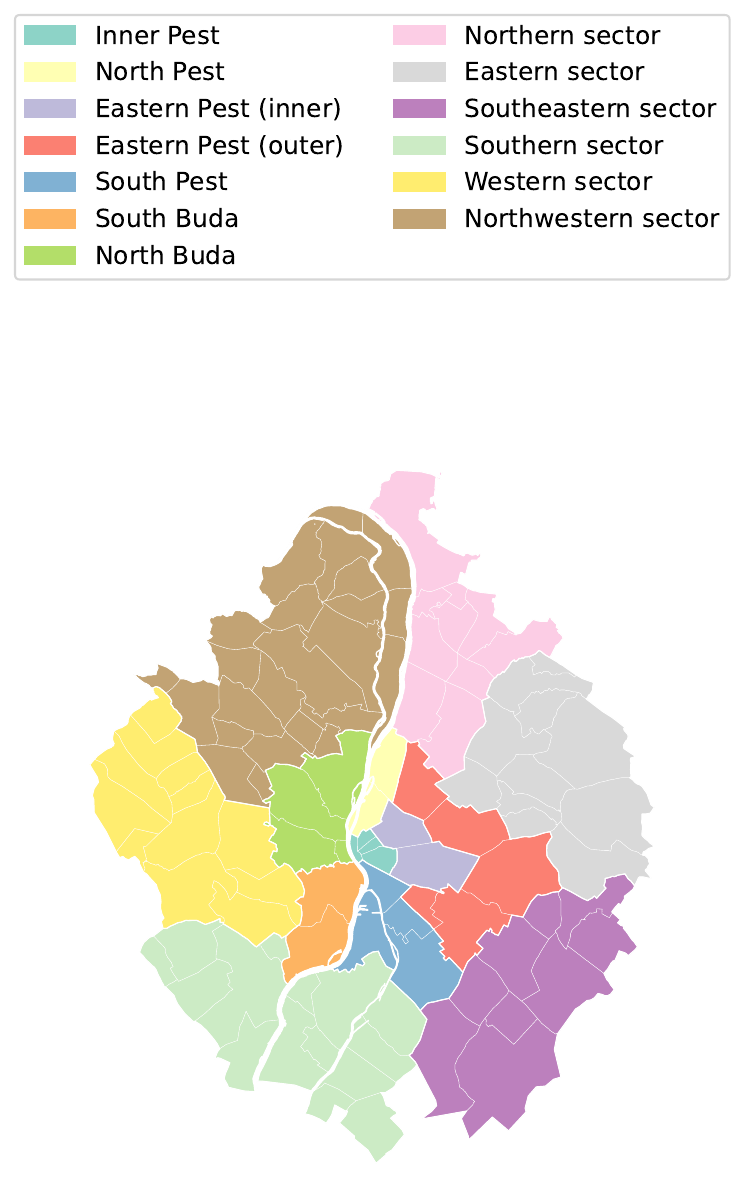}
        \captionsetup{position=bottom,justification=centering}
        \caption{}
        \label{si:fig:map_orig_comm}
    \end{subfigure}
    \hfill
    \begin{subfigure}[t]{0.7\linewidth}
        \includegraphics[width=\linewidth]{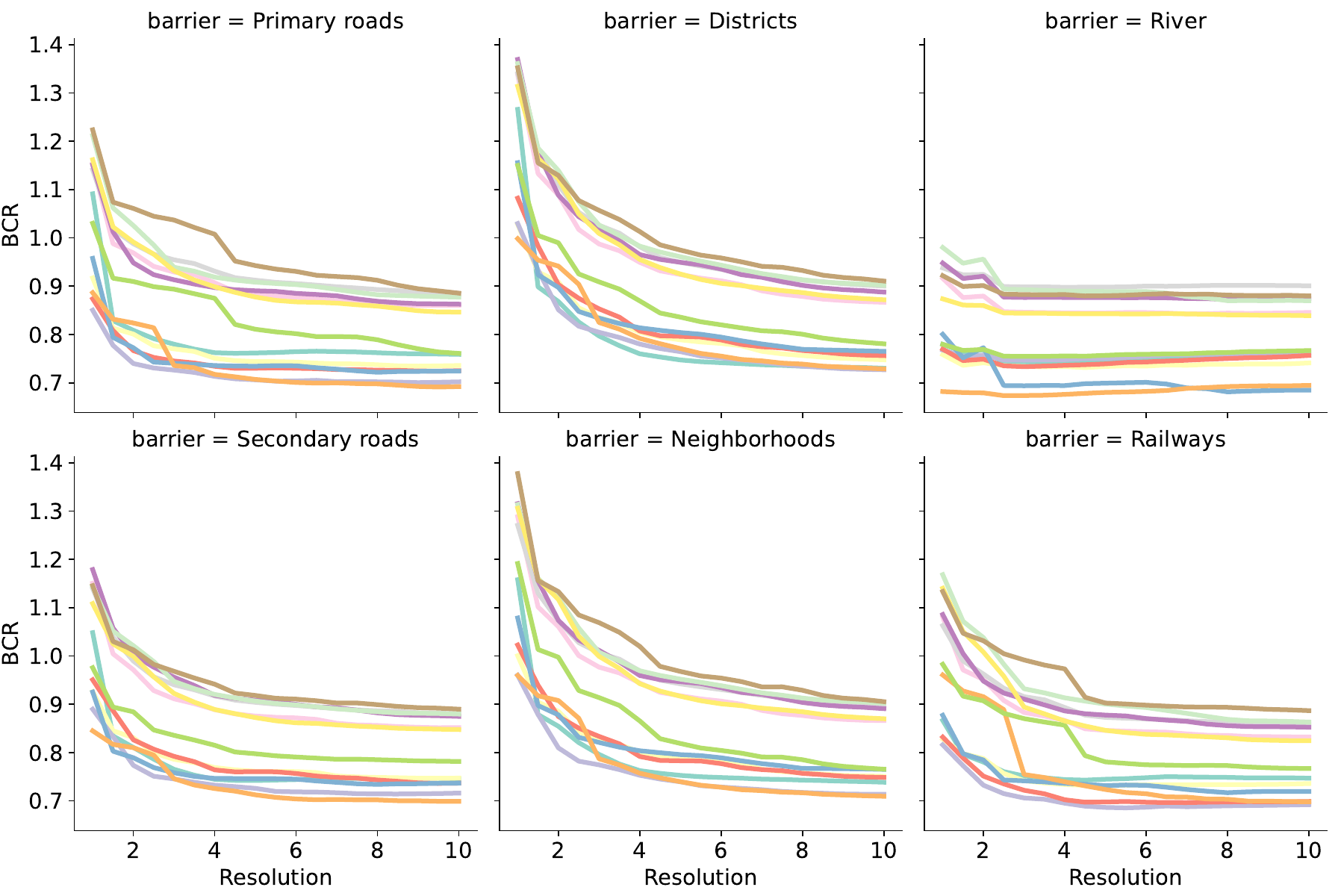}
        \captionsetup{position=bottom,justification=centering}
        \caption{}
        \label{si:fig:by_barrier_orig_comm}
    \end{subfigure}
    \caption{
        Barrier effect based on the users' home location groups compared to the communities of the original network.
        Figure~\textbf{\subref{si:fig:map_orig_comm}} shows the seven district groups in Budapest and six sectors of the agglomeration, defined by \acrshort{HCSO}.
        Figure \textbf{\subref{si:fig:by_barrier_orig_comm}} shows the changes of the crossing ratio regarding the type of the barrier.
    }
    \label{si:fig:coeff_by_home_orig_comm}
\end{figure}

Figure~\ref{si:fig:coeff_by_home_orig_comm} shows the same as Figure~\ref{fig:coeff_by_home} of the main text, but instead of highlighting the Budapest--agglomeration difference, it uses the different colors for every sector of the agglomeration and district groups of Budapest.

This representation can reveal additional details.
In the case of the North Buda regarding the primary roads, the \acrshort{BCR} behaves more like the agglomeration at small $\gamma$, but more like the rest of Budapest at higher resolution.
The reason may be that the Buda-side districts are not divided by primary roads (see Figure~\ref{si:fig:trunk_vs_motorway}), so primary roads cannot form the communities in that area.

\begin{figure}[!ht]
    \centering
    \begin{subfigure}[t]{0.285\linewidth}
        \includegraphics[width=\linewidth]{13/map_with_legend}
        \captionsetup{position=bottom,justification=centering}
        \caption{}
        \label{si:fig:map_own_comm}
    \end{subfigure}
    \hfill
    \begin{subfigure}[t]{0.7\linewidth}
        \includegraphics[width=\linewidth]{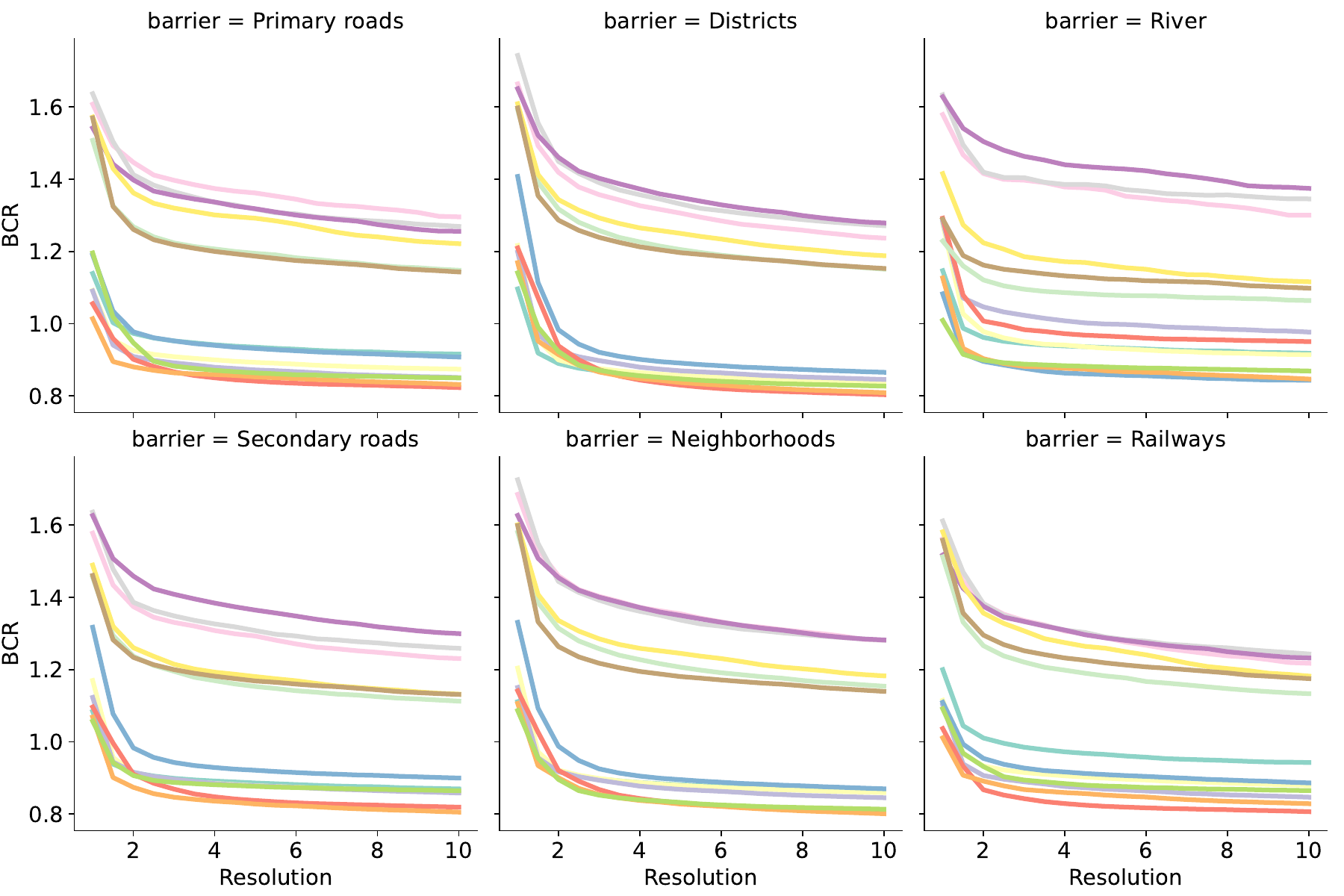}
        \captionsetup{position=bottom,justification=centering}
        \caption{}
        \label{si:fig:by_barrier_own_comm}
    \end{subfigure}
    \caption{
        Barrier effect based on the users' home location groups compared to the communities of each subnetwork.
        Figure \textbf{\subref{si:fig:by_barrier_own_comm}} shows the changes of the crossing ratio regarding the type of the barrier.
    }
    \label{si:fig:coeff_by_home_own_comm}
\end{figure}

The pre-pandemic network has been divided based on the home locations of the users to thirteen networks.
However, the \acrshort{BCR} was compared to the original network.
Another approach is to determine communities for the new networks and compare the \acrshort{BCR} to the those communities.
Figure~\ref{si:fig:coeff_by_home_own_comm} show the \acrshort{BCR} compared to the communities of each subnetwork, constructed for the seven district groups of Budapest and six sectors of the agglomeration (defined by  \acrshort{HCSO}).

The result also shows clear distinction between the Budapest and the agglomeration areas, except the river, where the western sectors of the agglomeration behave much closer to the city than to the eastern sectors.
The city center is more complex and attracts more diverse people \cite{juhasz2023amenity}.
Crossing the river Danube seems more probable from the western agglomeration than from the eastern sectors.
This assumption is supported by the visitation analysis, presented in Section~\ref{si:sec:stop_distribution}.
Furthermore, the order of the sectors remains practically the same for each barrier type.
The \acrshort{BCR} value of the eastern sectors are higher than the western sectors' for every $\gamma$ value.

\clearpage
\subsection{Home-location-based Network Details}
\label{si:sec:home_group_network_details}

\begin{figure}[!ht]
    \centering
    \begin{subfigure}[t]{0.300\linewidth}
        \includegraphics[width=\linewidth]{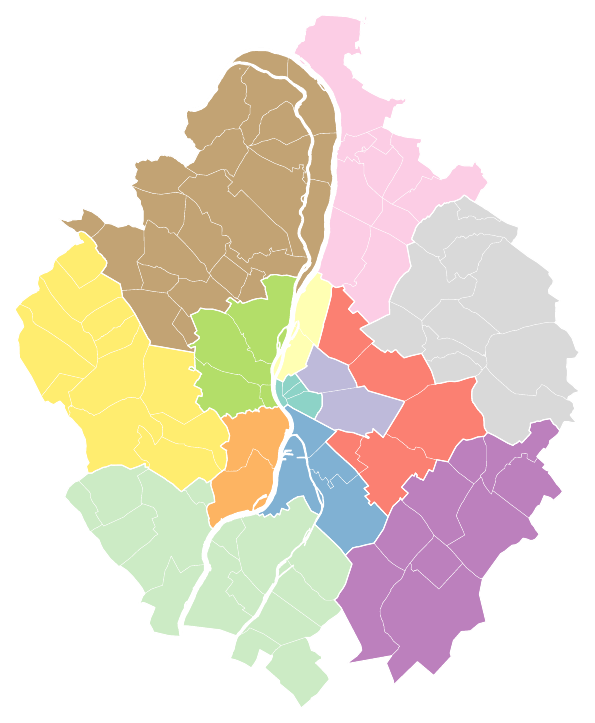}
        \captionsetup{position=bottom,justification=centering}
        \caption{}
        \label{si:fig:map_without_legend}
    \end{subfigure}
    \hfill
    \begin{subfigure}[t]{0.6\linewidth}
        \includegraphics[width=\linewidth]{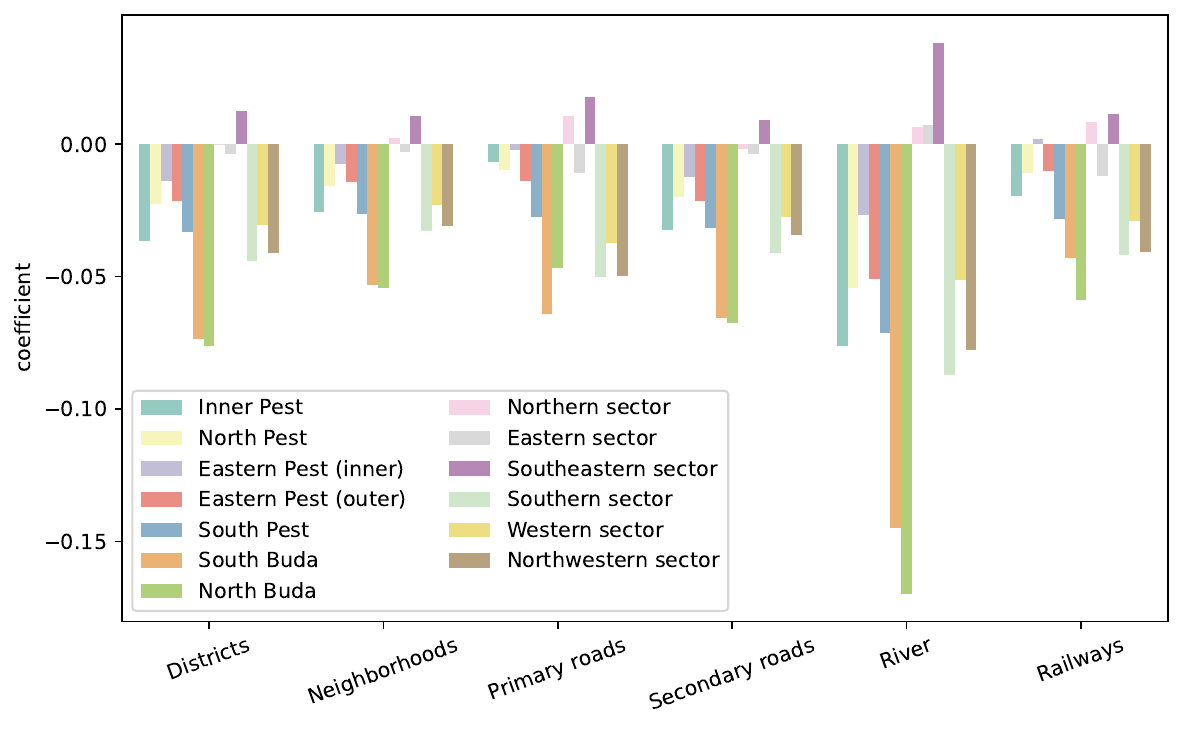}
        \captionsetup{position=bottom,justification=centering}
        \caption{}
        \label{si:fig:coefficients}
    \end{subfigure}

    \vfill

    \begin{subfigure}[t]{0.45\linewidth}
        \includegraphics[width=\linewidth]{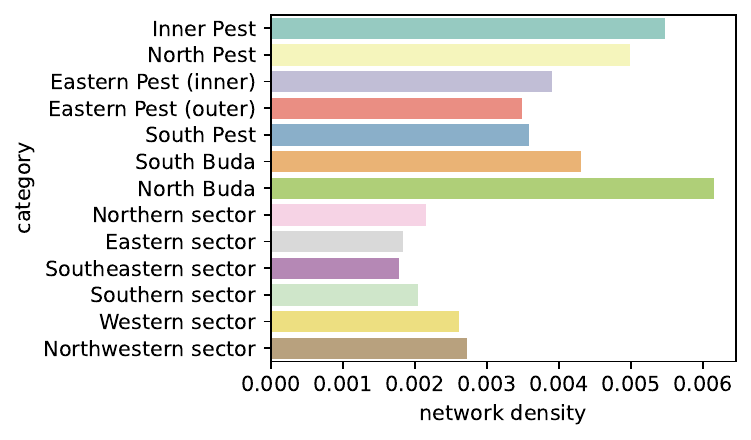}
        \captionsetup{position=bottom,justification=centering}
        \caption{}
        \label{si:fig:density}
    \end{subfigure}
    \hfill
    \begin{subfigure}[t]{0.45\linewidth}
        \includegraphics[width=\linewidth]{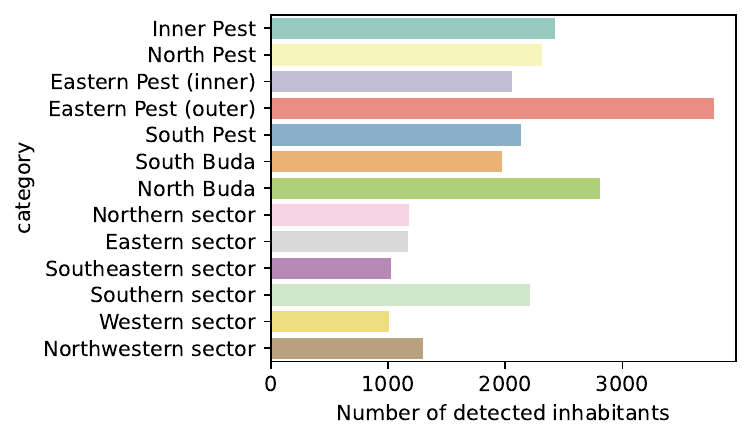}
        \captionsetup{position=bottom,justification=centering}
        \caption{}
        \label{si:fig:inhabitants}
    \end{subfigure}

    \caption{
        Network details and the barrier effect based on the users' home location groups.
        The \acrshort{HCSO} defines seven district groups in Budapest and six sectors of the agglomeration (\textbf{\subref{si:fig:map_without_legend}}).
        From the Budapest mobility of the inhabitants of these areas, 13 networks was built and evaluated.
        Network density shown in \textbf{\subref{si:fig:density}}, the detected number of inhabitant in \textbf{\subref{si:fig:inhabitants}}, and the gravity model coefficients are displayed in \textbf{\subref{si:fig:coefficients}}.
    }
    \label{si:fig:coeff_by_home_details}
\end{figure}

Figure~\ref{si:fig:density} shows the density of the network.
People living in Budapest create more dense networks than the people living in the agglomeration.
Figure~\ref{si:fig:inhabitants} shows the number of the detected inhabitant to give some context to the network density.

The coefficients of the gravity model are displayed in Figure~\ref{si:fig:coefficients} by barrier categories and home groups.
North and South Buda stand closely together in almost every barrier category and differ from the Pest-side district groups.
Among the agglomeration sectors there are also some clustering: Northern, Eastern and Southeastern sectors (East of the capital) significantly differ from the Southern, Western and Southwestern sectors.

\clearpage
\section{Budapest Stops of Inhabitants Living in the Agglomeration}
\label{si:sec:stop_distribution}

\begin{figure}[!ht]
    \centering
    \includegraphics[width=\textwidth]{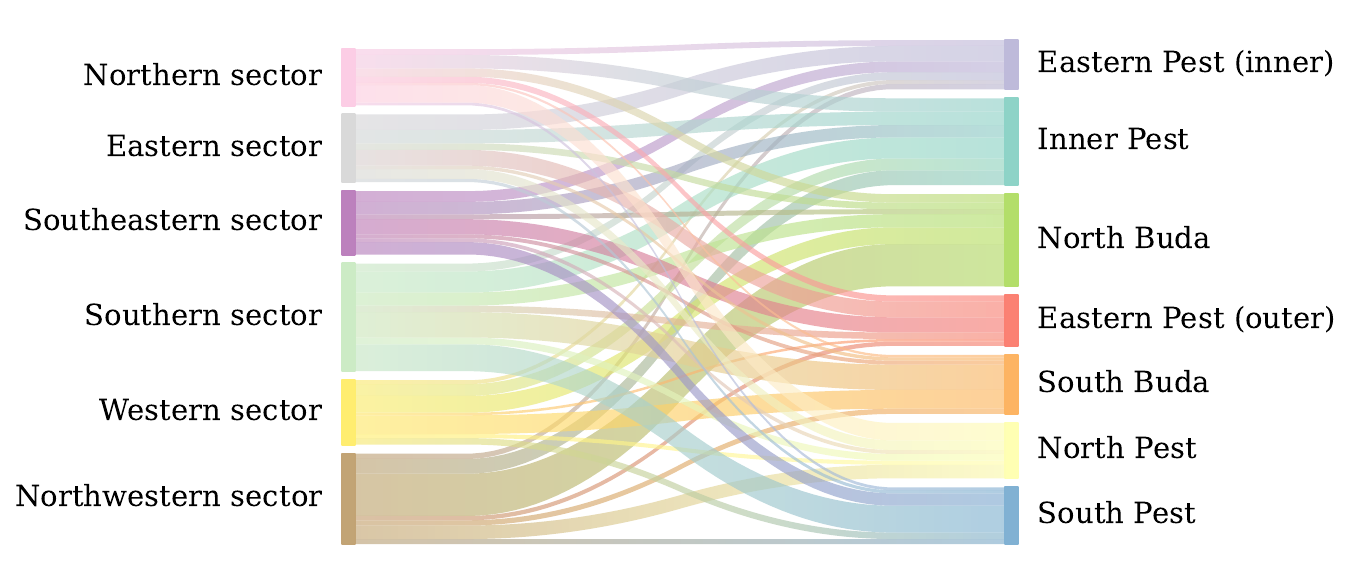}
    \caption{Visitation (stop) distribution within Budapest of the people living in the agglomeration.
    }
    \label{si:fig:stop_distribution}
\end{figure}

The commuting patterns from the agglomeration sectors to the district groups of Budapest are regularly analyzed by the census data and already studied using mobile network data \cite{pinter2022commuting}.
Based on the mobile positioning data, utilized in this study, the visitation patterns of Budapest areas from the agglomeration is determined (Figure~\ref{si:fig:stop_distribution} and Table~\ref{tab:stop_distribution}).

Figure~\ref{si:fig:stop_distribution} shows the distribution of visitations from the agglomeration sectors to the Budapest district groups, using the same colors as in Figure~\ref{si:fig:map_own_comm}.
It is also in numerical format in Table~\ref{tab:stop_distribution}, where the rows represent the district groups and the columns represent the agglomeration sectors.

As expected, most of the visit from the eastern sectors target the Pest-side (East of the river) and the city center, while the majority of the visitations from the western sectors trend to the Buda-side (West of the river) and also to the city center.

\begin{table}[ht]
\caption{Stop distribution in the district groups (rows) of inhabitants living in the agglomeration sectors (column).}
\label{tab:stop_distribution}
\begin{tabular}{lrrrrrr}
\toprule
sector & Eastern & Northern & Northwestern & Southeastern & Southern & Western \\
district group &  &  &  &  &  &  \\
\midrule
Eastern Pest (inner) & 0.2284 & 0.1031 & 0.0579 & 0.1697 & 0.0734 & 0.0609 \\
Eastern Pest (outer) & 0.2338 & 0.1086 & 0.0478 & 0.2399 & 0.0606 & 0.0395 \\
Inner Pest & 0.1978 & 0.2385 & 0.1637 & 0.1959 & 0.1954 & 0.1838 \\
North Buda & 0.0982 & 0.1484 & 0.4664 & 0.0780 & 0.1223 & 0.2606 \\
North Pest & 0.1434 & 0.3093 & 0.1494 & 0.0603 & 0.0634 & 0.0635 \\
South Buda & 0.0508 & 0.0419 & 0.0576 & 0.0616 & 0.2354 & 0.2917 \\
South Pest & 0.0475 & 0.0504 & 0.0571 & 0.1946 & 0.2495 & 0.1000 \\
\bottomrule
\end{tabular}
\end{table}

\clearpage
\section{Aggregated Barrier Crossing Ratio}

The Section~\nameref{sec:crossing_ratio} of the main text introduces the \acrfull{BCR}, which measures the relationship between those mobility events that cross barriers and those that cross barriers and detected communities as well.

\begin{figure}[!ht]
    \begin{subfigure}[t]{0.485\linewidth}
        \includegraphics[width=\linewidth]{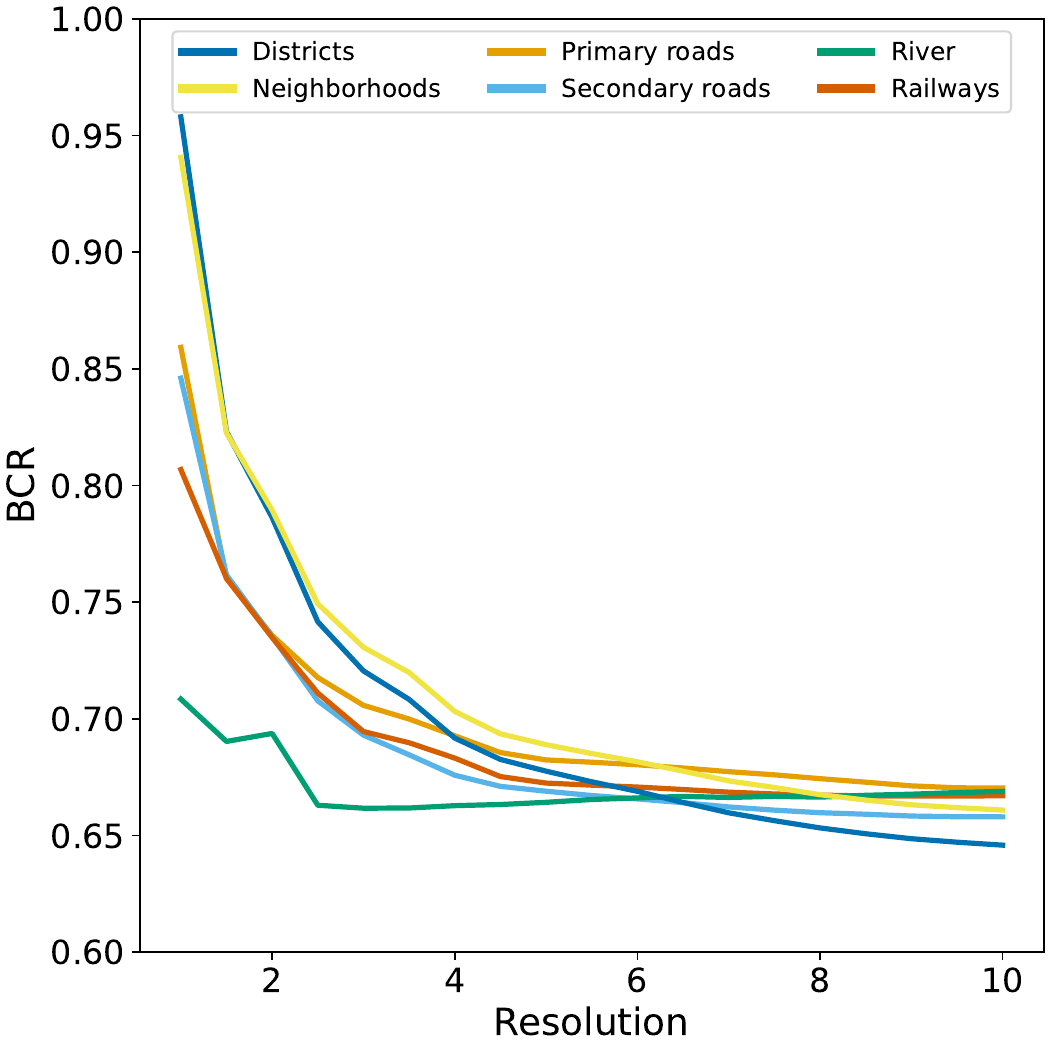}
        \captionsetup{position=bottom,justification=centering}
        \caption{}
        \label{si:fig:crossing_ratio}
    \end{subfigure}
    \hfill
    \begin{subfigure}[t]{0.485\linewidth}
        \includegraphics[width=\linewidth]{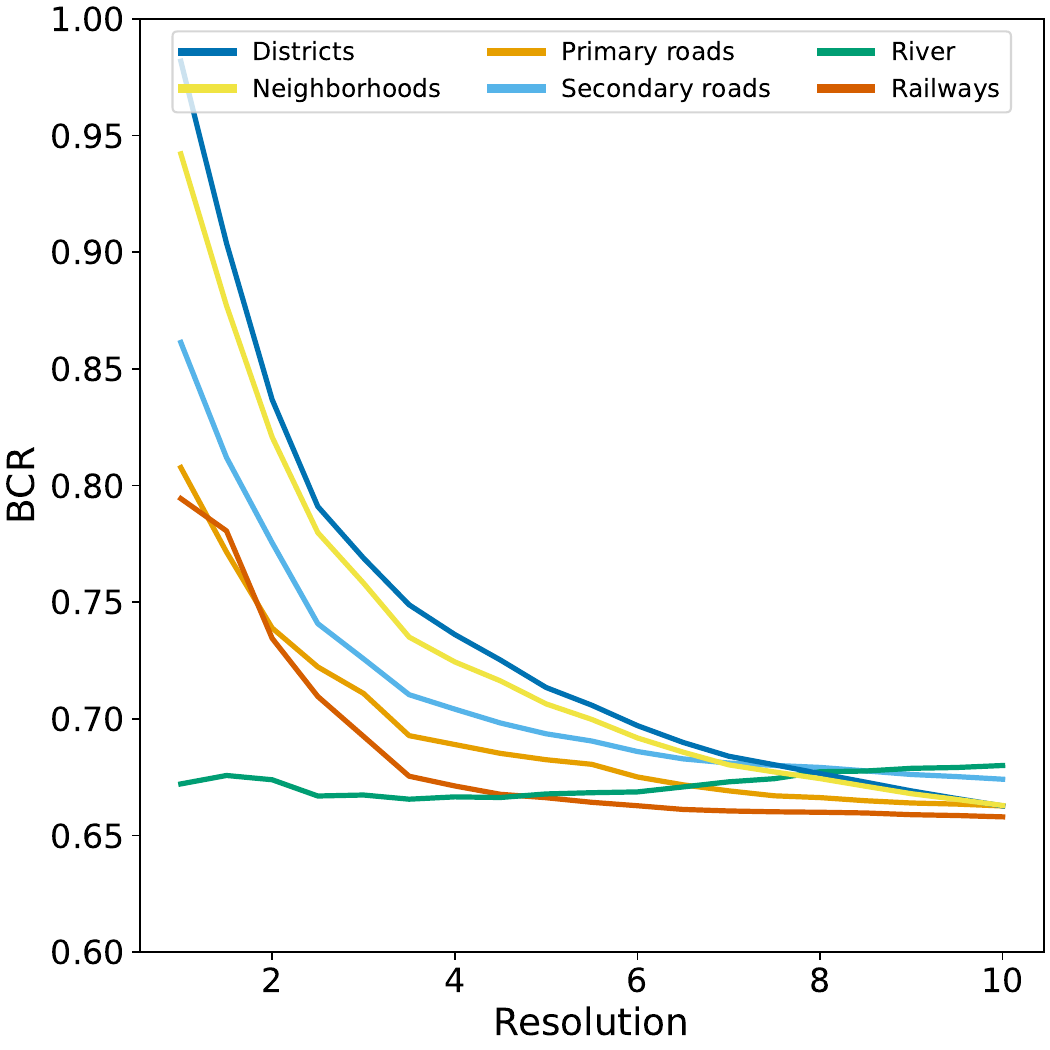}
        \captionsetup{position=bottom,justification=centering}
        \caption{}
        \label{si:fig:crossing_ratio_covid}
    \end{subfigure}
    \label{si:fig:bcr}
    \caption{
        Aggregated \acrfull{BCR} for the pre-pandemic (\textbf{\subref{si:fig:crossing_ratio}}) and the pandemic data sets (\textbf{\subref{si:fig:crossing_ratio_covid}}).
    }
\end{figure}

Figure~\ref{si:fig:crossing_ratio} shows the change of the \acrshort{BCR} changes according to the resolution ($\gamma$) during the pre-pandemic interval while Figure~\ref{si:fig:crossing_ratio_covid} shows the same during the pandemic.
The two intervals are described in Section~\ref{si:sec:covid}.

The \acrshort{BCR} reaches roughly the same value at higher $\gamma$ in both cases, however in the pre-pandemic data set that happens faster.
This might be the result of the decreased mobility during the pandemic.

\clearpage
\section{Complexity OLS tables}
\label{si:sec:complexity_ols_tables}

The \acrfull{BCR} is regressed with an \acrfull{OLS} linear regression, using the formula (\ref{eq:bcr}) from the main text.
The explanatory variables are the geographical distance of the user's home location from the city center and the complexity of amenity portfolio of the visited location as proposed by \cite{juhasz2023amenity}.
The \acrshort{BCR} is explained separately by the barrier types of primary roads (Table~\ref{tab:road1}), secondary roads (Table~\ref{tab:road2}), districts (Table~\ref{tab:districts}), neighborhoods (Table~\ref{tab:neighborhoods}),  rivers (Table~\ref{tab:river}), and railways (Table~\ref{tab:railways}), and for increasing resolution ($\gamma$) values.

\newcolumntype{L}{>{\centering\tiny\arraybackslash}l}
\newcolumntype{C}{>{\centering\tiny\arraybackslash}c}

\begingroup
\renewcommand{\arraystretch}{0.85} 
\begin{table}[ht]
\caption{Primary roads}
\label{tab:road1}
\begin{tabularx}{\textwidth}{LLLLLLLLLLL}
\toprule
\multicolumn{11}{C}{dependent variable: $BCR^{\text{Primary roads}}$} \\
 & $\gamma = 1$ & $\gamma = 2$ & $\gamma = 3$ & $\gamma = 4$ & $\gamma = 5$ & $\gamma = 6$ & $\gamma = 7$ & $\gamma = 8$ & $\gamma = 9$ & $\gamma = 10$ \\
\midrule
intercept & 2.085$^{\ast\ast\ast}$ & 0.669$^{\ast\ast\ast}$ & -0.009 & -0.014 & 0.089 & -0.639 & -0.124 & 0.395$^{\ast\ast\ast}$ & 0.311$^{\ast\ast\ast}$ & 0.419$^{\ast\ast\ast}$ \\
 & (0.319) & (0.184) & (0.137) & (0.131) & (0.120) & (0.644) & (0.202) & (0.088) & (0.115) & (0.087) \\
distance (log10) & -0.117 & 0.180$^{\ast\ast\ast}$ & 0.338$^{\ast\ast\ast}$ & 0.329$^{\ast\ast\ast}$ & 0.288$^{\ast\ast\ast}$ & 0.489$^{\ast\ast\ast}$ & 0.337$^{\ast\ast\ast}$ & 0.188$^{\ast\ast\ast}$ & 0.208$^{\ast\ast\ast}$ & 0.177$^{\ast\ast\ast}$ \\
 & (0.084) & (0.048) & (0.036) & (0.035) & (0.032) & (0.170) & (0.053) & (0.023) & (0.030) & (0.023) \\
complexity & 0.150$^{\ast\ast\ast}$ & -0.137$^{\ast\ast\ast}$ & -0.094$^{\ast\ast\ast}$ & -0.097$^{\ast\ast\ast}$ & -0.077$^{\ast\ast\ast}$ & -0.081 & -0.086$^{\ast\ast\ast}$ & -0.092$^{\ast\ast\ast}$ & -0.076$^{\ast\ast\ast}$ & -0.089$^{\ast\ast\ast}$ \\
 & (0.027) & (0.016) & (0.012) & (0.011) & (0.010) & (0.054) & (0.017) & (0.007) & (0.010) & (0.007) \\
\hline
observations & 15697 & 15957 & 16126 & 16263 & 16343 & 16418 & 16472 & 16500 & 16521 & 16549 \\
R$^2$ & 0.003 & 0.007 & 0.013 & 0.014 & 0.012 & 0.001 & 0.006 & 0.018 & 0.009 & 0.017 \\
adjusted R$^2$ & 0.002 & 0.007 & 0.013 & 0.014 & 0.012 & 0.001 & 0.006 & 0.018 & 0.009 & 0.017 \\
\bottomrule
\end{tabularx}
\end{table}
\begin{table}[ht]
\caption{Secondary roads}
\label{tab:road2}
\begin{tabularx}{\textwidth}{LLLLLLLLLLL}
\toprule
\multicolumn{11}{C}{dependent variable: $BCR^{\text{Secondary roads}}$} \\
 & $\gamma = 1$ & $\gamma = 2$ & $\gamma = 3$ & $\gamma = 4$ & $\gamma = 5$ & $\gamma = 6$ & $\gamma = 7$ & $\gamma = 8$ & $\gamma = 9$ & $\gamma = 10$ \\
\midrule
intercept & 2.358$^{\ast\ast\ast}$ & 1.044$^{\ast\ast\ast}$ & 0.093 & -0.029 & 0.271$^{\ast}$ & 0.171 & 0.137 & 0.256$^{\ast\ast}$* & 0.300$^{\ast\ast\ast}$ & 0.192$^{\ast}$ \\
 & (0.237) & (0.217) & (0.199) & (0.428) & (0.143) & (0.125) & (0.131) & (0.129) & (0.102) & (0.105) \\
distance (log10) & -0.169$^{\ast\ast\ast}$ & 0.128$^{\ast\ast}$* & 0.345$^{\ast\ast\ast}$ & 0.367$^{\ast\ast\ast}$ & 0.269$^{\ast\ast\ast}$ & 0.287$^{\ast\ast\ast}$ & 0.288$^{\ast\ast\ast}$ & 0.252$^{\ast\ast\ast}$ & 0.236$^{\ast\ast\ast}$ & 0.261$^{\ast\ast\ast}$ \\
 & (0.062) & (0.057) & (0.052) & (0.113) & (0.038) & (0.033) & (0.034) & (0.034) & (0.027) & (0.028) \\
complexity & -0.087$^{\ast\ast\ast}$ & -0.224$^{\ast\ast\ast}$ & -0.142$^{\ast\ast\ast}$ & -0.188$^{\ast\ast\ast}$ & -0.150$^{\ast\ast\ast}$ & -0.156$^{\ast\ast\ast}$ & -0.143$^{\ast\ast\ast}$ & -0.156$^{\ast\ast\ast}$ & -0.151$^{\ast\ast\ast}$ & -0.143$^{\ast\ast\ast}$ \\
 & (0.020) & (0.018) & (0.017) & (0.036) & (0.012) & (0.010) & (0.011) & (0.011) & (0.009) & (0.009) \\
\hline
observations & 15747 & 15951 & 16139 & 16248 & 16339 & 16404 & 16453 & 16473 & 16501 & 16525 \\
R$^2$ & 0.001 & 0.012 & 0.010 & 0.003 & 0.017 & 0.024 & 0.020 & 0.021 & 0.031 & 0.029 \\
adjusted R$^2$ & 0.001 & 0.011 & 0.010 & 0.003 & 0.017 & 0.024 & 0.020 & 0.021 & 0.031 & 0.028 \\
\bottomrule
\end{tabularx}
\end{table}
\begin{table}[ht]
\caption{Districts}
\label{tab:districts}
\begin{tabularx}{\textwidth}{LLLLLLLLLLL}
\toprule
\multicolumn{11}{C}{dependent variable: $BCR^{\text{Districts}}$} \\
 & $\gamma = 1$ & $\gamma = 2$ & $\gamma = 3$ & $\gamma = 4$ & $\gamma = 5$ & $\gamma = 6$ & $\gamma = 7$ & $\gamma = 8$ & $\gamma = 9$ & $\gamma = 10$ \\
\midrule
intercept & 3.022$^{\ast\ast\ast}$ & 1.132$^{\ast\ast\ast}$ & -0.097 & 0.092 & 0.191 & 0.052 & -0.007 & 0.143 & 0.105 & 0.065 \\
 & (0.275) & (0.205) & (0.170) & (0.159) & (0.139) & (0.133) & (0.130) & (0.114) & (0.115) & (0.109) \\
distance (log10) & -0.225$^{\ast\ast\ast}$ & 0.191$^{\ast\ast\ast}$ & 0.469$^{\ast\ast\ast}$ & 0.395$^{\ast\ast\ast}$ & 0.346$^{\ast\ast\ast}$ & 0.374$^{\ast\ast\ast}$ & 0.378$^{\ast\ast\ast}$ & 0.328$^{\ast\ast\ast}$ & 0.335$^{\ast\ast\ast}$ & 0.339$^{\ast\ast\ast}$ \\
 & (0.072) & (0.054) & (0.045) & (0.042) & (0.037) & (0.035) & (0.034) & (0.030) & (0.030) & (0.029) \\
complexity & -0.113$^{\ast\ast\ast}$ & -0.335$^{\ast\ast\ast}$ & -0.201$^{\ast\ast\ast}$ & -0.242$^{\ast\ast\ast}$ & -0.209$^{\ast\ast\ast}$ & -0.220$^{\ast\ast\ast}$ & -0.217$^{\ast\ast\ast}$ & -0.218$^{\ast\ast\ast}$ & -0.220$^{\ast\ast\ast}$ & -0.215$^{\ast\ast\ast}$ \\
 & (0.023) & (0.017) & (0.014) & (0.013) & (0.012) & (0.011) & (0.011) & (0.009) & (0.010) & (0.009) \\
\hline
observations & 15831 & 16035 & 16202 & 16321 & 16400 & 16467 & 16516 & 16538 & 16562 & 16585 \\
R$^2$ & 0.002 & 0.029 & 0.026 & 0.034 & 0.033 & 0.041 & 0.042 & 0.050 & 0.050 & 0.055 \\
adjusted R$^2$ & 0.002 & 0.028 & 0.026 & 0.034 & 0.033 & 0.040 & 0.042 & 0.050 & 0.050 & 0.054 \\
\bottomrule
\end{tabularx}
\end{table}
\begin{table}[ht]
\caption{Neighborhoods}
\label{tab:neighborhoods}
\begin{tabularx}{\textwidth}{LLLLLLLLLLL}
\toprule
\multicolumn{11}{C}{dependent variable: $BCR^{\text{Neighborhoods}}$} \\
 & $\gamma = 1$ & $\gamma = 2$ & $\gamma = 3$ & $\gamma = 4$ & $\gamma = 5$ & $\gamma = 6$ & $\gamma = 7$ & $\gamma = 8$ & $\gamma = 9$ & $\gamma = 10$ \\
\midrule
intercept & 1.466$^{\ast\ast\ast}$ & 0.404$^{\ast}$ & -0.522$^{\ast\ast\ast}$ & -0.209 & 0.035 & -0.071 & -0.044 & 0.201$^{\ast}$ & 0.183 & 0.162 \\
 & (0.266) & (0.219) & (0.182) & (0.179) & (0.161) & (0.158) & (0.138) & (0.115) & (0.118) & (0.114) \\
distance (log10) & 0.150$^{\ast\ast}$* & 0.359$^{\ast\ast\ast}$ & 0.566$^{\ast\ast\ast}$ & 0.462$^{\ast\ast\ast}$ & 0.376$^{\ast\ast\ast}$ & 0.396$^{\ast\ast\ast}$ & 0.376$^{\ast\ast\ast}$ & 0.301$^{\ast\ast\ast}$ & 0.303$^{\ast\ast\ast}$ & 0.304$^{\ast\ast\ast}$ \\
 & (0.070) & (0.058) & (0.048) & (0.047) & (0.042) & (0.042) & (0.036) & (0.030) & (0.031) & (0.030) \\
complexity & -0.048$^{\ast\ast}$* & -0.244$^{\ast\ast\ast}$ & -0.155$^{\ast\ast\ast}$ & -0.202$^{\ast\ast\ast}$ & -0.172$^{\ast\ast\ast}$ & -0.190$^{\ast\ast\ast}$ & -0.187$^{\ast\ast\ast}$ & -0.193$^{\ast\ast\ast}$ & -0.195$^{\ast\ast\ast}$ & -0.193$^{\ast\ast\ast}$ \\
 & (0.023) & (0.018) & (0.015) & (0.015) & (0.013) & (0.013) & (0.012) & (0.010) & (0.010) & (0.009) \\
\hline
observations & 15831 & 16035 & 16202 & 16321 & 16400 & 16467 & 16516 & 16538 & 16562 & 16585 \\
R$^2$ & 0.001 & 0.018 & 0.021 & 0.023 & 0.020 & 0.025 & 0.030 & 0.040 & 0.038 & 0.041 \\
adjusted R$^2$ & 0.001 & 0.018 & 0.020 & 0.023 & 0.020 & 0.025 & 0.030 & 0.039 & 0.038 & 0.041 \\
\bottomrule
\end{tabularx}
\end{table}
\begin{table}[ht]
\caption{River}
\label{tab:river}
\begin{tabularx}{\textwidth}{LLLLLLLLLLL}
\toprule
\multicolumn{11}{C}{dependent variable: $BCR^{\text{River}}$} \\
 & $\gamma = 1$ & $\gamma = 2$ & $\gamma = 3$ & $\gamma = 4$ & $\gamma = 5$ & $\gamma = 6$ & $\gamma = 7$ & $\gamma = 8$ & $\gamma = 9$ & $\gamma = 10$ \\
\midrule
intercept & -0.064 & -0.328$^{\ast\ast\ast}$ & -0.344$^{\ast\ast\ast}$ & -0.397$^{\ast\ast\ast}$ & -0.404$^{\ast\ast\ast}$ & -0.381$^{\ast\ast\ast}$ & -0.375$^{\ast\ast\ast}$ & -0.334$^{\ast\ast\ast}$ & -0.328$^{\ast\ast\ast}$ & -0.329$^{\ast\ast\ast}$ \\
 & (0.106) & (0.078) & (0.101) & (0.061) & (0.063) & (0.061) & (0.064) & (0.061) & (0.063) & (0.065) \\
distance (log10) & 0.242$^{\ast\ast\ast}$ & 0.299$^{\ast\ast\ast}$ & 0.299$^{\ast\ast\ast}$ & 0.300$^{\ast\ast\ast}$ & 0.300$^{\ast\ast\ast}$ & 0.293$^{\ast\ast\ast}$ & 0.291$^{\ast\ast\ast}$ & 0.280$^{\ast\ast\ast}$ & 0.278$^{\ast\ast\ast}$ & 0.278$^{\ast\ast\ast}$ \\
 & (0.028) & (0.021) & (0.027) & (0.016) & (0.017) & (0.016) & (0.017) & (0.016) & (0.017) & (0.017) \\
complexity & 0.102$^{\ast\ast\ast}$ & 0.092$^{\ast\ast\ast}$ & 0.117$^{\ast\ast\ast}$ & 0.121$^{\ast\ast\ast}$ & 0.127$^{\ast\ast\ast}$ & 0.125$^{\ast\ast\ast}$ & 0.126$^{\ast\ast\ast}$ & 0.126$^{\ast\ast\ast}$ & 0.122$^{\ast\ast\ast}$ & 0.124$^{\ast\ast\ast}$ \\
 & (0.009) & (0.007) & (0.009) & (0.005) & (0.005) & (0.005) & (0.005) & (0.005) & (0.005) & (0.005) \\
\hline
observations & 15788 & 15994 & 16102 & 16311 & 16390 & 16457 & 16506 & 16525 & 16548 & 16574 \\
R$^2$ & 0.010 & 0.019 & 0.015 & 0.042 & 0.042 & 0.044 & 0.040 & 0.043 & 0.038 & 0.037 \\
adjusted R$^2$ & 0.010 & 0.019 & 0.015 & 0.042 & 0.042 & 0.043 & 0.040 & 0.043 & 0.037 & 0.037 \\
\bottomrule
\end{tabularx}
\end{table}
\begin{table}[ht]
\caption{Railways}
\label{tab:railways}
\begin{tabularx}{\textwidth}{LLLLLLLLLLL}
\toprule
\multicolumn{11}{C}{dependent variable: $BCR^{\text{Railways}}$} \\
 & $\gamma = 1$ & $\gamma = 2$ & $\gamma = 3$ & $\gamma = 4$ & $\gamma = 5$ & $\gamma = 6$ & $\gamma = 7$ & $\gamma = 8$ & $\gamma = 9$ & $\gamma = 10$ \\
\midrule
intercept & 0.004 & -0.021 & -0.548$^{\ast\ast\ast}$ & -0.450$^{\ast\ast\ast}$ & -0.081 & -0.175 & -0.157$^{\ast}$ & 0.345$^{\ast\ast\ast}$ & 0.426$^{\ast\ast\ast}$ & 0.326$^{\ast\ast\ast}$ \\
 & (0.217) & (0.183) & (0.187) & (0.151) & (0.121) & (0.137) & (0.089) & (0.076) & (0.078) & (0.070) \\
distance (log10) & 0.404$^{\ast\ast\ast}$ & 0.376$^{\ast\ast\ast}$ & 0.499$^{\ast\ast\ast}$ & 0.453$^{\ast\ast\ast}$ & 0.334$^{\ast\ast\ast}$ & 0.357$^{\ast\ast\ast}$ & 0.341$^{\ast\ast\ast}$ & 0.199$^{\ast\ast\ast}$ & 0.175$^{\ast\ast\ast}$ & 0.198$^{\ast\ast\ast}$ \\
 & (0.057) & (0.048) & (0.049) & (0.040) & (0.032) & (0.036) & (0.023) & (0.020) & (0.020) & (0.018) \\
complexity & -0.128$^{\ast\ast\ast}$ & -0.188$^{\ast\ast\ast}$ & -0.138$^{\ast\ast\ast}$ & -0.127$^{\ast\ast\ast}$ & -0.095$^{\ast\ast\ast}$ & -0.100$^{\ast\ast\ast}$ & -0.097$^{\ast\ast\ast}$ & -0.108$^{\ast\ast\ast}$ & -0.112$^{\ast\ast\ast}$ & -0.107$^{\ast\ast\ast}$ \\
 & (0.019) & (0.015) & (0.016) & (0.013) & (0.010) & (0.011) & (0.007) & (0.006) & (0.006) & (0.006) \\
\hline
observations & 15697 & 15923 & 16117 & 16259 & 16347 & 16428 & 16465 & 16506 & 16536 & 16559 \\
R$^2$ & 0.009 & 0.018 & 0.015 & 0.019 & 0.017 & 0.015 & 0.032 & 0.031 & 0.029 & 0.036 \\
adjusted R$^2$ & 0.009 & 0.018 & 0.015 & 0.019 & 0.017 & 0.015 & 0.032 & 0.031 & 0.029 & 0.036 \\
\bottomrule
\end{tabularx}
\end{table}

\endgroup






\clearpage
\section{YJMob100K}
\label{si:sec:yjmob100k}

\begin{figure}[th]
    \centering
    \begin{subfigure}[t]{0.32\linewidth}
        \includegraphics[width=\linewidth]{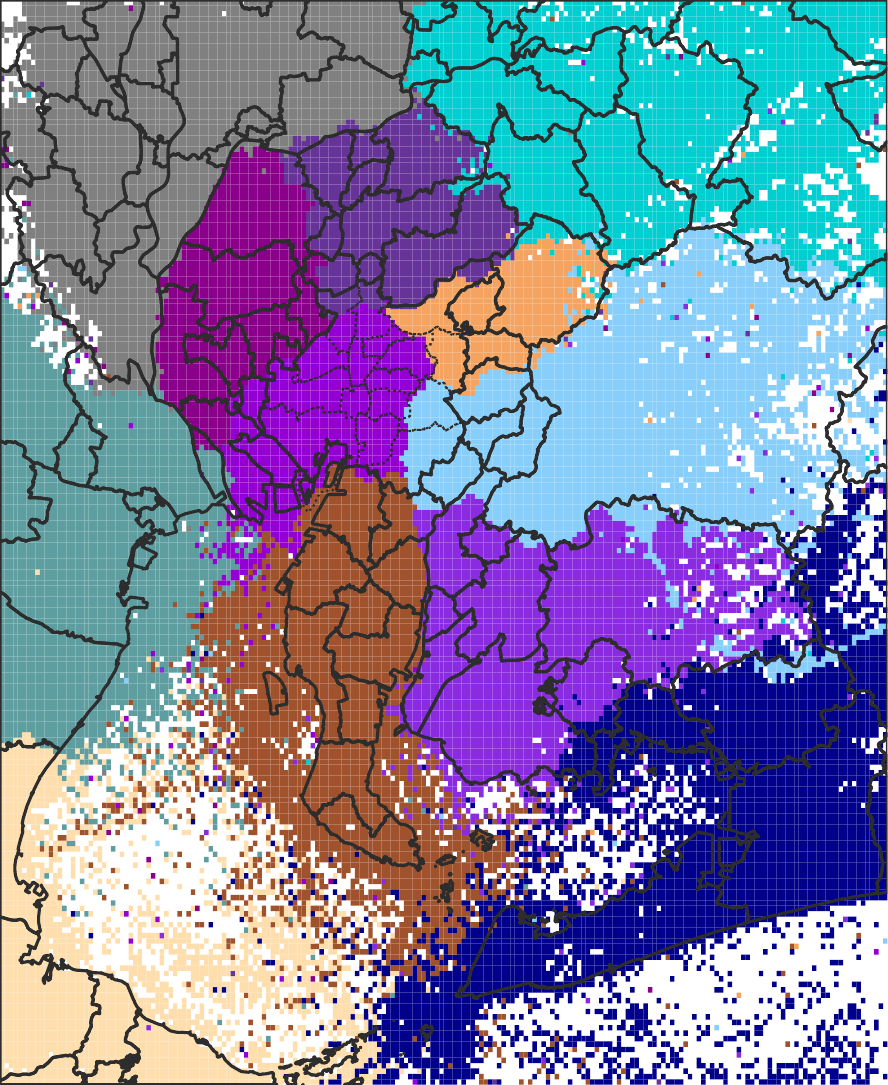}
        \captionsetup{position=bottom,justification=centering}
        \caption{}
        \label{si:fig:nagoya_communities_res_1}
    \end{subfigure}
    \enspace
    \begin{subfigure}[t]{0.32\linewidth}
        \includegraphics[width=\linewidth]{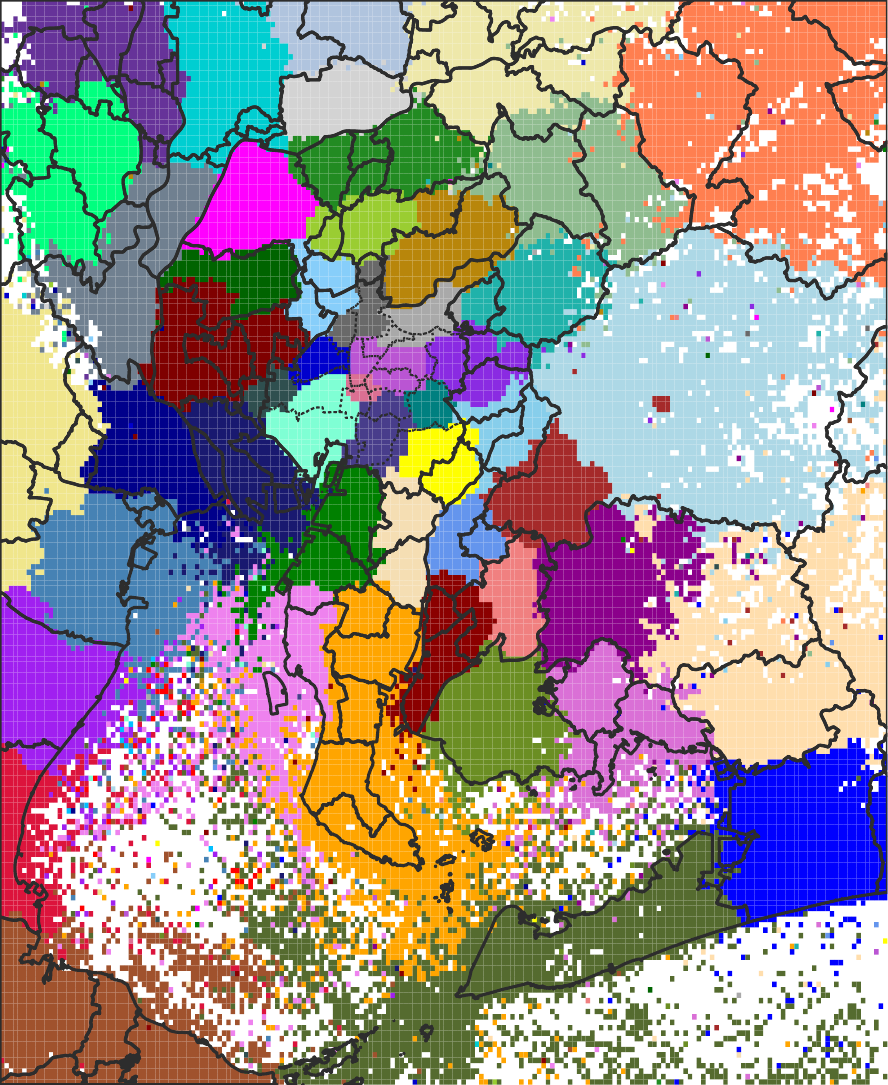}
        \captionsetup{position=bottom,justification=centering}
        \caption{}
        \label{si:fig:nagoya_communities_res_5}
    \end{subfigure}
    \enspace
    \begin{subfigure}[t]{0.32\linewidth}
        \includegraphics[width=\linewidth]{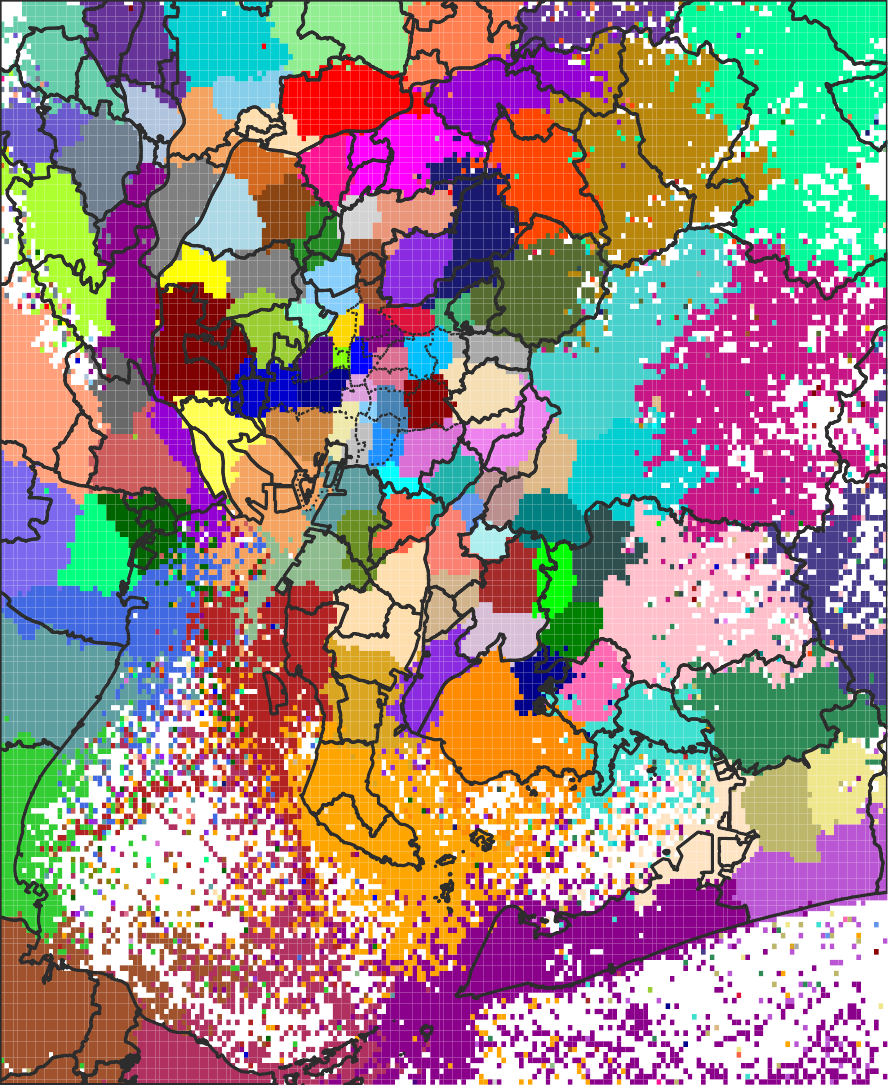}
        \captionsetup{position=bottom,justification=centering}
        \caption{}
        \label{si:fig:nagoya_communities_res_12}
    \end{subfigure}

    \begin{subfigure}[t]{0.32\linewidth}
        \includegraphics[width=\linewidth]{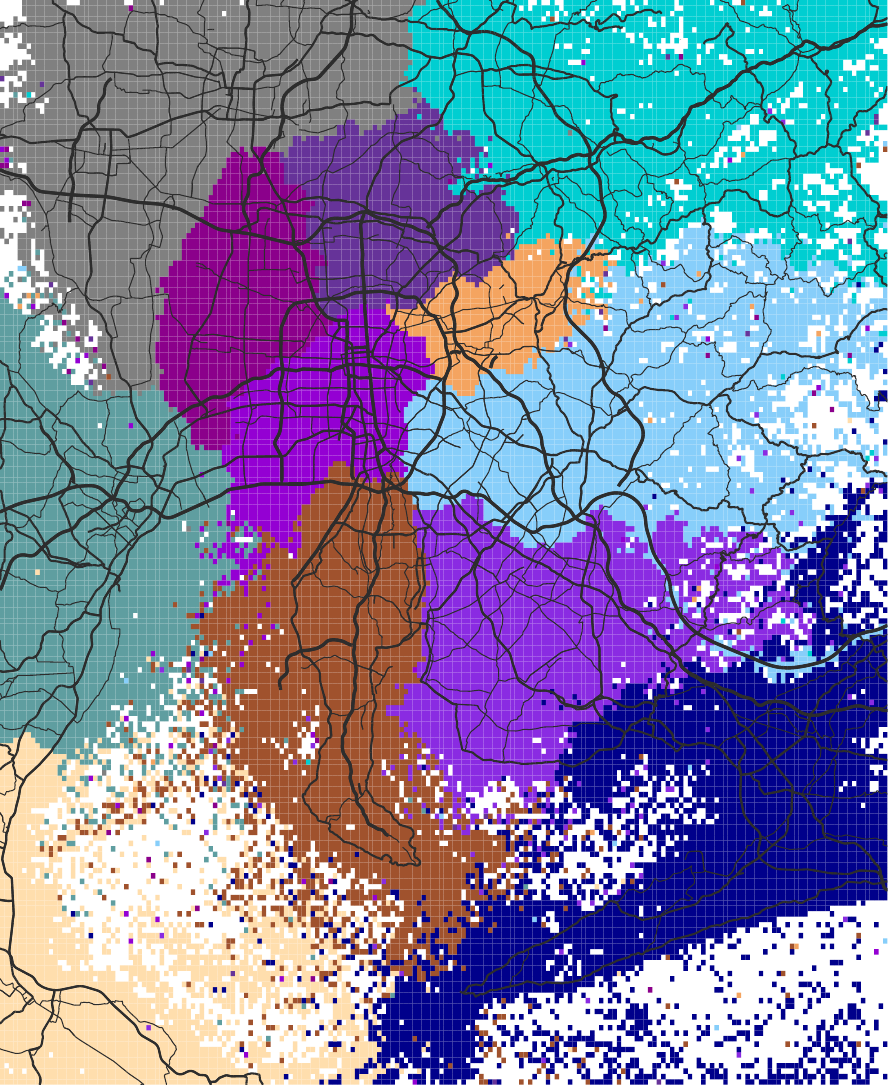}
        \captionsetup{position=bottom,justification=centering}
        \caption{}
        \label{si:fig:nagoya_communities_roads_res_1}
    \end{subfigure}
    \enspace
    \begin{subfigure}[t]{0.32\linewidth}
        \includegraphics[width=\linewidth]{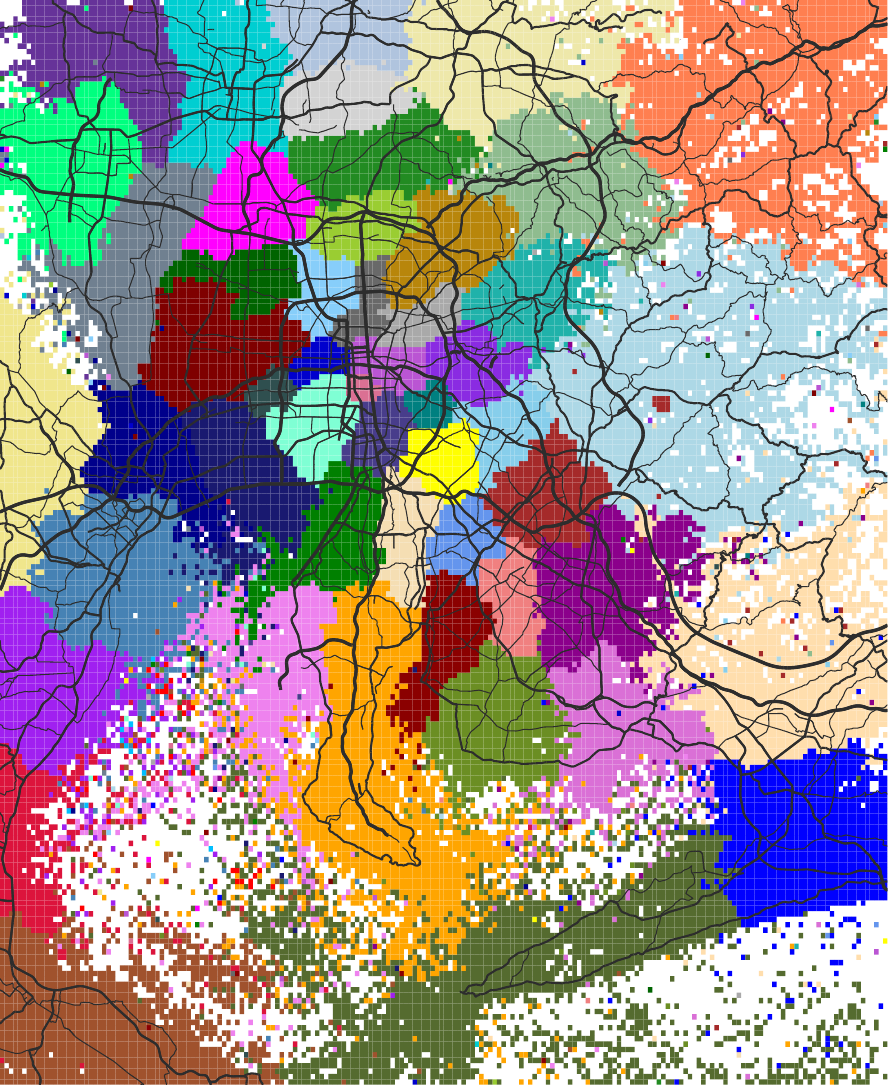}
        \captionsetup{position=bottom,justification=centering}
        \caption{}
        \label{si:fig:nagoya_communities_roads_res_5}
    \end{subfigure}
    \enspace
    \begin{subfigure}[t]{0.32\linewidth}
        \includegraphics[width=\linewidth]{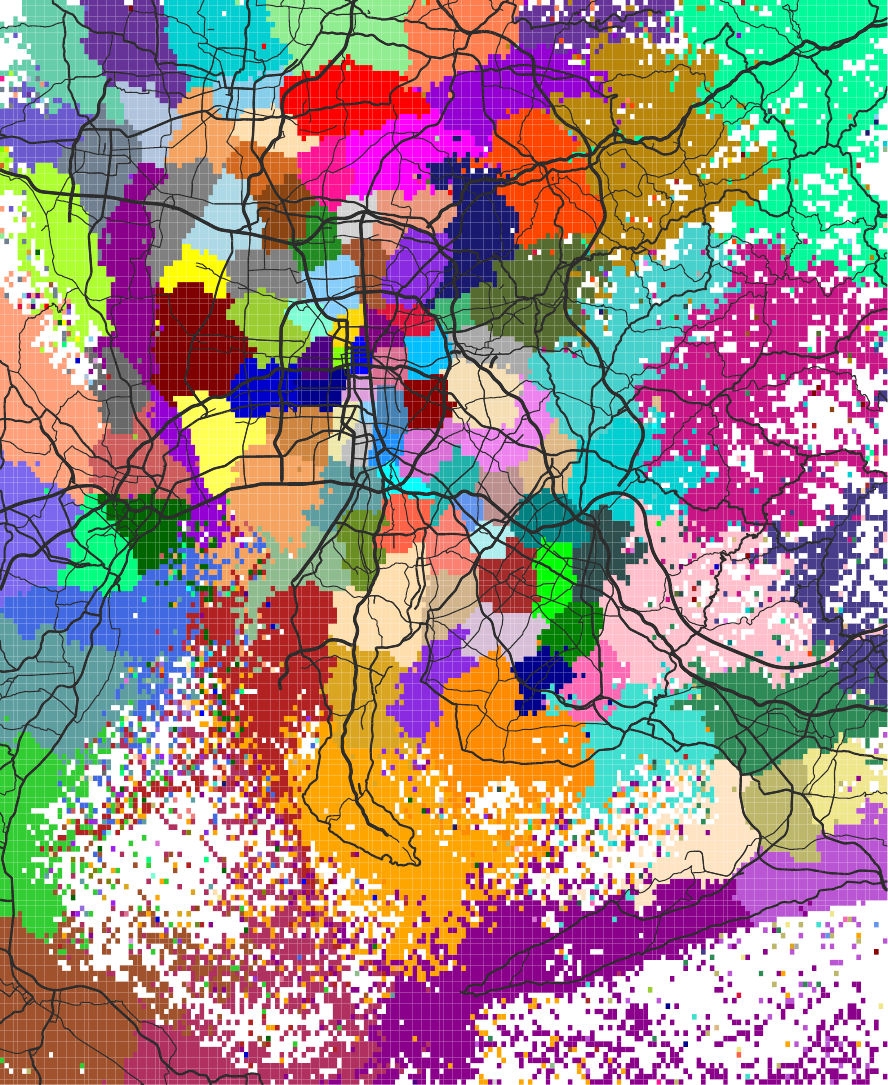}
        \captionsetup{position=bottom,justification=centering}
        \caption{}
        \label{si:fig:nagoya_communities_roads_res_12}
    \end{subfigure}

    \begin{subfigure}[t]{0.32\linewidth}
        \includegraphics[width=\linewidth]{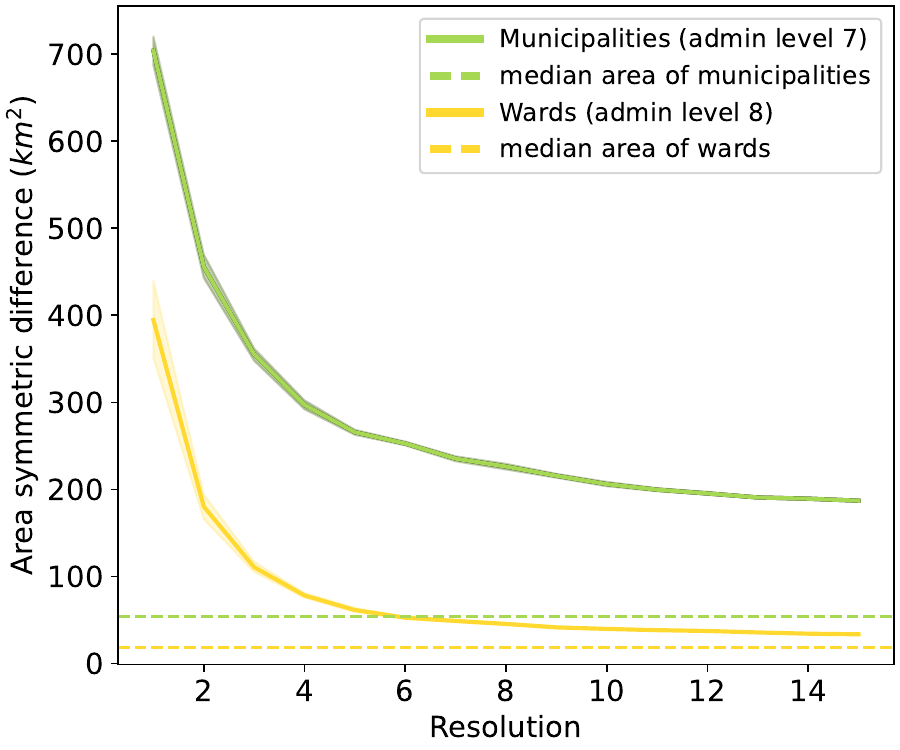}
        \captionsetup{position=bottom,justification=centering}
        \caption{}
        \label{si:fig:yjmob100k_sad}
    \end{subfigure}
    \enspace
    \begin{subfigure}[t]{0.32\linewidth}
        \includegraphics[width=\linewidth]{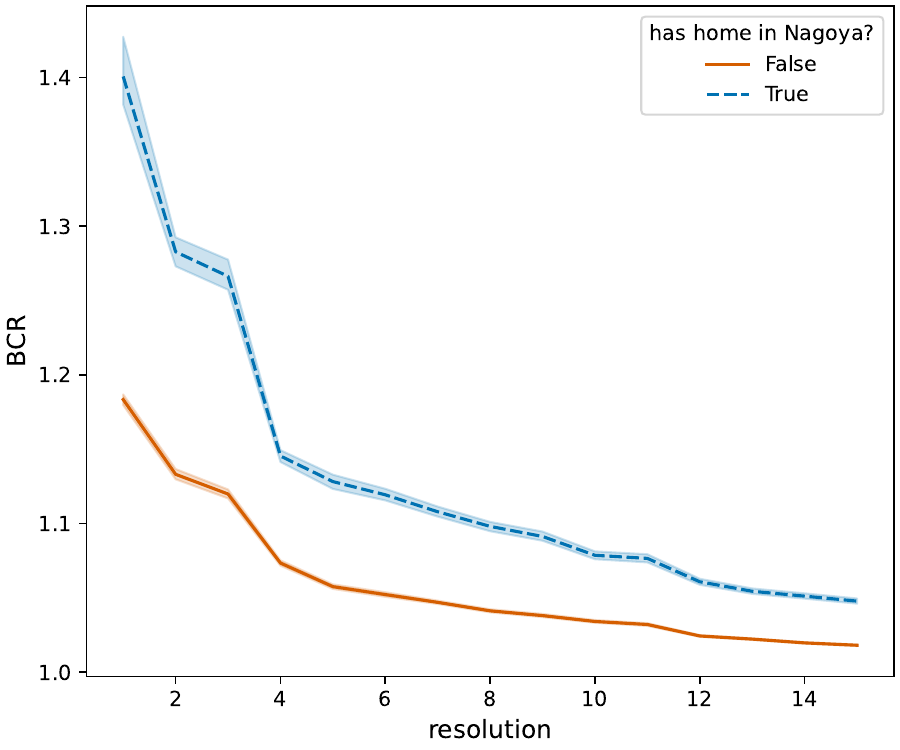}
        \captionsetup{position=bottom,justification=centering}
        \caption{}
        \label{si:fig:yjmob100k_admn7_bcr}
    \end{subfigure}
    \enspace
    \begin{subfigure}[t]{0.32\linewidth}
        \includegraphics[width=\linewidth]{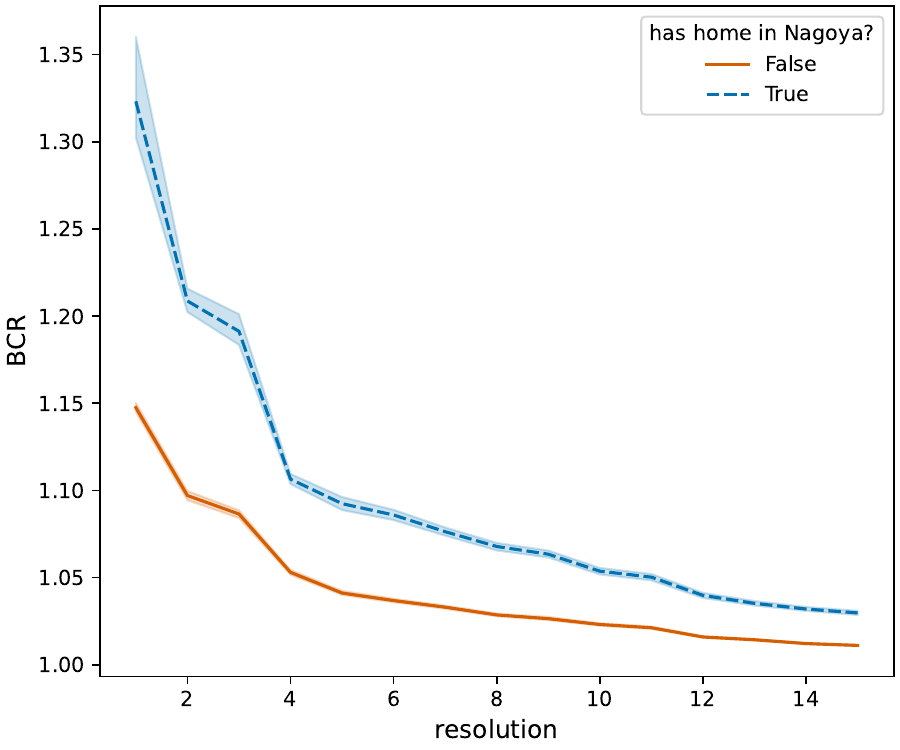}
        \captionsetup{position=bottom,justification=centering}
        \caption{}
        \label{si:fig:yjmob100k_road1_bcr}
    \end{subfigure}
    \caption{
        Louvain communities compared to administrative boundaries at resolution 1 (\textbf{\subref{si:fig:nagoya_communities_res_1}}), 5 (\textbf{\subref{si:fig:nagoya_communities_res_5}}), and 12 (\textbf{\subref{si:fig:nagoya_communities_res_12}}), and compared to the road network at resolution 1,5, and 12 (\textbf{\subref{si:fig:nagoya_communities_roads_res_1}}, \textbf{\subref{si:fig:nagoya_communities_roads_res_5}}, and \textbf{\subref{si:fig:nagoya_communities_roads_res_12}} respectively).
        The decreasing \acrlong{SAD} shows an improving fit of mobility clusters to the administrative boundaries (\textbf{\subref{si:fig:yjmob100k_sad}}).
        The individual \acrlong{BCR} shows that the barriers like municipality boundaries (\textbf{\subref{si:fig:yjmob100k_admn7_bcr}}) and primary roads (\textbf{\subref{si:fig:yjmob100k_road1_bcr}}) affect people less if they have home locations in Nagoya compared to those who live in its agglomeration.
        }
    \label{si:fig:yjmob100k_communities}
\end{figure}

A metropolitan scale, longitudinal, and anonymized mobility trajectory data was released \cite{yabe2024yjmob100k}, which aims to be a benchmark dataset of human mobility \cite{yabe2024enhancing}.
The presented approach was applied to the `YJMob100K' data set, which already contains trajectories in their raw form.
The exact procedure that made it possible to use the `YJMob100K' data for this study can be found in \cite{pinter2024revealing}.
The difference is that the trajectories connect 500 meters by 500 meters grid cells, instead of blocks extracted from the road network.
Figure~\ref{si:fig:nagoya_communities_res_5} and \ref{si:fig:nagoya_communities_res_12} illustrate the results of the community detection procedure at resolution ($\gamma$) $1$, $5$, and $12$ in contrast of the municipalities (solid line) and wards (dotted line) extracted from \acrfull{OSM}.
The mobility clusters match the administrative boundaries, but a community often matches multiple municipalities.
Interestingly, this effect does not apply to the road network (Figure~\ref{si:fig:nagoya_communities_roads_res_1}, \ref{si:fig:nagoya_communities_roads_res_5}, \ref{si:fig:nagoya_communities_roads_res_12}) as in the case of Budapest (Figure~\ref{fig:stream} of the main text).
Note that the water surfaces are excluded from the municipality boundaries.
However, activities occur there (e.g., people using ferries), which can be observed in the data, so some of those cells are associated with communities.

The \acrfull{SAD} is calculated (Figure~\ref{si:fig:yjmob100k_sad}) compared to the municipality boundaries (admin level 7 in \acrshort{OSM}) and the wards of Nagoya (admin level 8 in \acrshort{OSM}).
In both cases, the \acrshort{SAD} approximates the administrative boundaries as the resolution increases.

The individual \acrfull{BCR} is also determined based on the `YJMob100K' data.
Similarly to the Budapest case, people with estimated home locations within Nagoya (\num{18699} individuals) are distinguished from the rest of the users.
Just as in the case of Budapest, the two groups show different exposure to the barrier effect (Figure~\ref{si:fig:yjmob100k_admn7_bcr} and \ref{si:fig:yjmob100k_road1_bcr}), however, as opposed to Budapest the relation of the two trends are the opposite.

In this section, we demonstrated our approach on a different data set covering a much larger area from a different part of the world.
The `YJMob100K' data set follows \num{100000} people in about a 100 km by 100 km area, while Budapest and its agglomeration is about \num{2540} km$^2$.
The administrative area of Nagoya as a principal city is \num{326.45} km$^2$, which is about the \num{62}\% of Budapest (\num{525.14} km$^2$), however, Nagoya is the center of the Chūkyō metropolitan area with the area of \num{7072} km$^2$.
In our study, we focused on mobility that took place within the administrative boundaries of Budapest, a smaller area but with more users.
Due to the high number of differences not every result shows exactly the expected finding.
Further study is required, involving data about more cities, to explore the barrier effect in more types of cities.

    \clearpage
    \setglossarystyle{list}
    \printglossary[title=Abbreviations, toctitle=Abbreviations, nogroupskip=true, nonumberlist, nopostdot]

    \printbibliography[title=Supplementary References]
\end{refsection}

\end{document}